\pdfoutput=1
\documentclass{article}
\usepackage{amsmath}
\usepackage{amssymb}
\usepackage{amsfonts}
\usepackage{latexsym}
\usepackage{graphicx}

\begin{document}

\newtheorem{thm}{Theorem}
\newtheorem{lem}[thm]{Lemma}
\newtheorem{cor}[thm]{Corollary}
\newtheorem{ex}{Example}
\newtheorem{deff}[thm]{Definition}
%\newproof{pf}{Proof}

\begin{center}
{\Large Minimum firing times of 
firing squad synchronization problems 
for paths in grid spaces}
\end{center}

\begin{center}
{\large Kojiro Kobayashi}
\end{center}
\vspace*{-2.0em}
\begin{center}
\texttt{kojiro@gol.com}
\end{center}

\begin{center}
{September 5, 2019}
\end{center}

\medskip

\noindent
{\bf Abstract}

\medskip

We consider the firing squad synchronization problems 
for paths in the two and the three-dimensional 
grid spaces. 
Minimal-time solutions of these problems are not 
known and are unlikely to exist. 
However, at present we have no proofs of their nonexistence. 
In this paper we show one result that suggests 
what type of study is necessary in order to prove 
their nonexistence. 

\medskip


\medskip

\section{Introduction}
\label{section:introduction}

\subsection{The problem and its history}
\label{subsection:problem_and_history}

The firing squad synchronization problem (FSSP) 
is a problem that was proposed by J. Myhill in 1957 and became 
widely known to researchers in automata theory by 
an article by E. F. Moore (\cite{Moore}) that gave a concise description of 
the problem.

\begin{figure}[htbp]
\begin{center}
\includegraphics[scale=1.0]{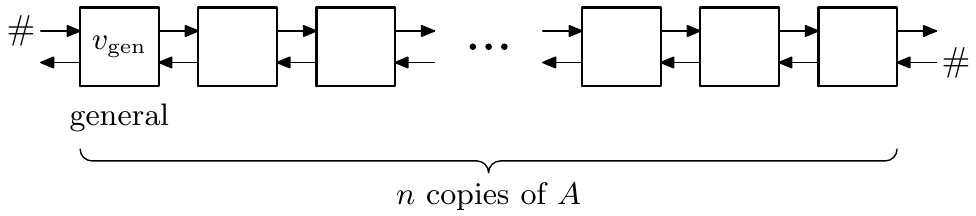}
\end{center}
\caption{A network $C_{n}$ consisting of $n$ copies of 
a finite automaton $A$.}
\label{figure:fig003}
\end{figure}

The problem is to design a finite automaton $A$.  $A$ has 
two inputs (one from the left and another from the right) and 
two outputs (one to the left and another to the right).  
The value of each output of $A$ at a time is the state of 
$A$ at the time.
The set of the states of $A$ contains 
at least three distinct states: 
the ``general'' state {\rm G}, the quiescent state ${\rm Q}$, 
and the firing state ${\rm F}$.
For each $n$, let $C_{n}$ denote the network of $n$ copies 
of $A$ that are connected as shown in Fig. \ref{figure:fig003}.  
We call each copy of $A$ of $C_{n}$ a {\it node}\/ of $C_{n}$.  
We call the leftmost node of $C_{n}$ the ``general'' of $C_{n}$.
The value of the left input of the general 
and the right input of the rightmost node are a special symbol ``$\#$'' 
that means that there are no nodes there.
If the state of $A$ is ${\rm Q}$ and the values 
of its two inputs are either ${\rm Q}$ or ``$\#$'' at 
a time $t$, the state of $A$ at time $t+1$ is ${\rm Q}$.
At time $0$ we set the general in the general 
state ${\rm G}$ and all other nodes in the quiescent state ${\rm Q}$.
Then, the state of a node of $C_{n}$ at a time is uniquely 
determined by the state transition function of $A$.  
$A$ must satisfy the condition: for any $n$ there exists a time $t_{n}$ 
(that may depend on $n$) such that any node of $C_{n}$ enters 
the firing state ${\rm F}$ at the time $t_{n}$ for the first time.
We call a finite automaton $A$ satisfying the above condition 
a {\it solution}\/ of FSSP, and call the time $t_{n}$ 
the {\it firing time}\/ of the solution $A$ for $C_{n}$.

The problem to find a solution of FSSP 
is a typical example of applications of 
``divide and conquer'' programming 
technique and a student who knows it 
usually finds the basic idea for 
a solution within a few hours.
It needs much more time to determine necessary states and 
define the state transition function of the solution.

The original FSSP and many of its variations have been 
extensively studied 
and the study has focused mainly on two problems:
finding solutions with small firing time and 
finding ones with small numbers of states.

FSSP and its variations are practically important because 
we can use their solutions to synchronize 
large-scale networks consisting of identical processors 
quickly.
They are also theoretically important because they are 
a mathematical formulation of one of the most basic 
protocols of distributed computing, that is, 
to realize global synchronization using only 
local information exchanges.
In this paper we study how small the firing time of solutions of FSSP 
can be.
By this we study the inherent limit of the time necessary 
for this protocol.

By $v_{\rm gen}$ we denote the general node of $C_{n}$.
For a node $v$ in $C_{n}$ and a time $t$, by 
${\rm st}(v, t, C_{n}, A)$ we denote the state of 
the node $v$ at the time $t$ when copies of $A$ are placed 
at nodes of $C_{n}$.
Using this notation, the condition for $A$ to be a solution 
is expressed as follows.
\begin{quote}
For any $n$ 
there exists a time $t_{n}$ such that, 
for any $v \in C_{n}$, ${\rm st}(v, t, C_{n}, A)$ $\not= {\rm F}$ 
for $t < t_{n}$ and ${\rm st}(v, t_{n}, C_{n}, A) = {\rm F}$.
\end{quote}
By ${\rm ft}(C_{n}, A)$ we denote the firing time $t_{n}$.
We also say that the solution $A$ fires $v$ at time $t_{n}$.

As we mentioned previously, to construct a solution is easy 
and we usually construct a solution $A$ such that 
${\rm ft}(C_{n}, A) = 3n + O(\log n)$ (\cite{Minsky}).
An interesting problem is how small ${\rm ft}(C_{n}, A)$ can be.

We can show a lower bound ${\rm ft}(C_{n}, A) \geq 2n - 2$.
Intuitively the general cannot know the position of the rightmost 
node before time $2n - 2$ (the time for a signal to go from the 
general to the rightmost node and go back to the general) and 
cannot fire before that time.
We can modify this intuitive argument to a rigorous proof of the 
lower bound.

In 1962 Goto constructed a solution $\tilde{A}$ with the firing 
time ${\rm ft}(C_{n}, \tilde{A}) = 2n - 2$ (\cite{Goto_1962}).
Using the above mentioned lower bound we know that the firing 
time of $\tilde{A}$ is less than or equal to the firing time 
of any solution $A$ for any $C_{n}$, that is, 
\begin{equation}
(\forall A) (\forall C_{n}) [ {\rm ft}(C_{n}, \tilde{A}) 
\leq {\rm ft}(C_{n}, A) ].
\label{equation:eq000}
\end{equation}
We call a solution $\tilde{A}$ that satisfies the condition 
(\ref{equation:eq000}) a {\it minimal-time solution}\/.
Hence the Goto's solution is a minimal-time solution and 
the problem to find a best solution with respect to 
firing time was essentially solved.
Later Waksman (\cite{Waksman_1966}) and 
Balzer (\cite{Balzer_1967}) constructed minimal-time solutions 
with different ideas.
Still remains the problem to construct simpler solutions 
(solutions with small number of states, or easy to understand).

The FSSP has many variations.  
For a variation $\Gamma$ of FSSP, we call a problem instance of $\Gamma$ 
(a network in which copies of a finite automaton are placed) 
a {\it configuration}\/ of $\Gamma$.
One of the variations is the {\it generalized FSSP}\/.  
A configuration of this variation is a linear array of nodes 
of the form shown in Fig. \ref{figure:fig003} and 
the general may be an arbitrary node of the array.
For this variation too a minimal-time solution 
was found (\cite{Moore_Langdon_1968}) and its firing time is 
\begin{equation}
\max \{ 2n - 2 - i, n - 1 + i \}, 
\label{equation:eq001}
\end{equation}
where $i$ ($0 \leq i \leq n-1$) is the position of the general 
(the leftmost node has the number $0$).

Examples of other variations are FSSP for rings, 
one-way rings, squares, rectangles, cubes, cuboids 
(rectangular parallelepipeds), undirected networks 
and directed networks.
For the variations for squares, rectangles, cubes, 
and cuboids, the general may be 
either one of the corner nodes or an arbitrary node.
For many of these variations minimal-time 
solutions are known.
For surveys on these results 
we refer readers to 
\cite{Goldstein_Kobayashi_SIAM_2005,Goldstein_Kobayashi_SIAM_2012,
Kobayashi_TCS_2014,Mazoyer_Survey,Napoli_Parente,
Umeo_2005_Survey,Umeo_2012_ACRI}.

There are several variations for which we do not know minimal-time 
solutions.
For some of them the author has no intuitive feeling 
as for existence or nonexistence of minimal-time solutions.
One example is the variation such that configurations are squares and 
the general may be an arbitrary node 
(\cite{Kobayashi_TCS_2014, Umeo_2012_ACRI}).
However, there are also variations that seem to have no 
minimal-time solutions 
(\cite{Goldstein_Kobayashi_SIAM_2005, 
Goldstein_Kobayashi_SIAM_2012, Kobayashi_TCS_2001}).
The problems we consider in this paper are such variations.
Before we explain them, we need a definition of 
minimal-time solutions of a variation $\Gamma$ that is 
equivalent to the one shown previously but is more useful 
than that for our purpose.

A minimal-time solution of a variation $\Gamma$ of FSSP 
was a solution 
$\tilde{A}$ of $\Gamma$ such that
\begin{equation}
(\forall A) (\forall C) [ {\rm ft}(C, \tilde{A}) \leq 
{\rm ft}(C, A)],
\label{equation:eq002}
\end{equation}
where $A$ and $C$ range over all solutions and all 
configurations, respectively 
(see (\ref{equation:eq000})).

Suppose that the variation $\Gamma$ has a solution.
For each configuration $C$ of $\Gamma$, we define 
the {\it minimum firing time}\/ ${\rm mft}_{\Gamma}(C)$ 
of $C$ by 
\begin{equation}
{\rm mft}_{\Gamma}(C) = \min_{A} {\rm ft}(C, A),
\label{equation:eq003}
\end{equation}
where $A$ ranges over all solutions of $\Gamma$.
This value is well-defined because the variation $\Gamma$ 
has at least one solution.
We may define a minimal-time solution of $\Gamma$ to 
be a solution $\tilde{A}$ of $\Gamma$ such that 
\begin{equation}
(\forall C) [ {\rm ft}(C, \tilde{A}) = {\rm mft}_{\Gamma}(C) ], 
\label{equation:eq004}
\end{equation}
where $C$ ranges over all configurations of $\Gamma$.

We can show that the two definitions of minimal-time solutions 
by (\ref{equation:eq002}) and (\ref{equation:eq004}) 
are equivalent.
Suppose that $\tilde{A}$ is a minimal-time solution by 
the definition using (\ref{equation:eq002}).
For any configuration $C$ and solution $A$ 
we have ${\rm ft}(C, \tilde{A}) \leq 
{\rm ft}(C, A)$.
Hence for any configuration $C$ we have 
${\rm ft}(C, \tilde{A}) \leq \min_{A} {\rm ft(C, A)} 
= {\rm mft}_{\Gamma}(C) \leq {\rm ft}(C, \tilde{A})$
and ${\rm ft}(C, \tilde{A}) = {\rm mft}_{\Gamma}(C)$.
Therefore $\tilde{A}$ is a minimal-time solution by 
(\ref{equation:eq004}).
Suppose that $\tilde{A}$ is a minimal-time solution by 
the definition using (\ref{equation:eq004}).
For any configuration $C$ and solution $A$ we have 
${\rm ft}(C, \tilde{A}) = {\rm mft}_{\Gamma}(C) \leq 
{\rm ft}(C, A)$.
Therefore $\tilde{A}$ is a minimal-time solution by 
(\ref{equation:eq002}).

The first definition has the merit that its intuitive 
meaning is clear (``a best solution with respect to 
firing time'').
The second definition is rather technical and its 
intuitive meaning is not clear.
However it has one merit.
It comes from the fact: if a variation $\Gamma$ has 
a minimal-time solution $\tilde{A}$ then 
we can compute the function ${\rm mft}_{\Gamma}(C)$ 
in polynomial time.
To determine the value ${\rm mft}_{\Gamma}(C)$, 
it is sufficient to simulate the behavior of the 
configuration $C$ that 
is given copies of $\tilde{A}$ and see the firing time.
This firing time is, by definition, 
the value ${\rm mft}_{\Gamma}(C)$.
This simulation can be performed in polynomial time of 
$n$ because ${\rm mft}_{\Gamma}(C) = O(n)$ 
($n$ is the number of nodes in $C$).
This fact means that, if we have proved that 
the function ${\rm mft}_{\Gamma}(C)$ of $\Gamma$ cannot be 
computed in polynomial time 
then we have proved that the variation $\Gamma$ has 
no minimal-time solutions.

In this subsection on a survey on FSSP's we considered only variations 
that are modifications of the original FSSP with respect to 
shapes of configurations and the position of the general.
However, if we relax this restriction there are proofs of 
nonexistence of minimal-time solutions for some variations.
Schmid and Worsch (\cite{Schmid_Worsch_IFIP_2004}) 
considered the variation of 
the original FSSP such that there may be more than one general 
and they may be activated independently in different times.
For this variation they showed that there are no 
minimal-time solutions.
(Compare this result with Jiang's result (\cite{Jiang1,Jiang2})
that if more than one general are allowed then FSSP for rings 
has no solutions even if we require all generals to be 
activated simultaneously.)
Yamashita et al (\cite{Yamashita}) considered 
the variation of the original FSSP such that ``sub-generals'' 
are allowed.
A sub-general node behaves as in the quiescent state until 
an active signal from the general arrives.
When an active signal arrives, it can act differently from 
a node in the quiescent state.
They showed that this variation also has no minimal-time solutions.

\subsection{Six variations $k$PATH, g-$k$PATH, $k$REG}
\label{subsection:six_variations}

In this paper we consider the following six variations 
$k$PATH, g-$k$PATH, $k$REG ($k = 2, 3$).

\medskip

\noindent
(1) 2PATH: The variation such that a configuration is a path 
in the two-dimensional grid space and any of the 
two end nodes may be the general.  

\noindent
(2) g-2PATH: The variation such that a configuration is a path 
in the two-dimensional grid space and any node may be 
the general.
(``g'' in g-2PATH is for ``{\it g}\/eneralized'').

\noindent
(3) 2REG: The variation such that a configuration is a 
connected finite region in the two-dimensional grid space and 
any node may be the general.

\noindent
(4 - 6) 3PATH, g-3PATH, 3REG: 
The three-dimensional analogues of 2PATH, 
g-2PATH and 2REG, respectively.

\medskip

\noindent
In the four variations $k$PATH, g-$k$PATH, we require 
a path not to cross and not to touch itself.

\medskip

\begin{figure}[htbp]
\begin{center}
\includegraphics[scale=1.0]{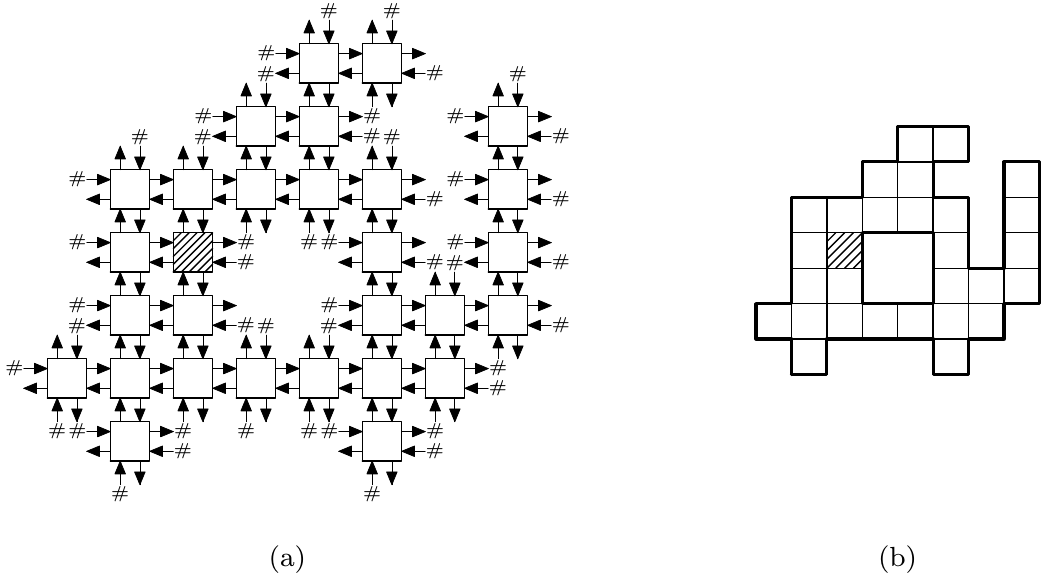}
\end{center}
\caption{(a) A configuration of 2REG. (b) A simplified 
representation of the configuration.}
\label{figure:fig000}
\end{figure}
In Fig. \ref{figure:fig000}(a) we show an example of configurations of 
2REG.  
The hatched box is the general.
We represent this configuration by a simplified figure shown 
in Fig. \ref{figure:fig000}(b).

\begin{figure}[htbp]
\begin{center}
\includegraphics[scale=1.0]{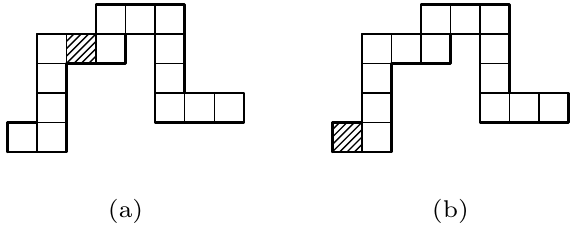}
\end{center}
\caption{(a) A configuration of g-2PATH. 
(b) A configuration of 2PATH.}
\label{figure:fig001}
\end{figure}

\begin{figure}[htbp]
\begin{center}
\includegraphics[scale=1.0]{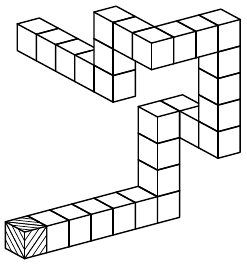}
\end{center}
\caption{A configuration of 3PATH.}
\label{figure:fig041}
\end{figure}

Fig. \ref{figure:fig001}(b) we show examples 
of configurations of g-2PATH and 2-PATH respectively.
In Fig. \ref{figure:fig041} we show an example of 
configurations of 3PATH.
The variation $k$PATH is a natural modification 
of the original FSSP 
for the $k$-dimensional grid space.
Both of g-$k$PATH and $k$REG are 
modifications of the generalized FSSP for the $k$-dimensional 
grid space.

All of these six variations have solutions 
because configurations of these variations are 
directed graphs and we know that 
FSSP for directed graphs has a solution 
(\cite{Kobayashi_JCSS_1978}).
However, at present 
we do not know whether they have minimal-time solutions or 
not and we have no proofs of their nonexistence.

In \cite{Goldstein_Kobayashi_SIAM_2005, Kobayashi_TCS_2001} 
we showed some circumstantial evidences that these 
six variations have no minimal-time solutions.
To present these results and discuss on them we need 
some basic notions and notations from 
complexity theory 
(the theory of computational complexity), 
such as complexity classes 
${\rm P}$, ${\rm NP}$, ${\rm coNP}$, 
$\Sigma_{i}^{\rm p}$, $\Pi_{i}^{\rm p}$, $\Delta_{i}^{\rm p}$, 
${\rm PH}$, ${\rm PSPACE}$, ${\rm EXP}$, 
complete sets, reducibilities 
$\leq_{\rm m}^{\rm p}$, 
$\leq_{\rm T}^{\rm p}$, and so on.
In \ref{section:complexity} we summarize them.
There we also explain some notations that are not standard, 
such as ${\rm P}_{\rm F}$, 
$\Delta_{i, {\rm F}}^{\rm p}$, 
${\rm PH}_{\rm F}$, ${\rm PSPACE}_{\rm F}$, ${\rm EXP}_{\rm F}$, 
and nonstandard usage of the reducibility ``$\leq_{\rm T}^{\rm p}$''.

In \cite{Kobayashi_TCS_2001} we introduced a problem which we called 
the {\it two-dimensional path extension problem}\/ (abbreviated as 2PEP).
\begin{figure}[htbp]
\begin{center}
\includegraphics[scale=1.0]{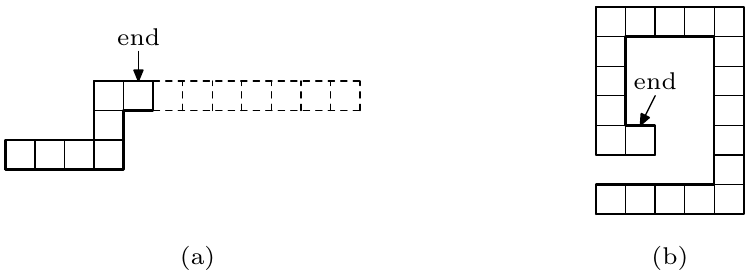}
\end{center}
\caption{Two examples of 2PEP. The answer is YES for (a) and 
NO for (b).}
\label{figure:fig002}
\end{figure}
In the problem we are given a path in the two-dimensional grid space 
with one of its terminal positions specified as the end position. 
We are to decide whether or not we can extend the path 
from the end position 
to a path whose length is the double of that of the original path.  
Here, by the {\it length}\/ of a path we mean the number of positions 
in it.
In Fig. \ref{figure:fig002} we show two paths for 2PEP.
For the path in (a) the answer is YES and the dashed boxes show 
one of the desired extensions.
For the path in (b) the answer is obviously NO.
The problem is difficult when the end position is surrounded by the 
path itself and the extension cannot escape from the 
surrounded region as is in (b).
It is easy to see that  ${\rm 2PEP} \in {\rm NP}$.

At present the exhaustive search is the only algorithm we know 
for 2PEP and 2PEP seems to be a very difficult problem.
However, we cannot prove its NP-completeness.

In \cite{Goldstein_Kobayashi_SIAM_2005,Kobayashi_TCS_2001} 
We showed the following results.

\medskip

\noindent
{\bf Result 1}.  If $\text{2PEP} \not\in \text{P}$ then 
$\text{2PATH}$ has no minimal-time solutions 
(\cite{Kobayashi_TCS_2001}).

\medskip

\noindent
{\bf Result 2}.  If $\text{2PEP} \not\in \text{P}$ then 
$\text{g-2PATH}$ has no minimal-time solutions 
(\cite{Kobayashi_TCS_2001}).

\medskip

\noindent
{\bf Result 3}.  If $\text{2PEP} \not\in \text{P}$ then 
$\text{2REG}$ has no minimal-time solutions 
(implicit in \cite{Kobayashi_TCS_2001}).

\medskip

\noindent
{\bf Result 4}.  If ${\rm P} \not= {\rm NP}$ then 
$\text{3PATH}$ has no minimal-time solutions 
(\cite{Goldstein_Kobayashi_SIAM_2005}).

\medskip

\noindent
{\bf Result 5}.  If ${\rm P} \not= {\rm NP}$ then 
$\text{g-3PATH}$ has no minimal-time solutions 
(\cite{Goldstein_Kobayashi_SIAM_2005}).
\medskip

\noindent
{\bf Result 6}.  If ${\rm P} \not= {\rm NP}$ then 
$\text{3REG}$ has no minimal-time solutions 
(\cite{Goldstein_Kobayashi_SIAM_2005}).

\medskip

These results were obtained as follows.  
Let ${\rm MFT}_{\Gamma}$ 
denote the computation problem to compute the 
value of ${\rm mft}_{\Gamma}(C)$. 
As we noted previously, 
if ${\rm MFT}_{\Gamma} \not\in {\rm P}_{\rm F}$ then 
the variation $\Gamma$ has no minimal-time solutions.
(${\rm P}_{\rm F}$ denotes the class of 
computation problems that are solvable in 
polynomial time.  
See \ref{section:complexity}.)

In \cite{Kobayashi_TCS_2001} we obtained one 
simple characterization of the function 
${\rm mft}_{\rm 2PATH}(C)$.
Using this characterization we could show that 
we can simulate decision of the set ${\rm 2PEP}$ by 
computation of the function ${\rm mft}_{\rm 2PEP}(C)$ 
in polynomial time, and hence 
${\rm 2PEP} \leq_{\rm T}^{\rm p} {\rm MFT}_{\rm 2PATH}$.
(This means that the decision problem of 
the set ${\rm 2PEP}$ can be solved by a polynomial-time 
deterministic oracle Turing machine that uses the oracle 
for the computation problem ${\rm MFT}_\text{2PATH}$.
See \ref{section:complexity}.)
Consequently, 
if ${\rm 2PEP} \not\in {\rm P}$ then 
${\rm MFT}_{\rm 2PATH} \not\in {\rm P}_{\rm F}$ and hence 
the variation ${\rm 2PATH}$ has no minimal-time solutions.
Thus we obtained Result 1.

As for Result 4, 
in \cite{Goldstein_Kobayashi_SIAM_2005} we could show that 
${\rm HAMPATH} \leq_{\rm T}^{\rm p} {\rm 3PEP} 
\leq_{\rm T}^{\rm p} {\rm MFT}_{\rm 3PATH}$, 
where ${\rm HAMPATH}$ is the Hamiltonian path problem 
(\cite{Garey_Johnson}) known to be ${\rm NP}$-complete 
and ${\rm 3PEP}$ is the three-dimensional analogue of 
${\rm 2PEP}$.
From this we obtained Result 4.

As for Results 2, 3, 5, 6, in \cite{Goldstein_Kobayashi_SIAM_2005} 
we could show that 
${\rm mft}_\text{$k$PATH}(C) = {\rm mft}_\text{g-$k$PATH}(C)$ 
$=$ ${\rm mft}_\text{$k$REG}(C)$ for any configuration $C$ of 
$\text{$k$PATH}$, 
and hence 
${\rm MFT}_\text{$k$PATH}$ $\leq_{\rm T}^{\rm p}$ 
${\rm MFT}_\text{g-$k$PATH}$, 
${\rm MFT}_\text{$k$PATH}$ $\leq_{\rm T}^{\rm p}$ ${\rm MFT}_\text{$k$REG}$.
Thus we obtained Results 2, 3, 5, 6.

We may summarize our strategy to obtain Results 1 -- 6 as follows.
\begin{description}
\item[Step 1.] We investigate the function 
${\rm mft}_{\Gamma}(C)$.
\item[Step 2.] Based on our understanding of 
${\rm mft}_{\Gamma}(C)$ obtained in Step 1, we show 
that we can simulate decision of a set of words $X$ by 
computation of the function ${\rm mft}_{\Gamma}(C)$ in 
polynomial time, and hence 
$X \leq_{\rm T}^{\rm p} {\rm MFT}_{\Gamma}$.
\end{description}
Then we obtain a result ``if $X \not\in {\rm P}$ then 
the variation $\Gamma$ has no minimal-time solutions.''
Moreover, if $X$ is a $\mathcal{C}$-complete set for 
a class of sets $\mathcal{C}$ ($\supseteq {\rm P}$), 
we obtain a result 
``if ${\rm P} \not= \mathcal{C}$ then 
the variation $\Gamma$ has no minimal-time solutions.''

\subsection{The problems considered in this paper and their motivations}
\label{subsection:problems_and_motivations}

The motivation of this paper is to improve Results 1 -- 6 
by replacing assumptions in them with some 
weaker assumptions, for example, by replacing 
the assumption 
``${\rm P} \not= {\rm NP}$'' in Result 4 with a 
weaker assumption 
``${\rm P} \not= \Sigma_{2}^{\rm p}$.''

Strictly speaking, we should say 
that the assumption ${\rm P} \not= \Sigma_{2}^{\rm p}$ is 
``seemingly weaker'' than the assumption 
${\rm P} \not= {\rm NP}$ 
instead of simply saying ``weaker'' because 
at present we know that 
${\rm P} \not= {\rm NP} \Longrightarrow {\rm P} \not= \Sigma_{2}^{\rm p}$ is 
true but we do not know that 
${\rm P} \not= \Sigma_{2}^{\rm p} 
\Longrightarrow 
{\rm P} \not= {\rm NP}$ 
is false.
However, to simplify the description we continue 
to say that 
``$\mathcal{A}$ is stronger than $\mathcal{B}$'' 
and 
``$\mathcal{B}$ is weaker than $\mathcal{A}$'' 
only to mean 
that $\mathcal{A} \Longrightarrow \mathcal{B}$ is true.

Examples of such improvements are to replace 
the assumption ``${\rm 2PEP} \not\in {\rm P}$'' 
in Results 1 -- 3 with 
the weaker assumption ``${\rm P} \not= {\rm NP}$.''
These are very interesting and important problems.
However, in this paper we are not concerned with 
these improvements.
In this paper we are concerned with 
improving Results 2, 3, 5, 6 by replacing the 
assumption ``${\rm 2PEP} \not\in {\rm P}$'' or 
``${\rm P} \not= {\rm NP}$'' in them with weaker 
assumptions 
``${\rm P} \not= \Sigma_{2}^{\rm p}$'', 
``${\rm P} \not= \Sigma_{3}^{\rm p}$'', \ldots, 
``${\rm P} \not= {\rm PSPACE}$'' using our strategy 
explained in the previous subsection.

For such improvements we must study 
the functions ${\rm mft}_\Gamma(C)$ as the first step 
(Step 1 of our strategy) 
($\Gamma$ $=$ $\text{g-2PATH}$, 
$\text{2REG}$, $\text{g-3PATH}$, $\text{3REG}$).
In this paper we study the two functions 
${\rm mft}_\text{g-2PATH}(C)$, 
${\rm mft}_\text{g-3PATH}(C)$ 
to deepen our understanding of their properties.

We briefly explain why we exclude Results 1, 4 
from our project of improvements.
There is an essential difference between the two variations 
$k$PATH and the four variations g-$k$PATH, $k$REG.
For $k$PATH, 
by the characterization of ${\rm mft}_\text{$k$PATH}(C)$ 
obtained in \cite{Kobayashi_TCS_2001} 
we know that ${\rm MFT}_\text{$k$PATH} \in \Delta_{2, \rm{F}}^{\rm p}$.
($\Delta_{2, \rm{F}}^{\rm p}$ denotes the class of functions 
that can be computed by polynomial-time deterministic oracle 
Turing machines with ${\rm NP}$-set oracles. 
See \ref{section:complexity}.) 
However, the functions 
${\rm mft}_\text{g-$k$PATH}(C)$ and 
${\rm mft}_\text{$k$REG}(C)$ seem to be very 
difficult to compute and at present 
we cannot show 
${\rm MFT}_\text{g-$k$PATH} \in \Delta_{2, \rm{F}}^{\rm p}$ and 
${\rm MFT}_\text{$k$REG} \in \Delta_{2, \rm{F}}^{\rm p}$.

For example, consider to improve Result 1 
by replacing the assumption 
${\rm 2PEP} \not\in {\rm P}$ in it with the weaker assumption 
${\rm P} \not= \Sigma_{2}^{\rm p}$ by our strategy.
For this we must show 
$\tilde{L} \leq_{\rm T}^{\rm p} {\rm MFT}_\text{$2$PATH}$ 
for a $\Sigma_{2}^{\rm p}$-complete set $\tilde{L}$.
However, if we succeeded in it we have shown that 
$\Sigma_{2}^{\rm p} \subseteq \Delta_{2}^{\rm p}$ because 
${\rm MFT}_\text{$2$PATH} \in \Delta_{2, \rm{F}}^{\rm p}$, 
and this is a breakthrough in complexity theory.
The same is true for 3PATH.
Therefore, the improvement we are 
considering here is 
extremely difficult for $k$PATH.
In other words, if we are to use our strategy 
to improve Results 1, 4 we should be satisfied 
with the version of Result 1 with its assumption replaced with 
``${\rm P} \not= {\rm NP}$'' and the present Result 4.

We also exclude improvements that use the assumption 
``${\rm P} \not= {\rm PH}$'' because PH is unlikely to 
have complete problems (see \ref{section:complexity}).

\subsection{The main result and its implications}
\label{subsection:main_result_and_implication}

Now we are ready to explain our main result.
We define a condition on configurations of $\text{g-$k$PATH}$ 
($k = 2, 3$) 
which we call the {\it condition of noninterference of 
extensions}\/ (abbreviated as CNI) and show 
a simple characterization (Theorem \ref{theorem:th005}) of 
${\rm mft}_\text{g-$k$PATH}(C)$ for configurations $C$ 
that satisfy CNI.
Using this characterization we show that 
we can compute the value of ${\rm mft}_\text{g-$k$PATH}(C)$ 
for CNI-satisfying configurations $C$ 
in polynomial time using an ${\rm NP}$ set as an oracle.
In other words, if we denote the computation problem 
to compute ${\rm mft}_\text{g-$k$PATH}(C)$ for CNI-satisfying 
configurations $C$ by ${\rm MFT}_\text{g-$k$PATH,CNI}$, then 
${\rm MFT}_\text{g-$k$PATH,CNI} \in \Delta_{2,{\rm F}}^{\rm p}$.

CNI is the conjunction of three conditions which 
we call the conditions for $I$, for $J$, and for $K$.
Here we will explain only the condition for $K$ 
because the three conditions are alike 
and the condition for $K$ is the main part of CNI.

Let $C = p_{r} p_{r+1} \ldots p_{-1} p_{0} p_{1} \ldots 
p_{s-1} p_{s}$ be a configuration of $\text{g-$k$PATH}$ 
($r \leq 0 \leq s$).
All of $p_{r}, \ldots, p_{s}$ are positions in the 
$k$-dimensional grid space (elements of $\mathbf{Z}^{k}$).
$p_{0}$ is the origin $(0, 0)$ or $(0, 0, 0)$ and is the position of 
the general.

Suppose that $r < i \leq 0 \leq j < s$.
If $q_{u} q_{u+1} \ldots q_{i-2}$ ($u \leq i-1$) is 
a (possibly empty) sequence of 
positions such that 
$q_{u} q_{u+1} \ldots q_{i-2} p_{i-1} p_{i} \ldots p_{j} p_{j+1}$ is 
a configuration of g-$k$PATH, we say that 
the (nonempty) sequence $q_{u} q_{u+1} \ldots q_{i-2} p_{i-1}$ 
is a {\it  consistent left extension}\/ of 
$p_{i} \ldots p_{j}$.
Similarly, 
if $q_{j+2} \ldots q_{v-1} q_{v}$ ($j+1 \leq v$) is 
a (possibly empty) sequence of 
positions such that 
$p_{i-1} p_{i} \ldots p_{j} p_{j+1} q_{j+2} \ldots q_{v}$ is 
a configuration of g-$k$PATH, we say that 
the (nonempty) sequence $p_{j+1} q_{j+2} \ldots q_{v}$ 
is a {\it consistent right extension}\/ of 
$p_{i} \ldots p_{j}$.
We say that these consistent left and right 
extensions {\it interfere}\/ if 
the sequence 
$q_{u} \ldots q_{i-2} p_{i-1} p_{i} \ldots 
p_{j} p_{j+1} q_{j+2} \ldots q_{v}$ is not a configuration 
of g-$k$PATH (because the two extensions either overlap or touch).

\begin{figure}
\begin{center}
\includegraphics[scale=1.0]{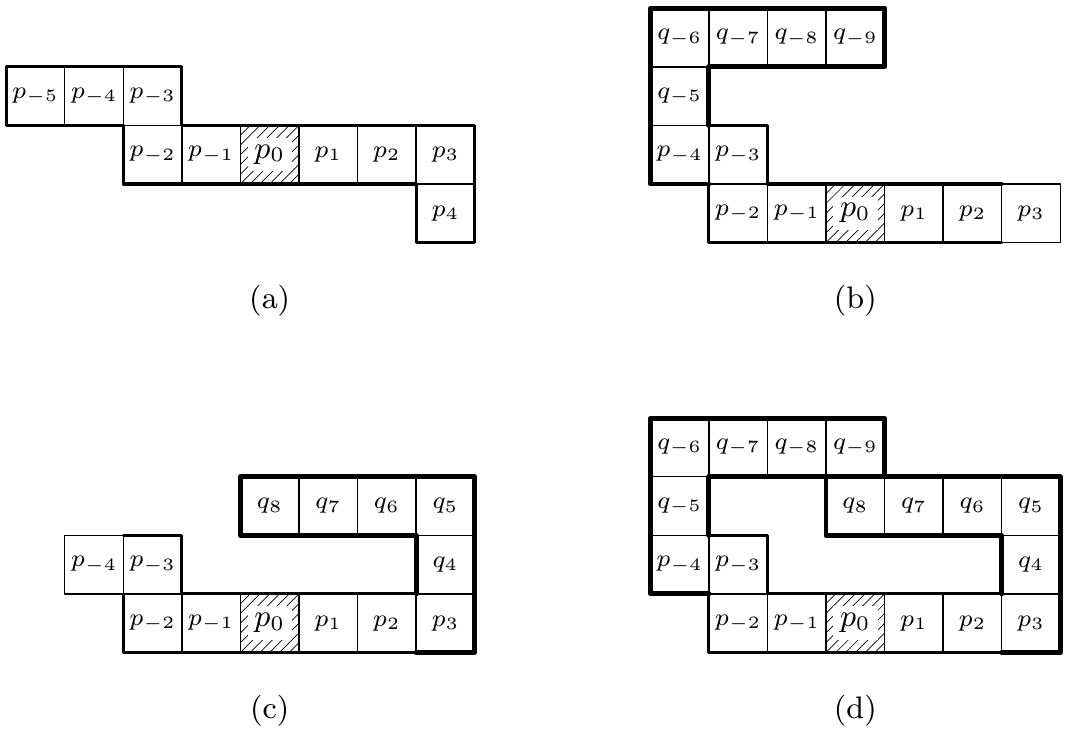}
\end{center}
\caption{Consistent left and right extensions and their 
interference.}
\label{figure:fig050}
\end{figure}

In the remainder of this subsection we show examples only for 
g-2PATH.
We show an example in Fig. \ref{figure:fig050}.
In (a) we show a configuration $p_{-5} \ldots p_{4}$ 
of g-2PATH ($r = -5$, $s = 4$).
We consider the case $i = -3$, $j = 2$.
In (b) and (c) we show a consistent left extension 
$q_{-9} \ldots q_{-5} p_{-4}$ and 
a consistent right extension 
$p_{3} q_{4} \ldots q_{8}$ of 
$p_{-3} \ldots p_{2}$ ($u = -9$, $v = 8$).
These two extensions interfere because 
$q_{-9} \ldots q_{-5} p_{-4} p_{-3} \ldots p_{2} p_{3} q_{4} \ldots q_{8}$ 
is not a configuration as is shown in (d).

The condition for $K$ is that, 
for each $i, j$ such that $r < i \leq 0 \leq j < s$, 
if the number of consistent left extensions and that of consistent 
right extensions of $p_{i} \ldots p_{j}$ are finite 
then consistent left extensions and 
consistent right extensions of $p_{i} \ldots p_{j}$ 
do not interfere.
\begin{figure}
\begin{center}
\includegraphics[scale=1.0]{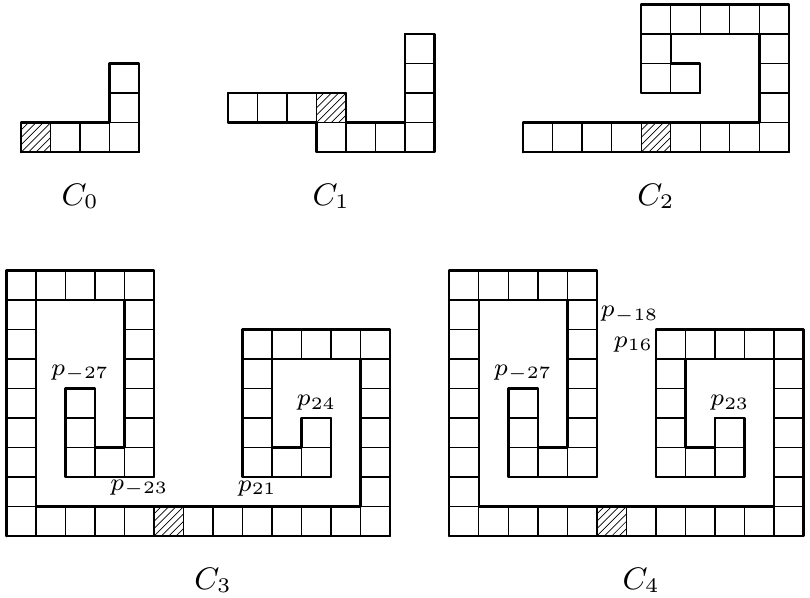}
\end{center}
\caption{Five examples of g-$2$PATH. 
$C_{0}, C_{1}, C_{2}, C_{3}$ satisfy the $K$-condition and 
$C_{4}$ does not.}
\label{figure:fig051}
\end{figure}

For the five configurations $C_{0}$, \ldots, $C_{4}$ 
shown in Fig. \ref{figure:fig051}, 
$C_{0}$, $C_{1}$, $C_{2}$, $C_{3}$ satisfy 
the condition for $K$ and $C_{4}$ does not.
$C_{0}$ satisfies the condition for $K$ 
because $r = 0$ and there is no $i$ such that 
$r < i \leq 0$.
$C_{1}$ satisfies the condition for $K$ because for any $i, j$ 
such that $r < i \leq 0 \leq j < s$, both of consistent left 
extensions and consistent right extensions are infinitely many.
$C_{2}$ satisfies the condition for $K$ because for any $i$ 
such that $r < i \leq 0$, consistent left extensions 
are infinitely many.
$C_{3}$ and $C_{4}$ are alike.
The essential difference is the following.
In both of them, the left hand 
(the part $p_{r} \ldots p_{0}$) 
and the right hand (the part $p_{0} \ldots p_{s}$) 
construct a corridor.
However, in $C_{3}$ the width of the corridor is $3$ 
and a path can pass through it and in $C_{4}$ 
the width is $2$ and a path cannot pass through it.
In $C_{3}$ both of consistent left extensions and 
consistent right extensions of $p_{i} \ldots p_{j}$ 
are finitely many if and only if $j \leq -23$ and 
$21 \leq i$, and for such $i$, $j$ consistent left 
extensions and consistent right extensions enter 
two different ``bottles'' and they cannot interfere.
In $C_{4}$ the similar condition is $j \leq -18$ and 
$16 \leq i$ and consistent left and right extensions 
enter one large ``bottle'' and they can freely interfere.
Hence $C_{3}$ satisfies the condition for $K$ and $C_{4}$ 
does not.
The configurations $C_{0}$, $C_{1}$, $C_{2}$, $C_{3}$ 
satisfy the conditions for $I$, $J$ also, and 
consequently satisfy CNI.
$C_{4}$ does not satisfy CNI because it does not 
satisfy the condition for $K$.

In many cases the condition for $K$ determines whether $C$ 
satisfies CNI or not.
However, there are cases where the conditions for $I$, $J$  
play essential roles.
The two configurations $C_{0}$, $C_{1}$ shown in 
Fig. \ref{figure:fig060}, Fig. \ref{figure:fig061} are 
similar.
However, $C_{0}$ satisfies CNI 
but $C_{1}$ does not because $C_{1}$ does not satisfy 
the condition for $I$.

We summarize implications of our main result.
Here we assume the following three statements 
which we intuitively believe to be true. 
\begin{enumerate}
\item[(1)] The computation of 
${\rm mft}_\text{g-$k$PATH}(C)$ is harder than that of 
${\rm mft}_\text{$k$PATH}(C)$, 
that is, 
${\rm MFT}_\text{$k$PATH} <_{\rm T}^{\rm p} 
{\rm MFT}_\text{g-$k$PATH}$.
\item[(2)] The computation of 
${\rm mft}_\text{g-$k$PATH}(C)$ is harder than decision 
of ${\rm NP}$ sets, that is, 
${\rm MFT}_\text{g-$k$PATH} \not\leq_{\rm T}^{\rm p} L$ 
for any ${\rm NP}$ set $L$.
\item[(3)] $\Sigma_{2}^{\rm p}$ includes 
$\Delta_{2}^{\rm p}$ properly.
\end{enumerate}
From our main result 
``${\rm MFT}_\text{g-$k$PATH,CNI} \in \Delta_{2,{\rm F}}^{p}$'' 
and the assumptions (1), (2) 
we can derive the following two results.
\begin{enumerate}
\item[(4)] For $k = 2$, 
${\rm MFT}_\text{$2$PATH} \leq_{\rm T}^{\rm p} 
{\rm MFT}_\text{g-$2$PATH,CNI} <_{\rm T}^{\rm p} 
{\rm MFT}_\text{g-$2$PATH}$. 
\item[(5)] For $k = 3$, 
${\rm MFT}_\text{$3$PATH} \equiv_{\rm T}^{\rm p} 
{\rm MFT}_\text{g-$3$PATH,CNI} <_{\rm T}^{\rm p} 
{\rm MFT}_\text{g-$3$PATH}$.
\end{enumerate}
The derivation is simple and we show it in 
\ref{section:facts_for_implication}.
(In the appendix we also show that the assumption (1) 
implies the assumption (2) for the case $k = 3$.)

For both of $k=2, 3$ we have 
${\rm MFT}_\text{g-$k$PATH,CNI} <_{\rm T}^{\rm p} 
{\rm MFT}_\text{g-$k$PATH}$.
Hence we have the following implication of our result.
Suppose that $\mathcal{C}$ is one of 
$\Sigma_{2}^{\rm p}, \Sigma_{3}^{\rm p}, \ldots, {\rm PSPACE}$, 
$\tilde{L}$ is a $\mathcal{C}$-complete set, 
and we try to simulate the decision of $\tilde{L}$ 
by the computation of ${\rm mft}_\text{g-$k$PATH}(C)$ 
to improve Result 2 and Result 5 so that 
the assumption ${\rm P} \not= \mathcal{C}$ is used 
instead of stronger assumptions 
${\rm 2PEP} \not\in {\rm P}$ and 
${\rm P} \not= {\rm NP}$ respectively.
\begin{description}
\item[Implication 1.]
The simulation must use computation of 
${\rm mft}_\text{g-$k$PATH}(C)$ for CNI-nonsatisfying 
configurations $C$.
\end{description}
For, otherwise we have 
$\tilde{L} \leq_{\rm T}^{\rm p} {\rm MFT}_\text{g-$k$PATH,CNI} 
\leq_{\rm T}^{\rm p} L$ for an ${\rm NP}$ set $L$ and consequently 
$\Sigma_{2}^{\rm p} \subseteq 
\mathcal{C} \subseteq \Delta_{2}^{\rm p}$, 
contradicting our assumption (3).

\medskip

Next, we consider what makes the computation of 
${\rm mft}_\text{g-$k$PATH}(C)$ 
harder than that of ${\rm mft}_\text{$k$PATH}(C)$.
For the case $k = 3$, 
the result (5) means that CNI-satisfying configurations 
do not make the computation of ${\rm mft}_\text{g-$3$PATH}(C)$ 
harder than that of ${\rm mft}_\text{$3$PATH}(C)$, 
but CNI-nonsatisfying configurations make the computation 
harder.
For the case $k = 2$, it is possible that 
${\rm MFT}_\text{$2$PATH} <_{\rm T}^{\rm p} 
{\rm MFT}_\text{g-$2$PATH,CNI} <_{\rm T}^{\rm p} 
{\rm MFT}_\text{g-$2$PATH}$.
If this is true, CNI-satisfying configurations of g-$2$PATH 
make the computation of ${\rm mft}_\text{g-$2$PATH}(C)$ 
harder than that of ${\rm mft}_\text{$2$PATH}(C)$, 
and CNI-nonsatisfying configurations make the computation 
furthermore harder.
Therefore, we have the following implication of 
our result.
\begin{description}
\item[Implication 2.]
The interference of left and right consistent extensions 
of parts of configurations is one of the factors 
that make the computation of ${\rm mft}_\text{g-$k$PATH}(C)$ 
harder than that of ${\rm mft}_\text{$k$PATH}(C)$.
Moreover, for the case $k = 3$, it is the factor 
that makes the computation harder.
\end{description}

Our result implies that 
for our project of improvements of Result 2 and Result 5 
it is essential to understand properties of ${\rm mft}_\text{g-$k$PATH}(C)$ 
for CNI-nonsatisfying configurations $C$.

\subsection{The organization of the paper}
\label{subsection:organization}

This paper is divided into three parts.
The first part consists of Sections 2, 3, 4 and 
in it we present technical materials that are 
necessary for presenting the main result of 
the paper.
The second part consists of Sections 5, 6, 7 and in it 
we present the main result of the paper, its applications 
and its implications for our project to improve 
Results 2, 5.
The third part is Section 8.
In it we consider a problem that is different 
from the problems considered in previous sections.
The problem is to construct small solutions 
of $\Gamma=\text{$k$PATH}, \text{g-$k$PATH}, \text{$k$REG}$ 
that fire $C$ at time ${\rm mft}_\Gamma(C)$ 
for a given specific configuration $C$.
In Section 9 we present conclusions.
In Appendix A we summarize basic notions and notations 
from complexity theory 
used in this paper.
We also explain some nonstandard notions and notations 
that are used in this paper.

Almost all results in this paper on $k$PATH, g-$k$PATH, $k$REG 
are true for both of 
the two-dimensional and the three-dimensional variations 
with obvious modifications of constants and orders of 
the power in expressions. 
For such results we show them only for the two-dimensional variations 
$2$PATH, g-$2$PATH, $2$REG.
When straightforward modifications are not sufficient for 
formulating the three-dimensional results from the two-dimensional 
results, we explain what modifications are 
necessary.
When results for the two-dimensional and the three-dimensional 
variations are essentially different, we state it explicitly.

\section{Basic notions and notations}
\label{section:basic_notions_and_notations}

In this section we explain basic notions and notations 
on FSSP.
The variations considered in this paper are 
$k$PATH, g-$k$PATH, $k$REG ($k = 2, 3$).
However, as we wrote in Subsection 
\ref{subsection:organization} 
we explain notions, notations, definitions and results 
only for the two-dimensional variations ($k = 2$).

In the variations 2PATH, g-2PATH and 2REG, 
a copy of a finite automaton 
is placed at a position $(x, y)$ 
in the two-dimensional grid space identified with 
$\mathbf{Z}^{2}$ ($\mathbf{Z}$ denotes 
the set of all integers).
We call a copy of a finite automaton a {\it node}\/.
Let $\epsilon_{0}, \epsilon_{1}, \epsilon_{2}, \epsilon_{3}$ 
be defined by 
$\epsilon_{0} = (1, 0)$, 
$\epsilon_{1} = (0, 1)$, 
$\epsilon_{2} = (-1, 0)$, 
$\epsilon_{3} = (0, -1)$.
A position $(x, y)$ is {\it adjacent}\/ to the four 
positions $(x, y) + \epsilon_{i}$ ($i = 0, 1, 2, 3$).
We call $(x, y) + \epsilon_{i}$ the position that 
is adjacent to the position $(x, y)$ in the direction $i$.
We understand this number $i$ with ``modulo $4$'' 
(for example, the direction $6$ means the direction $2$).
We identify the directions $0$, $1$, $2$, and $3$ respectively 
with the east, the north, the west, and the south, 
respectively.
A node has four inputs and four outputs, for each of the 
four directions.
The input from the direction $i$ of a node at $(x, y)$ is 
connected with the output to the direction $i + 2$ of the 
node at $(x, y) + \epsilon_{i}$.
If there is no node at $(x, y) + \epsilon_{i}$, 
the value of the input is a special symbol {\rm \#}.

A {\it path}\/ is a nonempty sequence 
$X = p_{0} p_{1} \ldots p_{n-1}$ of positions such that 
$p_{i}$ and $p_{i+1}$ are adjacent for each $0 \leq i \leq n-2$.
We call the value $n$ the {\it length}\/ of the path $X$ 
and denote it by $|X|$. 
We call the positions $p_{0}$ and $p_{n-1}$ the {\it start position}\/ 
and the {\it end position}\/ of the path respectively, 
and call both of them the {\it terminal positions} of the path.
We say that the path is {\it between}\/ $p_{0}$ 
{\it and}\/ $p_{n-1}$ or {\it from}\/ $p_{0}$ {\it to}\/ 
$p_{n-1}$.

Let $C$ be a set of positions.
We say that $C$ is {\it connected} if there is a path 
from $p$ to $p'$ in $C$ for any nodes $p, p'$ in $C$.
Suppose that there is a path from $p$ to $p'$ in $C$ and that $X$ 
is a shortest path from $p$ to $p'$ in $C$.
Then, by the {\it distance} between $p$ and $p'$ in $C$ 
we mean the value 
$|X| - 1$ and denote it by ${\rm d}_{C}(p, p')$, or simply 
by ${\rm d}(p, p')$ when $C$ is understood.

A {\it configuration}\/ (a problem instance) of 2REG 
is a connected finite set of positions $C$ 
that contains the origin $(0, 0)$.
The origin is the position of the general $v_{\rm gen}$.
A configuration of g-2PATH is a configuration $C$ of 
2REG that satisfies the conditions:
\begin{enumerate}
\item[$\bullet$] $C$ is a path $p_{0} \ldots p_{n-1}$.
\item[$\bullet$] $p_{0}, \ldots p_{n-1}$ are different.
\item[$\bullet$] There is no pair $i, j$ such that 
$i + 2 \leq j$ and $p_{i}$, $p_{j}$ are adjacent.
\end{enumerate}
In other words, $C$ is a path that contains $(0, 0)$ and 
that neither crosses nor touches itself.
A configuration of 2PATH is a configuration 
$p_{0} \ldots p_{n-1}$ of g-2PATH such that 
either $p_{0}$ or $p_{n-1}$ is $(0, 0)$.

For a configuration $C$ of 2REG, we call the value 
$\max \{ {\rm d}_{C}((0, 0), p) ~|~ $ $p$ is a node in $C\}$
the {\it radius}\/ of $C$ and denote it by ${\rm rad}(C)$.

Although a position is an element of $\mathbf{Z}^{2}$ and 
a node is a copy of a finite automaton placed at a position, 
we use these terms interchangeably. 
For example, we say ``a position $(x, y)$ fires'' 
or ``a node $(x, y)$ fires'' instead of 
``the node at a position $(x, y)$ fires.''

For a position $p$ in a configuration $C$ of a variation $\Gamma$, 
by the {\it boundary condition}\/ of $p$ in $C$ 
we mean the vector $\mathbf{b} = (b_{0}, b_{1}, b_{2}, b_{3}) 
\in \{0,1\}^{4}$ 
such that $b_{i}$ is $1$ if the position $p + \epsilon_{i}$ 
is in the configuration $C$ 
(and hence there is a node at the position) and is $0$ otherwise 
($0 \leq i \leq 3$).
We denote this vector ${\mathbf b}$ by ${\rm bc}_{C}(p)$, or 
${\rm bc}(p)$ when $C$ is understood.

In Subsection \ref{subsection:problem_and_history} we defined the FSSP.
We call this definition the {\it traditional model}\/ of FSSP.
In this paper we use a slightly modified definition which we 
call the {\it boundary-sensitive model}\/ of FSSP.
There are two modifications.

First, in the traditional model there is one unique general state ${\rm G}$.
In the boundary-sensitive model there may be more than one 
general states ${\rm G}_{0}$, ${\rm G}_{1}$, $\ldots$, ${\rm G}_{m-1}$.
The general state to be used for a configuration $C$ is 
uniquely determined by the boundary condition of the general 
$v_{\rm gen}$ in $C$.
Formally, a mapping $\tau$ from $\{0, 1\}^{4}$ to $\{0, \ldots, m-1\}$ 
is specified and in a configuration $C$ the general 
state ${\rm G}_{\tau({\rm bc}_{C}(v_{\rm gen}))}$ is used as the state of 
the general at time $0$.

Second, in the traditional model there is one unique firing state ${\rm F}$.
In the boundary-sensitive model there is a set $\mathcal{F}$ of 
firing states and any state in $\mathcal{F}$ may be used 
as a firing state.
We require the quiescent state ${\rm Q}$ not to be in $\mathcal{F}$ 
but some of the general states may be in $\mathcal{F}$.
The condition for a finite automaton $A$ to be a solution is as follows.
\begin{quote}
For any configuration $C$ of $\Gamma$ 
there exists a time $t_{C}$ such that, for any $v \in C$, 
${\rm st}(v, t, C, A) \not\in \mathcal{F}$ for $t < t_{C}$ and 
${\rm st}(v, t_{C}, C, A) \in \mathcal{F}$.
\end{quote}

The boundary-sensitive model has two merits.
Suppose that $v$, $v'$ are nodes in a configuration $C$ 
and consider the minimum time for $v'$ to know the boundary 
condition of $v$. 
In the boundary-sensitive model this time is simply 
$d(v_{\rm gen}, v) + d(v, v')$.
However, in the traditional model this time is 
$1 + d(v_{\rm gen}, v) + d(v, v')$ for $v = v_{\rm gen}$ and 
$d(v_{\rm gen}, v) + d(v, v')$ for $v \not= v_{\rm gen}$.
This irregularity in the traditional model 
makes the study of minimal-time solutions complex and tedious 
in many inessential ways.

In the traditional model a node cannot fire at time $0$ but 
in the boundary-sensitive model this is possible.
Because of this, in the boundary-sensitive model no special care is 
necessary for the configuration that has only one position 
(the ``singleton'' configuration).
As an example, consider the original FSSP.
In the traditional model, ${\rm mft}(C_{n})$ is $1$ for $n = 1$ and 
$2n - 2$ for $n \geq 2$. However, in the boundary-sensitive model, 
${\rm mft}(C_{n})$ is $2n - 2$ for any $n$.

The traditional model is simple and is suited to present 
FSSP as an interesting problem in automata theory, 
and the boundary-sensitive model is suited to the theoretical 
study of minimal-time solutions.
For more details on the boundary-sensitive model, 
see \cite{Kobayashi_TCS_2014,Kobayashi_Goldstein_UC_2005}.

We call a finite automaton $A$ that satisfies 
the following modified condition 
a {\it partial solution}\/ of a variation $\Gamma$ of FSSP.%
\footnote{The term ``a partial solution of a variation of 
FSSP'' is also used for a different meaning (\cite{Umeo_2015_Survey}).}
\begin{quote}
For any configuration $C$ of $\Gamma$, 
either 
(1) there is a time $t_{C}$ such that, for any node 
$v$ in $C$, ${\rm st}(v, t, C, A) \not\in {\mathcal F}$ 
for any time $t < t_{C}$ and 
${\rm st}(v, t_{C}, C, A) \in {\mathcal F}$, 
or (2) 
${\rm st}(v, t, C, A) \not\in {\mathcal F}$ for any node $v$ 
in $C$ and any time $t$.
\end{quote}
We call the set of configurations $C$ for which the case (1) 
is true the {\it domain}\/ of the partial solution $A$.
For a configuration $C$ in the domain, by 
${\rm ft}(C, A)$ we denote the time $t_{C}$ mentioned in (1).
For a configuration $C$ not in the domain, 
${\rm ft}(C, A)$ is undefined.

Suppose that $A$ is a partial solution and $A'$ is a solution.
Let $A''$ be the finite automaton that simulates both of 
$A$, $A'$ simultaneously and fires as soon as at least one 
of $A$, $A'$ fires.
Then $A''$ is a solution and we have 
${\rm ft}(C, A'') \leq {\rm ft}(C, A)$ 
and hence ${\rm mft}(C) \leq {\rm ft}(C, A)$ for any $C$ in 
the domain of $A$.
Therefore partial solutions are useful to derive 
upper bounds of ${\rm mft}(C)$.

Let $\Gamma$, $\Gamma'$ be variations of FSSP.
We say that $\Gamma'$ is a {\it conservative super-variation}\/ 
of $\Gamma$ and $\Gamma$ is a {\it conservative sub-variation}\/ 
of $\Gamma'$ 
if all configurations of $\Gamma$ are configurations of $\Gamma'$ 
and 
${\rm mft}_{\Gamma'}(C) = {\rm mft}_{\Gamma}(C)$ 
for any configuration $C$ of $\Gamma$.
For example, the generalized FSSP having the minimum firing 
time (\ref{equation:eq001}) is a conservative 
super-variation of the original FSSP having the 
minimum firing time $2n - 2$ 
because the value of (\ref{equation:eq001}) is $2n - 2$ 
when $i = 0$.
Let $\Gamma'$ be the FSSP for rectangles and $\Gamma$ be 
the FSSP for squares. 
Then $\Gamma'$ is not a conservative super-variation of $\Gamma$.
The minimum firing time of $\Gamma'$ for 
a rectangle of size $m \times n$ is 
$m + n + \max\{m, n\} - 3$ and 
the minimum firing time of $\Gamma$ for 
a square of size $n \times n$ is $2n - 2$.
The former value for $m = n$ is $3n - 3$ and 
is larger than the latter value.

In (\cite{Goldstein_Kobayashi_SIAM_2005, Kobayashi_TCS_2001}) 
we showed that both of g-2PATH and 2REG are 
conservative super-variations of 2PATH.
By this result, Results 2 and 3 followed immediately from Result 1.
However, at present we do not know whether 2REG is a conservative 
super-variation of g-2PATH or not.
Hence, even if we could improve Result 2, it does not automatically 
improve Result 3.
We cannot exclude the possibility that the computation of 
${\rm mft}_\text{2REG}(C)$ is easier than that of 
${\rm mft}_\text{g-2PATH}(C)$.

\section{An algorithm for computing ${\rm mft}_{\Gamma}(C)$}
\label{section:local_map_algorithm}

\subsection{Computability of ${\rm mft}_{\Gamma}(C)$}
\label{subsection:computability}

In \cite{Kobayashi_TCS_1978} the author showed that the function 
${\rm mft}_{\Gamma}(C)$ is a computable function if the 
variation $\Gamma$ has a natural definition 
and showed an algorithm to compute ${\rm mft}_{\Gamma}(C)$.
However that algorithm was not intended to be used practically. 
In \cite{Kobayashi_TCS_2014} we reformulated the algorithm 
with the intention to use it practically.
We reformulated it only for the variation of FSSP 
such that configurations are squares and 
the position of the general may be at any position.
However it can be easily modified for any sub-variation of 2REG, 
including 2REG itself, g-2PATH, and 2PATH.

We need some details of this reformulated algorithm 
for presenting and discussing the main result of this paper.
Hence we briefly explain the reformulated algorithm 
to make this paper self-contained.
We use the variation 2REG for the explanation.
When $\Gamma$ is a sub-variation of 2REG, simply 
replace ``configurations of 2REG'' with 
``configurations of $\Gamma$'' in definitions, 
statements of results, and proofs.

\subsection{Available information}
\label{subsection:available_information}

The basis of our algorithm is a notion ``available information 
at a node in a configuration at a time.''
This is usually called ``a local map'' and is one of the basic tools 
in the design of distributed computing algorithms.

For a node $v$ in a configuration $C$ and a time $t$, 
we define the {\it available information}\/ at $v$ in $C$ at 
time $t$ as follows.
If ${\rm d}_{C}(v_{\rm gen}, v) > t$ then the available information 
is the letter ${\rm Q}$.
Otherwise, the available information is the triple 
$(t, v, X)$, 
where $X$ is the set 
\[
\{ (v', {\rm bc}_{C}(v')) ~|~ 
\text{$v'$ is a node in $C$ and 
${\rm d}_{C}(v_{\rm gen}, v') + {\rm d}_{C}(v', v) \leq t$} \}.
\]
We denote the available information of $v$ in $C$ at $t$ 
by ${\rm ai}(v, t, C)$.

\begin{figure}[htbp]
\begin{center}
\includegraphics[scale=1.0]{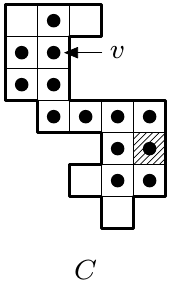}
\end{center}
\caption{
A configuration $C$ of $2$REG, a node $v$ of $C$, 
and the nodes (with dots) in the third component $X$ 
of ${\rm ai}(v, 8, C)$.}
\label{figure:fig005}
\end{figure}

For example, consider the configuration $C$ of 2REG 
and a node $v$ in $C$ shown in Fig. \ref{figure:fig005}.
Then ${\rm d}_{C}(v_{\rm gen}, v) = 6$ and hence 
${\rm ai}(v, t, C) = {\rm Q}$ for $0 \leq t \leq 5$ and 
${\rm ai}(v, t, C) \not= {\rm Q}$ for $6 \leq t$.
${\rm ai}(v, 8, C) = (8, (-3, 3), X)$, where $X$ is the set
\[
\{((0, 0), (0, 1, 1, 1)),
((0, 1), (0, 0, 1, 1)),
\ldots,
((-3, 3), (0, 1, 1, 1))\},
\]
and contains $(v', {\rm bc}_{C}(v'))$ 
for all nodes $v'$ with dots in Fig. \ref{figure:fig005}. 
The element $((0, 0), (0, 1, 1, 1))$ is for the general $v_{\rm gen}$, 
$((0, 1), (0, 0, 1, 1))$ is for the node north of the general, 
and 
$((-3, 3), (0, 1, 1, 1))$ is for the node $v$.

We have six facts on ${\rm ai}(v, t, C)$.
We showed proofs of these facts in the appendix of \cite{Kobayashi_TCS_2014}.

By an {\it infinite automaton} we mean a structure that is the same 
as a finite automaton except that the set of states may be 
an infinite set.

\medskip

\noindent
{\bf Fact 1}:\  There exists an infinite automaton $A_{0}$ 
such that ${\rm st}(v, t, C, A_{0}) = {\rm ai}(v, t$, $C)$ 
for any $v$, $C$, $t$.

\medskip

\noindent
{\bf Fact 2}:\  For any infinite automaton $A$ there exists a mapping 
$\varphi$ that maps states of $A_{0}$
 to states of $A$ so that 
$\varphi({\rm st}(v, t, C, A_{0})) = {\rm st}(v, t, C, A)$.

\medskip

\noindent
{\bf Fact 3}:\ If ${\rm ai}(v, t, C) = {\rm ai}(v', t, C')$ then 
${\rm st}(v, t, C, A) = {\rm st}(v', t, C', A)$ for any 
infinite automaton $A$ and any $v$, $v'$, $C$, $C'$, $t$.

\medskip

\begin{figure}[htbp]
\begin{center}
\includegraphics[scale=1.0]{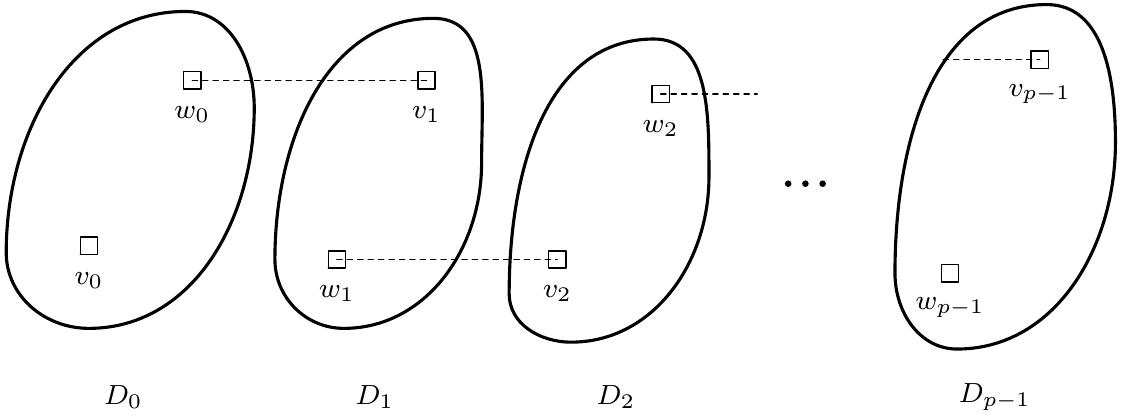}
\end{center}
\caption{Definition of safeness of 
available information $\sigma$ ($\not= {\rm Q}$).}
\label{figure:fig006}
\end{figure}

We say that available information $\sigma$ is {\it safe}\/ if 
either $\sigma = {\rm Q}$, 
or $\sigma \not= {\rm Q}$ and 
there exist
\begin{enumerate}
\item[$\bullet$] configurations $D_{0}$, $D_{1}$, \ldots, $D_{p-1}$ 
($p \geq 1$),
\item[$\bullet$] a time $t$,
\item[$\bullet$] positions $v_{0}$, $w_{0}$ in $D_{0}$, 
$v_{1}$, $w_{1}$ in $D_{1}$, 
\ldots, 
$v_{p-1}$, $w_{p-1}$ in $D_{p-1}$
\end{enumerate}
such that 
\begin{enumerate}
\item[(1)] $\sigma = {\rm ai}(v_{0}, t, D_{0}) \not= {\rm Q}$,
\item[(2)] ${\rm ai}(w_{l}, t, D_{l}) = {\rm ai}(v_{l+1}, t, D_{l+1}) 
\not= {\rm Q}$ for $0 \leq l \leq p-2$,
\item[(3)] ${\rm ai}(w_{p-1}, t, D_{p-1}) = {\rm Q}$
\end{enumerate}
(See Fig. \ref{figure:fig006}).

\medskip

By the {\it infinite state}\/ 2REG we mean 
the variation 2REG that is modified so that 
infinite state automata are allowed as solutions.

\medskip

\noindent
{\bf Fact 4}:\ If ${\rm ai}(v, t, C)$ is safe then any 
solution of the infinite state 2REG 
cannot fire $C$ at time $t$.

\medskip

\noindent
{\bf Fact 5}:\  For any $v$ in any $C$, ${\rm ai}(v, t, C)$ is unsafe 
for all sufficiently large $t$.

\medskip

\noindent
{\bf Fact 6}:\ For any $v$, $v'$ in $C$ and any $t$, if 
${\rm ai}(v, t, C)$ is safe then ${\rm ai}(v', t, C)$ is 
also safe.
(Hence, for any $C$ and $t$, either all of ${\rm ai}(v, t, C)$ are 
safe or all of ${\rm ai}(v, t, C)$ are unsafe.)

As we mentioned previously, 
we showed proofs of these six facts 
in the appendix of \cite{Kobayashi_TCS_2014}.
However, the proofs should be slightly modified 
for the present paper.
In that paper the position of the general may be an arbitrary 
position $v_{\rm gen} = (x_{\rm gen}, y_{\rm gen})$ and 
for any position $v = (x, y)$ we defined 
its {\it relative position}\/ ${\rm rp}(v)$ to be 
$v - v_{\rm gen} = (x - x_{\rm gen}, y - y_{\rm gen})$.
In the present paper we assume that $v_{\rm gen} = (0, 0)$ and 
hence ${\rm rp}(v) = v - (0, 0) = v$.
Therefore, all occurrences of ${\rm rp}(v)$ in the proofs 
should be replaced with $v$.

For configurations $C$, $C'$, a time $t$ and a node $v$, 
we define three relations.
$C \equiv_{t,v}' C'$ if and only if 
$v \in C \cap C'$ and 
${\rm ai}(v, t, C) = {\rm ai}(v, t, C') \not= {\rm Q}$.
$C \equiv_{t}' C'$ if and only if 
there exists $v$ such that 
$C \equiv_{t,v}' C'$.
$C \equiv_{t} C'$ if and only if there exists 
a sequence $C_{0}, \ldots, C_{m-1}$ of configurations 
($m \geq 1$) such that 
$C_{0} = C$, $C_{m-1} = C'$ and 
$C_{i} \equiv_{t}' C_{i+1}$ for $i$ ($0 \leq i \leq m-2$).
We say that a time $t$ is {\it safe}\/ for a configuration $C$ 
if there exists $C'$ such that $C \equiv_{t} C'$ and 
${\rm rad}(C') > t$, and 
{\it unsafe}\/ for $C$ otherwise 
(that is, ${\rm rad}(C') \leq t$ for any $C'$ such that 
$C \equiv_{t} C'$).
We need another fact.

\medskip

\noindent
{\bf Fact 7}:\  For any $C$ and $v$,
\[
\text{${\rm ai}(v, t, C)$ is safe} \Longleftrightarrow 
\text{$t$ is safe for $C$}.
\]

\noindent
We show the proof of Fact 7 in \ref{section:fact_7}.

There is a clear intuitive motivation for the definition of ``safeness'' 
and we explain it in \cite{Kobayashi_TCS_2014}.
There we also explain why we use the word ``safe.''  
(If $t$ is safe for $C$, a prisoner in front of a firing squad $C$ 
feels safe from shooting by soldiers at time $t$.)

\medskip

\subsection{The algorithm}
\label{subsection:algorithm}

Let ${\rm mft}_\text{$2$REG,inf}(C)$ denote 
the minimum firing time of a configuration $C$ 
of the infinite 2REG.
Using the seven facts given in the previous subsection, 
we show a characterization of both of 
${\rm mft}_\text{$2$REG,inf}(C)$ and 
${\rm mft}_\text{$2$REG}(C)$.
This characterization gives an algorithm to compute 
${\rm mft}_\text{$2$REG}(C)$.

We modify the infinite automaton $A_{0}$ mentioned 
in Fact 1 so that it can be used as a solution of 
the infinite 2REG.
States of $A_{0}$ are available information.
The available information ${\rm Q}$ plays the role 
of the quiescent state of $A_{0}$.
The state of the general $v_{\rm gen}$ in 
a configuration $C$ at time $0$ 
(the general state of $C$) is 
${\rm ai}(v_{\rm gen}, 0, C) 
= (0, (0, 0), \{((0, 0), {\rm bc}_{C}((0, 0)))\})$.
This depends on the boundary condition of $v_{\rm gen}$ 
in $C$.
However we use the boundary-sensitive model of FSSP 
and this dependency is allowed.
A state of $A_{0}$ is a firing state 
if and only if it is unsafe (as available information).

By Fact 6, $A_{0}$ is a partial solution of 
the infinite 2REG.
But by Fact 5, this partial solution is a solution.
Moreover, by Fact 4 this solution is 
a minimal-time solution.  
Therefore, the infinite 2REG has 
a minimal-time solution $A_{0}$.
The firing time of $v$ in $C$ of $A_{0}$ is 
the minimum value of $t$ such that 
${\rm ai}(v, t, C)$ is unsafe.
But by Fact 7 this is the minimum value of $t$ 
that is unsafe for $C$.
Therefore, the value ${\rm mft}_\text{$2$REG, inf}(C)$ 
is $\min \{ t ~|~ \text{$t$ is unsafe for $C$}\}$.

For any time $T$, let $A_{1,T}$ be the automaton that simulates 
$A_{0}$ and enters the quiescent state $Q$ at time $T+1$.
This $A_{1,T}$ is a finite automaton. 
(For the reason, see \cite{Kobayashi_TCS_2014}.
The number of the states of $A_{1,T}$ depends on $T$.
However, once the value of $T$ is fixed, 
$A_{1,T}$ is a fixed finite automaton.)
If $A_{0}$ fires $C$ before or at time $T$, $A_{1,T}$ fires $C$ 
at the same time.
If $A_{0}$ fires $C$ after time $T$, $A_{1,T}$ never fires any node of $C$.
Hence $A_{1,T}$ is a partial solution of 2REG such that 
its domain is 
$\{C ~|~ {\rm ft}(C, A_{0}) \leq T\}$ 
and 
${\rm ft}(C, A_{1,T}) = {\rm ft}(C, A_{0})$ 
for any $C$ in the domain.

Let $A_{2}$ be an arbitrary solution of 2REG and 
let $A_{3, T}$ be the finite automaton that simulates 
both of $A_{1, T}$, $A_{2}$ and fires if at least one of 
them fires.
Then $A_{3, T}$ is a solution of 2REG.
Let $C$ be an arbitrary configuration.
If $T$ is a value such that ${\rm ft}(C, A_{0}) \leq T$ then 
${\rm mft}_{\rm 2REG}(C) \leq {\rm ft}(C, A_{3, T}) 
\leq {\rm ft}(C, A_{1, T}) = {\rm ft}(C, A_{0}) 
= {\rm mft}_{\rm inf, 2REG}(C) \leq {\rm mft}_{\rm 2REG}(C)$.
Therefore 
${\rm mft}_\text{$2$REG,inf}(C) = 
{\rm mft}_\text{$2$REG}(C)$ and 
we have the following characterization of 
both of ${\rm mft}_\text{$2$REG,inf}(C)$ and ${\rm mft}_\text{$2$REG}(C)$:
\[
{\rm mft}_\text{$2$REG,inf}(C) = 
{\rm mft}_\text{$2$REG}(C) = 
\min \{ t ~|~ \text{$t$ is unsafe for $C$} \}.
\]
Moreover, $A_{3,T}$ is a solution of 2REG that fires $C$ 
at time ${\rm mft}_\text{$2$REG}(C)$ for any $T$ such that 
${\rm mft}_\text{$2$REG}(C) \leq T$.

The above characterization of ${\rm mft}_\text{$2$REG}(C)$ gives 
an algorithm for computing ${\rm mft}_\text{$2$REG}(C)$.
However, before explaining it we need a small result.
We defined ``$t$ is safe for $C$'' by the statement: 
there exists a sequence $C_{0}, \ldots, C_{m-1}$ of configurations 
($m \geq 1$) 
such that $C = C_{0}$, $C_{0} \equiv_{t}' C_{1} \equiv_{t}' \ldots 
\equiv_{t}' C_{m-1}$, and ${\rm rad}(C_{m-1}) \geq t + 1$.
However, when ${\rm rad}(C) \leq t$ we may restrict 
each $C_{i}$ to be such that ${\rm rad}(C_{i}) \leq t+1$.
(If $C_{i}'$ is the configuration obtained from $C_{i}$ by deleting 
all nodes $v$ in it such that ${\rm d}(v_{\rm gen}, v) > t + 1$, 
the new sequence $C_{0}', \ldots, C_{m-1}'$ can be used to show 
that $t$ is free for $C$ and ${\rm rad}(C_{i}') \leq t + 1$ 
for $0 \leq i \leq m-2$.)

Now we explain the algorithm to compute 
the value of ${\rm mft}_\text{$2$REG}(C)$.
This is the smallest value of $t$ that is unsafe for $C$.
Therefore it is sufficient to show an algorithm to decide 
whether $t$ is safe for $C$ or not for each $C$ and $t$.

If ${\rm rad}(C) \geq t + 1$ then $t$ is safe.
Suppose that ${\rm rad}(C) \leq t$.
Let $S$ be the set $\{ C \}$.
Starting with this set $S$, we perform 
the following sub-step repeatedly.
If all elements of $S$ are marked, then $t$ is unsafe.
Otherwise, we select one unmarked element $C'$ of $S$, 
mark it, list all configurations $C''$ such that 
$C' \equiv_{t, v}' C''$ for some $v \in C'$ and 
${\rm rad}(C'') \leq t + 1$.
If there is one configuration $C''$ such that 
${\rm rad}(C'') = t + 1$ in the list, then 
$t$ is safe.
If all configurations $C''$ in the list 
satisfy ${\rm rad}(C'') \leq t$, then 
add all configurations in the list that are 
not already in $S$ into $S$.

For each $C'$ and $v \in C'$, 
we can enumerate all $C''$ such that 
$C' \equiv_{t, v}' C''$ and ${\rm rad}(C'') \leq t + 1$ 
as follows.
We can enumerate all $C''$ such that $v \in C''$ and 
${\rm rad}(C'') \leq t + 1$.
For each such $C''$ we can compute both of 
${\rm ai}(t, v, C')$, ${\rm ai}(t, v, C'')$ 
and hence we can decide whether 
$C' \equiv_{t, v}' C''$ or not.

The algorithm terminates because 
the set $S$ contains only $C'$ such that 
${\rm rad}(C') \leq t+1$ and 
there are only finitely many such $C'$.
It is easy to show the correctness of this algorithm.

From now on, we call the algorithm to compute ${\rm mft}_{\Gamma}(C)$ 
explained above the {\it local map algorithm} 
because ${\rm ai}(v, t, C)$ 
is the most detailed map of the world $C$ in which 
a person $v$ is (the world view) 
based on the information that the person $v$ can 
collect at time $t$.
We call the partial solution $A_{1, T}$ 
the {\it local map partial solution}\/ for time $T$ and 
denote it by $A_{{\rm lm}, T}$.

We can modify the results obtained in 
Subsections \ref{subsection:computability} - 
\ref{subsection:algorithm} 
for sub-variations of 2REG and 3REG, the FSSP for bilateral rings, 
and the FSSP for unilateral rings.
For FSSP's for directed or undirected networks and their 
sub-variations, it is not immediately clear how to define 
${\rm ai}(v, t, C)$ because each node $v$ in a network 
has no unique coordinates $(x, y)$ to denote it.
However, for these variations too we can show these results 
by properly defining ${\rm ai}(v, t, C)$ 
(\cite{Goldstein_Kobayashi_SIAM_2012}).

\medskip

\subsection{The number of states of $A_{{\rm lm},T}$}
\label{subsection:number_of_states_of_lm_partial_solution}

In Section \ref{section:minimum_state_numbers} we need 
an estimation of the number of states of the local map 
partial solution $A_{{\rm lm},T}$ and we estimate it here.
If we want $A_{{\rm lm},T}$ to be as small as possible, 
a state of $A_{{\rm lm},T}$ is available information 
$\sigma$ such that there really exist $C$, $v \in C$ and $t (\leq T)$ 
such that $\sigma = {\rm ai}(v, t, C)$.
Let $N_{{\rm lm}, T}$ denote the number of such $\sigma$.
To determine the exact value of $N_{{\rm lm},T}$ is difficult.
However we can derive some upper bounds and lower bounds for the value.
By $T$ we denote ${\rm mft}_\text{g-$2$PATH}(C)$ and by $T'$ 
we denote $\lfloor T / 2 \rfloor$.

Suppose that we use $A_{{\rm lm},T}$ for 2REG.
Then ${\rm ai}(v, t, C)$ is either ${\rm Q}$ or an element 
$(t, v, X)$ of
\[
\{0, \ldots, T\} \times \{-T, \ldots, T\}^{2} \times 
\mathcal{P}(\{-T, \ldots, T\}^{2} \times \{0, 1\}^{4}),
\]
where $\mathcal{P}(S)$ denotes the power set of $S$ 
(that is, the set of all subsets of $S$).
Hence we have the following upper bound:
\begin{equation}
N_{{\rm lm},T} \leq 
1 + (T + 1)(2T + 1)^{2} 2^{16(2T + 1)^{2}} = 2^{64T^{2} + O(T)}.
\label{equation:eq005}
\end{equation}
As for lower bounds, 
the number of configurations $C$ such that 
${\rm rad}(C) \leq T'$ is a lower bound for $N_{{\rm lm},T}$.
This is because, for such $C$, 
${\rm ai}(v_{\rm gen}, T, C)$ completely determines $C$ and hence 
different $C$ give different states 
${\rm ai}(v_{\rm gen}, T, C)$ of $A_{{\rm lm},T}$.
It is easy to show that if we select a sufficiently small constant $c$ 
then $2^{c T^{2} - O(T)}$ is a lower bound for $N_{{\rm lm},T}$.
It is not difficult to show the following lower bound:
\begin{equation}
2^{(1/3) (T-3)^{2}} = 2^{(1/3) T^{2} - O(T)} \leq N_{{\rm lm},T}.
\label{equation:eq021}
\end{equation}

Next consider the case of g-2PATH.
In this case, $X$ is essentially a path of the form 
$p_{i} \ldots p_{0} \ldots p_{j}$ such that $-T \leq i \leq 0 \leq j \leq T$ 
and boundary conditions of $p_{i}$, $p_{j}$.
Hence we have the upper bound
\begin{equation}
N_{{\rm lm},T} \leq 
1 + (T + 1)(2T + 1)^{2}(1 + 4 + \ldots + 4^{T})^{2} (2^{4})^{2}
= 2^{4T + O(\log T)}.
\label{equation:eq018}
\end{equation}
As for lower bounds, suppose that $T'$ is even and 
consider configurations $C = p_{r} \ldots p_{0} \ldots p_{s}$ such that 
$r = -T'$, $s = T'$, $p_{r} = (-T'/2, -T'/2), p_{s} = (T'/2$, $T'/2)$.
These configurations satisfy ${\rm rad}(C) = T'$ and 
the number of such configurations is ${(}_{T'} C _{T'/2})^{2}$.
Using this fact, we have the following lower bound for even $T$:
\begin{equation}
({}_{T'} C _{T'/2})^{2} = 2^{T - O(\log T)} \leq N_{{\rm lm},T}.
\label{equation:eq022}
\end{equation}
Similarly, for odd $T'$ we have 
\begin{equation}
({}_{T'} C _{(T'-1)/2})^{2} = 2^{T - O(\log T)} \leq N_{{\rm lm},T}.
\label{equation:eq028}
\end{equation}

For 2PATH, we have an upper bound
\begin{equation}
N_{{\rm lm},T} \leq 
1 + (T + 1)(2T + 1)^{2}(1 + 4 + \ldots + 4^{T})(2^{4})
= 2^{2T + O(\log T)},
\label{equation:eq019}
\end{equation}
a lower bound for even $T'$
\begin{equation}
{}_{T'} C _{T'/2} = 2^{(1/2) T - O(\log T)} \leq N_{{\rm lm},T},
\label{equation:eq023}
\end{equation}
and a lower bound for odd $T'$
\begin{equation}
{}_{T'} C _{(T'-1)/2} = 2^{(1/2) T - O(\log T)} \leq N_{{\rm lm},T}.
\end{equation}
\label{equation:eq029}

\medskip

\subsection{Examples of applications of the local map algorithm}
\label{subsection:examples_of_local_map_algorithm}
 
We show two examples to determine the value of ${\rm mft}_{\Gamma}(C)$ 
with the local map algorithm.

We use one convention to show configurations of g-2PATH.
When we write a configuration of g-2PATH as 
$p_{r} \ldots p_{0} \ldots p_{s}$ ($r \leq 0 \leq s$) 
we assume that $p_{0}$ is the origin $(0, 0)$.
We call $p_{r} \ldots p_{0}$ and $p_{0} \ldots p_{s}$ 
the {\it left hand}\/ and the {\it right hand}\/ of 
the configuration respectively 
and call $p_{i-1}$, $p_{i+1}$ the position 
on the left and the position on the right of 
$p_{i}$ respectively.
Note that directions ``left,'' ``right'' are 
determined not by the configuration itself but 
by how we represent it as a path (a sequence) 
$p_{r} \ldots p_{0} \ldots p_{s}$.

\medskip

\begin{ex}
\label{example:ex001}
\end{ex}

\vspace*{-\medskipamount}

In the variation $\Gamma$ considered in this example, 
a configuration is 
a horizontal straight line 
of the form $p_{-a} p_{-a+1} \ldots p_{-1} p_{0} p_{1} 
\ldots p_{b-1} p_{b}$ 
($p_{k} = (k, 0)$ for $-a \leq k \leq b$) 
such that $0 \leq a \leq b \leq a + 2$.
We denote this configuration by $C_{a,b}$.

This is an artificial variation and is used only to 
show an example of applications of the local map 
algorithm.
At present we do not know whether this variation has 
a minimal-time solution or not.
However we can determine ${\rm mft}_\Gamma(C_{a,b})$ 
by the local map algorithm as follows:
\begin{equation*}
{\rm mft}_{\Gamma}(C_{a, a}) = 3 a, \\
{\rm mft}_{\Gamma}(C_{a, a + 1}) = 3 a + 2, \\
{\rm mft}_{\Gamma}(C_{a, a + 2}) = 3 a + 2.
\end{equation*}
We use the three examples $C_{3, 3}$, $C_{3, 4}$, $C_{3, 5}$ 
to show this result.

\medskip

\begin{figure}[htbp]
\begin{center}
\includegraphics[scale=1.0]{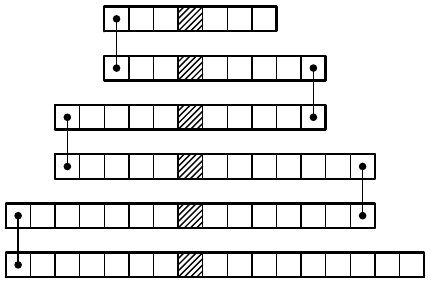}
\end{center}
\caption{A sequence of configurations that shows that 
time $8$ is safe for $C_{3,3}$.}
\label{figure:fig007}
\end{figure}

\begin{figure}[htbp]
\begin{center}
\includegraphics[scale=1.0]{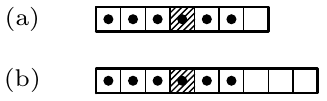}
\end{center}
\caption{Positions that appear in the third components of 
${\rm ai}((-3, 0), 8, C_{3, 3})$ (in (a)) and 
${\rm ai}((-3, 0), 8, C_{3, 5})$ (in (b)).}
\label{figure:fig008}
\end{figure}

\noindent
(1) Proof of ${\rm mft}_{\Gamma}(C_{3,3}) = 9$.
Fig. \ref{figure:fig007} shows that time $8$ is safe for $C_{3,3}$.
In this figure, a line between $v$ in $C$ and $v$ in $C'$ 
means that 
${\rm ai}(v, t, C) = {\rm ai}(v, t, C') \not= {\rm Q}$ 
and hence $C \equiv_{t}' C'$ ($t = 8$, in this case).
For example, the line between $(-3, 0)$ in $C_{3,3}$ and 
$(-3, 0)$ in $C_{3, 5}$ is justified by 
${\rm ai}((-3, 0), 8, C_{3,3}) = (8, (-3, 0), X)$, 
${\rm ai}((-3, 0), 8, C_{3,5}) = (8, (-3, 0), X')$, where 
\begin{eqnarray*}
X = \{(v', {\rm bc}_{C_{3,3}}(v')) ~|~ \text{$v'$ is a node with 
a dot in Fig. \ref{figure:fig008}(a)}\},\\
X' = \{(v', {\rm bc}_{C_{3,5}}(v')) ~|~ \text{$v'$ is a node with 
a dot in Fig. \ref{figure:fig008}(b)}\},
\end{eqnarray*}
and ${\rm bc}_{C_{3,3}}(v') = {\rm bc}_{C_{3,5}}(v')$ 
in the definitions of $X$, $X'$.
Fig. \ref{figure:fig007} shows that $C_{3,3} \equiv_{8}' C_{3,5} \equiv_{8}' 
C_{5,5} \equiv_{8}' C_{5, 7} \equiv_{8}' C_{7,7} \equiv_{8}' 
C_{7, 9}$.  However ${\rm rad}(C_{7,9}) = 9 > 8$.
Hence $8$ is safe for $C_{3,3}$.

We can show that time $9$ is unsafe for $C_{3,3}$ if we note 
that, for any node $v$ in $C_{3,3}$, 
the third component $X$ of ${\rm ai}(v, 9, C_{3,3})$ contains 
$(v', {\rm bc}(v'))$ for all $v'$ in $C_{3,3}$.
Thus ${\rm ai}(v, 9, C_{3,3})$ completely determines the 
structure of $C_{3,3}$ and 
if $C_{3,3} \equiv_{9}' C$ then $C = C_{3,3}$.
Therefore there does not exist $C$ such that 
$C_{3,3} \equiv_{9} C$ and ${\rm rad}(C) > 9$.

\medskip

\noindent
(2) Proof of ${\rm mft}_{\Gamma}(C_{3,4}) = 11$.
The proof is essentially the same as for the case (1) except that 
in Fig. \ref{figure:fig007} we replace the top 
configuration $C_{3,3}$ with $C_{3,4}$ and 
we add two more configurations 
$C_{9, 9}$ and $C_{9, 11}$ at the bottom.

\medskip

\begin{figure}[htbp]
\begin{center}
\includegraphics[scale=1.0]{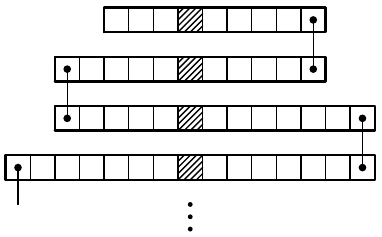}
\end{center}
\caption{A sequence of configurations that shows that 
time $10$ is safe for $C_{3,5}$.}
\label{figure:fig009}
\end{figure}

\noindent
(3) Proof of ${\rm mft}_{\Gamma}(C_{3,5}) = 11$.
The sequence of configurations shown in Fig. \ref{figure:fig009} 
shows that time $10$ is safe for $C_{3,5}$. 
In the figure we show only the beginning of the sequence.

\begin{figure}[htbp]
\begin{center}
\includegraphics[scale=1.0]{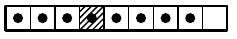}
\end{center}
\caption{Positions that appear in the third components of 
${\rm ai}((-3, 0), 11, C_{3, 5})$ and 
${\rm ai}((-2, 0), 11, C_{3, 5})$.}
\label{figure:fig010}
\end{figure}

The proof that time $11$ is unsafe for $C_{3,5}$ 
is a little different.
For any node $v$ in $C_{3,5}$ except $(-3, 0)$, $(-2, 0)$, 
${\rm ai}(v, 11, C_{3,5})$ completely determines 
the structure of $C_{3,5}$.
However, for $v = (-3, 0)$ and $v = (-2, 0)$, 
the third component $X$ of ${\rm ai}(v, 11, C_{3,5})$ 
contains $(v', {\rm bc}(v'))$ only for nodes with 
dots shown in Fig. \ref{figure:fig010}.
It does not contain the element for $(5, 0)$.
However it contains the element for $(4, 0)$ 
and its boundary condition $(1, 0, 1, 0)$ implies 
that the node $(5, 0)$ exists in the configuration.
Moreover the requirement $b \leq a + 2$ 
implies that the node $(6, 0)$ does not exist 
in the configuration.
Hence each of ${\rm ai}((-3, 0), 11, C_{3,5})$ and 
${\rm ai}((-2, 0), 11, C_{3,5})$ 
completely determines the structure of the configuration, 
and $C_{3,5} \equiv_{11} C$ implies 
$C = C_{3,5}$ and ${\rm rad}(C) \leq 11$.
Therefore, the time $11$ is unsafe for $C_{3,5}$.
\medskip

Let $\Gamma'$ be the generalized FSSP.
Then $\Gamma'$ is a super-variation of $\Gamma$ and 
${\rm mft}_{\Gamma'}(C_{a, b}) = a + 2b$ by (\ref{equation:eq001}).
Hence $\Gamma'$ is a nonconservative super-variation 
of $\Gamma$.
Let $\Gamma''$ be the sub-variation of $\Gamma$ 
such that only $C_{a,a}$ are configurations ($0 \leq a$).
Then we can easily show that 
${\rm mft}_{\Gamma''}(C_{a,a}) = 2a$.
Hence $\Gamma''$ is a nonconservative sub-variation of 
$\Gamma$.
We have minimal-time solutions of $\Gamma'$, $\Gamma''$ 
but at present we have not for $\Gamma$.
\hfill (End of Example \ref{example:ex001})

\begin{ex}
\label{example:ex002}
\end{ex}

\vspace*{-\medskipamount}

\begin{figure}[htbp]
\begin{center}
\includegraphics[scale=1.0]{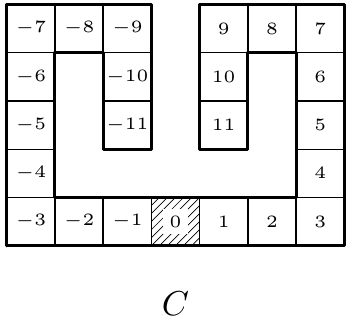}
\end{center}
\caption{The configuration $C$ of g-$2$PATH 
used in Example \ref{example:ex002}.}
\label{figure:fig011}
\end{figure}

In this example we consider the variation g-2PATH and 
show ${\rm mft}_\text{g-2PATH}(C)$ $= 30$ for the configuration 
$C = p_{-11} \ldots p_{-1} p_{0} p_{1} \ldots p_{11}$ 
shown in Fig. \ref{figure:fig011}.
In this figure (and in some later figures) 
we write only the index $i$ instead of $p_{i}$.
We proved this result in \cite{Kobayashi_TCS_1978}.
We use this example repeatedly in this paper.

\begin{figure}[htbp]
\begin{center}
\includegraphics[scale=1.0]{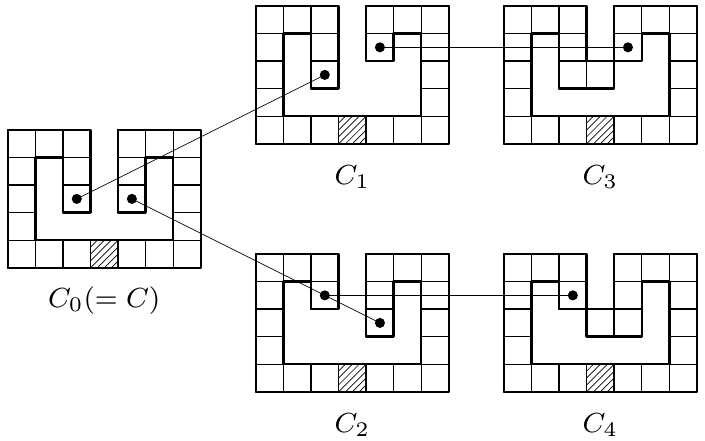}
\end{center}
\caption{All of the configurations $C'$ such that $C \equiv_{30} C'$.}
\label{figure:fig012}
\end{figure}

\begin{figure}
\begin{center}
\includegraphics[scale=1.0]{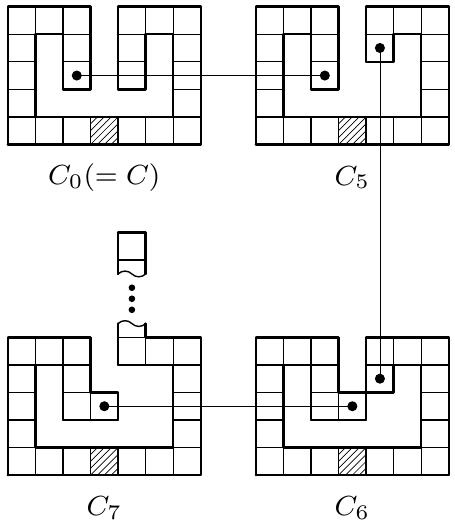}
\end{center}
\caption{Four configurations $C'$ such that $C \equiv_{29} C'$.}
\label{figure:fig013}
\end{figure}

Five configurations $C_{0}$ ($= C$), $C_{1}$, $C_{2}$, 
$C_{3}$, $C_{4}$ shown in Fig. \ref{figure:fig012} 
are all the configurations $C'$ such that $C \equiv_{30} C'$ 
(we give more details on this 
in Subsection \ref{subsection:example_application_consistency}) 
and all of them satisfy ${\rm rad}(C) \leq 30$.
Hence time $30$ is unsafe for $C$.
Fig. \ref{figure:fig013} shows a sequence of four 
configurations such that $C_{0} \equiv_{29}' C_{5} 
\equiv_{29}' C_{6} \equiv_{29}' C_{7}$ 
and ${\rm rad}(C_{7}) \geq 30$. 
(We select the extension of $C_{7}$ sufficiently long 
so that ${\rm rad}(C_{7}) = 30$.)
Therefore time $29$ is safe for $C$ and 
we have ${\rm mft}_\text{g-2PATH}(C) = 30$.

Suppose that $C'$ is the configuration that is obtained 
from $C$ by straightening it.
Then we can show 
${\rm mft}_\text{g-2PATH}(C') = 33$ 
by showing that time $32$ is safe and time $33$ is unsafe 
for $C'$.
Hence, by bending a straight line into the form of 
$C$ in Fig. \ref{figure:fig011}, the minimum firing time 
decreases by $3$.
\hfill (End of Example \ref{example:ex002})

\section{Consistent extensions}
\label{section:consistent_extensions}

\subsection{The basic step of the local map algorithm and 
consistent extensions}
\label{subsection:basic_step_and_consistent_extensions}

The basic step of the local map algorithm is 
the step to enumerate all configurations $C'$ 
such that $C \equiv_{t, v}' C'$ and 
${\rm rad}(C') \leq t+1$ for 
some given $t$, $C$, $v$ ($\in C$).
In Subsection \ref{subsection:algorithm} 
we gave the following 
algorithm for this step: enumerate all $C'$ such that 
${\rm rad}(C') \leq t+1$ and for each of them 
check whether 
${\rm ai}(v, t, C) = {\rm ai}(v, t, C') \not= {\rm Q}$ 
or not.
This algorithm is sufficient to show that 
${\rm mft}_\text{2REG}(C)$ is a computable function.
However, the algorithm is not practically feasible 
because to enumerate all $C'$ such that 
${\rm rad}(C') \leq t+1$ is impossible 
even for a small value of $t$.
In this section we show a more efficient algorithm 
for the enumeration.

Let $C$ be a configuration of a variation $\Gamma$, 
$M$ be a subset of $C$, and $C'$ be another configuration of $\Gamma$.
We say that $C'$ is a {\it consistent extension}\/ of a subset $M$ of $C$ 
if $M$ is a subset of $C'$ and 
${\rm bc}_{C}(v) = {\rm bc}_{C'}(v)$ for any node $v$ in $M$.
Although this is a relation among three objects $C$, $M$, $C'$, 
usually $C$ is a fixed configuration and 
only $M$, $C'$ vary.
In such cases we simply say that 
{\it $C'$ is a consistent extension of $M$}\/.
\begin{figure}[htbp]
\begin{center}
\includegraphics[scale=1.0]{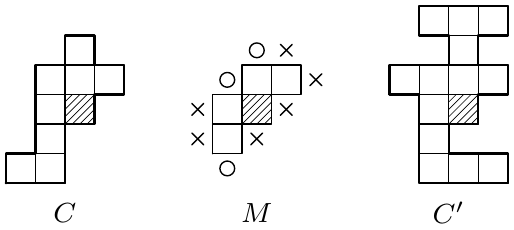}
\end{center}
\caption{An example of a consistent extension $C'$ of a 
subset $M$ of $C$.}
\label{figure:fig014}
\end{figure}
In Fig. \ref{figure:fig014} we show an example of 
$C$, $M$, $C'$ such that $C'$ is a 
consistent extension of a subset $M$ of $C$.
In the figure of $M$, a circle is a position which 
$C'$ should contain  
and a cross is a position which 
$C'$ should not contain for $C'$ to be a 
consistent extension of $M$.

For a configuration $C$, a node $v$ in $C$, and a time $t$ 
such that ${\rm d}_{C}(v_{\rm gen}, v) \leq t$, 
let $M(v, t, C)$ be the set 
\[
M(v, t, C) = \{v' \in C ~|~ {\rm d}_{C}(v_{\rm gen}, v') + 
{\rm d}_{C}(v', v)  \leq t \}.
\]
Note that 
${\rm ai}(v, t, C) \not= {\rm Q}$ and 
the third component of ${\rm ai}(v, t, C)$ is 
$\{(v', {\rm bc}_{C}(v')) ~|~$ $v' \in M(v, t, C)\}$.

We can show that if $C \equiv_{t, v}' C'$ then 
$C'$ is a consistent extension of 
$M(v, t, C)$ (as a subset of $C$).
However, the inverse is not necessarily true.
To show the inverse we need an additional condition.

\begin{thm}
\label{theorem:th000}
Let $C$, $C'$ be two configurations of $\Gamma$, 
$v$ be a node in $C$, and $t$ be a time such that 
${\rm d}_{C}(v_{\rm gen}, v) \leq t$.
Then $C \equiv_{t, v}' C'$ if and only if 
\begin{enumerate}
\item[{\rm (1)}] $C'$ is a consistent extension of 
$M(v, t, C)$, and
\item[{\rm (2)}] $M(v, t, C') - M(v, t, C) = \emptyset$.
\end{enumerate}
\end{thm}

%\begin{pf}
\noindent
{\it Proof}. 
The proof of the only if part ($\Longrightarrow$).

Suppose that $C \equiv_{t, v}' C'$.
Then $v \in C'$ and 
${\rm ai}(v, t, C) = {\rm ai}(v, t, C') 
\not= {\rm Q}$.
Let $X$, $X'$ be the third components of 
${\rm ai}(v, t, C)$ and ${\rm ai}(v, t, C')$ 
respectively.
Then $X = X'$ and hence 
$M(v, t, C) = M(v, t, C') \subseteq C'$.
If $v'$ is in $M(v, t, C)$ then $(v', {\rm bc}_{C}(v'))$  is 
in $X$ and hence in $X'$.
This means that ${\rm bc}_{C}(v') = {\rm bc}_{C'}(v')$ 
for any $v'$ in $M(v, t, C')$.
Therefore $C'$ is a consistent extension of $M(v, t, C')$.
Moreover $M(v, t, C') - M(v, t, C)$ is empty 
because $M(v, t, C) = M(v, t, C')$.
\medskip

\noindent
The proof of the if part ($\Longleftarrow$).

Suppose that (1), (2) are true.
Let $U$ be a shortest path from $v_{\rm gen}$ to $v$ in $C$.
Then $|U| - 1 = {\rm d}_{C}(v_{\rm gen}, v) \leq t$ 
and any node in $U$ is in $M(v, t, C)$ and consequently 
is in $C'$.
Hence $U$ is in $C'$ and 
${\rm d}_{C'}(v_{\rm gen}, v) \leq |U| - 1 \leq t$.
Let $v'$ be any element of $M(v, t, C)$ and 
let $V$, $W$ be a shortest path from $v_{\rm gen}$ to $v'$ 
and a shortest path from $v'$ to $v$ in $C$.
Then $(|V| - 1) + (|W| - 1) = {\rm d}_{C}(v_{\rm gen}, v') + 
{\rm d}_{C}(v', v) \leq t$ and 
any node in $V$ and $W$ is in $M(v, t, C)$ and hence 
is in $C'$.
Therefore, $V$, $W$ are in $C'$ and 
${\rm d}_{C'}(v_{\rm gen}, v') + {\rm d}_{C'}(v', v) 
\leq (|V| - 1) + (|W| - 1) \leq t$ 
and $v'$ is in $M(v, t, C')$.
Therefore $M(v, t, C) \subseteq M(v, t, C')$ 
and hence $M(v, t, C) = M(v, t, C')$ by (2).
By (1), for any $v'$ in $M(v, t, C)$, 
${\rm bc}_{C}(v') = {\rm bc}_{C'}(v')$.
Hence ${\rm ai}(v, t, C) = 
{\rm ai}(v, t, C') \not= {\rm Q}$, and 
$C \equiv_{t, v}' C'$.
\hfill $\Box$
%\end{pf}

\medskip

This theorem gives a new algorithm for 
the basic step of the local map algorithm, 
that is, to enumerate all 
configurations $C'$ that satisfy the 
condition ${\rm rad}(C') \leq t + 1$ and 
the two conditions (1), (2) mentioned in the theorem.
All of these conditions are restated as 
the following four conditions:
\begin{enumerate}
\item[(1')] ${\rm rad}(C') \leq t + 1$.
\item[(2')] $C'$ contains $M(v, t, C)$.
\item[(3')] For any position $v'$ that is not 
in $M(v, t, C)$ but is adjacent to a position in 
$M(v, t, C)$ then $v' \in C$ if and only if 
$v' \in C'$.
\item[(4')] There is not a node $v'$ in $C'$ that is not 
in $M(v, t, C)$ but is on a path in $C'$ from $v_{\rm gen}$ to 
$v$ of length at most $t$.
\end{enumerate}
Before we start the algorithm, we determine the 
set $M(v, t, C)$.
Moreover, for each position $v'$ that is not in 
$M(v, t, C)$ but is adjacent to a node in 
$M(v, t, C)$, we mark the position with a circle 
if $v' \in C$ and with a cross otherwise 
as is shown in Fig. \ref{figure:fig014}.
These marks simplify checking the condition (3').
In subsection 
\ref{subsection:example_application_consistency}, 
we show an example of applications of 
the new algorithm.

\subsection{Variations for which the condition (2) 
is not necessary in Theorem \ref{theorem:th000}}
\label{subsection:two_conditions}

\begin{figure}
\begin{center}
\includegraphics[scale=1.0]{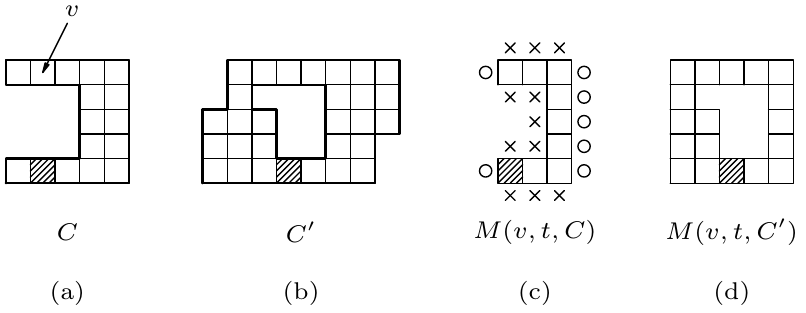}
\end{center}
\caption{An example of $C$, $C'$, $v \in C$ 
such that 
$C \equiv_{v,t}' C'$ is not true and 
$C'$ is a consistent extension of 
$M(v, t, C)$.}
\label{figure:fig015}
\end{figure}

The two statements 
``$C \equiv_{t,v}' C'$'' and 
``$C'$ is a consistent extension of 
$M(v, t, C)$'' are not equivalent.
We show an example for 2REG.
Suppose that $C$, $C'$ are the configurations of 2REG 
shown in the figures (a) and (b) 
of Fig. \ref{figure:fig015} respectively, 
$v$ is the node shown in the figure (a), 
and $t = 9$.
Then we have ${\rm d}_{C}(v_{\rm gen}, v) \leq t$.
In this case $M(v, t, C)$ and $M(v, t, C')$ are 
the sets of nodes shown in the figures (c) and (d) 
respectively.
Therefore $C \equiv_{v,t}' C'$ is not true 
because $M(v, t, C) \not= M(v, t, C')$ 
but 
$C'$ is a consistent extension of 
$M(v, t, C)$ 
(see the circles and the crosses in figure (c)).

For many variations of FSSP 
the condition (1) implies the condition (2) 
in Theorem \ref{theorem:th000} and hence 
$C \equiv_{t, v}' C'$ if and only if 
$C'$ is a consistent extension of $M(v, t, C)$.
We show two examples.

The first example is variations such 
that each configuration is a convex subset of 
the grid space $\mathbf{Z}^{k}$. 
For a subset $X$ of the grid space, 
we say that $X$ is {\it convex}\/ if for any 
two elements $v$, $v'$ of $X$, any shortest path 
between $v$, $v'$ in the grid space is in $X$.
Examples of variations having convex configurations are 
the FSSP for rectangles, 
the FSSP for 
cuboids (rectangular parallelepipeds), 
and their sub-variations, 
for example, the FSSP for rectangles of 
size $m \times n$ 
such that $m \leq n \leq 2m$.

\begin{thm}
\label{theorem:th001}
Suppose that all configurations of a variation $\Gamma$ are 
convex.
Then, under the assumptions of Theorem 
\ref{theorem:th000}, 
$C \equiv_{t,v}' C'$ if and only if 
$C'$ is a consistent extension of $M(v, t, C)$.
\end{thm}

%\begin{pf}
\noindent
{\it Proof}. 
It is sufficient to show that the condition (1) 
implies the condition (2) in Theorem 
\ref{theorem:th000}.
Suppose that $C'$ is a consistent extension of 
$M(v, t, C)$.
We assume that $M(v, t, C') - M(v, t, C)$ contains a 
node $v'$ and derive a contradiction.  
Note that $v$ is in both of $M(v, t, C)$ and $C'$ 
because ${\rm d}_{C}(v_{\rm gen}, v) \leq t$ and 
$M(v, t, C) \subseteq C'$.

\begin{figure}[htbp]
\begin{center}
\includegraphics[scale=1.0]{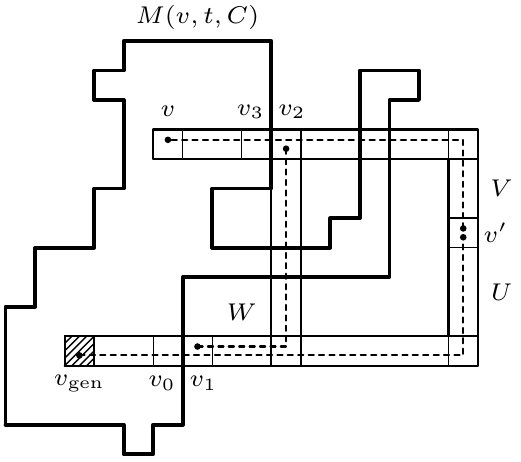}
\end{center}
\caption{Nodes, paths, and sets of nodes used in the proof of 
Theorem \ref{theorem:th001}.}
\label{figure:fig016}
\end{figure}

Let $U$ be a shortest path from $v_{\rm gen}$ to $v'$ in $C'$.
Let $v_{1}$ be the first node in $U$ that is out of $M(v, t, C)$ 
and $v_{0}$ be its preceding node in $U$.
Let $V$ be a shortest path from $v'$ to $v$ in $C'$.
Let $v_{2}$ be the last node in $V$ that is out of $M(v, t, C)$ 
and $v_{3}$ be the following node in $V$ 
(see Fig. \ref{figure:fig016}).
The part of $U$ from $v_{\rm gen}$ to $v_{0}$ is a shortest path in $C'$.
But it is also a path in $C$.  
Hence we have 
${\rm d}_{C}(v_{\rm gen}, v_{0}) \leq 
{\rm d}_{C'}(v_{\rm gen}, v_{0})$.
It is easy to see that 
${\rm d}_{C'}(v_{\rm gen}, v_{0}) + 1 = 
{\rm d}_{C'}(v_{\rm gen}, v_{1})$.
Similarly we have 
${\rm d}_{C}(v_{3}, v) \leq 
{\rm d}_{C'}(v_{3}, v)$ and 
$1 + {\rm d}_{C'}(v_{3}, v) = 
{\rm d}_{C'}(v_{2}, v)$.

Both of $v_{0}$, $v_{1}$ are in $C'$ and adjacent, 
$v_{0}$ is in $C$, and 
${\rm bc}_{C}(v_{0}) = {\rm bc}_{C'}(v_{0})$ 
because $C'$ is a consistent extension of 
$M(v, t, C)$.
Hence $v_{1}$ is in $C$.
Similarly, $v_{2}$ is in both of $C$, $C'$.
Let $W$ be a shortest path from $v_{1}$ to $v_{2}$ 
in $\mathbf{Z}^{2}$.
Then $W$ is in both of $C$, $C'$ because 
$C$, $C'$ are convex.
Therefore we have 
${\rm d}_{C}(v_{1}, v_{2}) = {\rm d}_{C'}(v_{1}, v_{2})$.

Finally we have 
\begin{eqnarray*}
\lefteqn{{\rm d}_{C}(v_{\rm gen}, v_{1}) + {\rm d}_{C}(v_{1}, v)} \\
& \leq & 
{\rm d}_{C}(v_{\rm gen}, v_{0}) + 1 + {\rm d}_{C}(v_{1}, v_{2}) + 1 + 
{\rm d}_{C}(v_{3}, v) \\
& \leq & 
{\rm d}_{C'}(v_{\rm gen}, v_{0}) + 1 + {\rm d}_{C'}(v_{1}, v_{2}) + 1 + 
{\rm d}_{C'}(v_{3}, v) \\
& = & 
{\rm d}_{C'}(v_{\rm gen}, v_{1}) + {\rm d}_{C'}(v_{1}, v_{2}) + 
{\rm d}_{C'}(v_{2}, v) \\
& \leq & 
{\rm d}_{C'}(v_{\rm gen}, v_{1}) + 
{\rm d}_{C'}(v_{1}, v') + 
{\rm d}_{C'}(v', v_{2}) + 
{\rm d}_{C'}(v_{2}, v) \\
& = & 
{\rm d}_{C'}(v_{\rm gen}, v') + {\rm d}_{C'}(v', v) \\
& \leq & t.
\end{eqnarray*}
The final step follows from 
$v' \in M(v, t, C')$.
This means that $v_{1} \in M(v, t, C)$ and this is a contradiction.
\hfill $\Box$
%\end{pf}

\medskip

The second example is sub-variations of g-$k$PATH, 
for example, 
the FSSP for configurations $p_{r} \ldots p_{0} \ldots p_{s}$ 
of g-$2$PATH such that the two end positions 
$p_{r}$, $p_{s}$ touch with corners 
(that is, $p_{s} - p_{r}$ is one of 
$(1, 1)$, $(-1, -1)$, $(1, -1)$, $(-1, 1)$).

\begin{thm}
\label{theorem:th002}
Suppose that $\Gamma$ is 
a sub-variation of ${\rm g}\text{-}k{\rm PATH}$ 
{\rm (}$k = 2, 3${\rm )}.
Then, under the assumptions of Theorem 
\ref{theorem:th000}, 
$C \equiv_{t,v}' C'$ if and only if 
$C'$ is a consistent extension of $M(v, t, C)$.
\end{thm}

%\begin{pf}
\noindent
{\it Proof}. 
It is sufficient to show that the condition (1) 
implies the condition (2) in Theorem 
\ref{theorem:th000}.
Let $C$ be of the form $p_{r} \ldots p_{s}$, 
$v$ be the position $p_{u}$ 
($\max \{r, -t\} \leq u \leq \min \{s, t\}$), 
and suppose that 
$C'$ is a consistent extension of $M(p_{u}, t, C)$.
We assume that $M(p_{u}, t, C') - M(p_{u}, t, C)$ contains a node $v'$ 
and derive a contradiction.

We first determine the set $M(p_{u}, t, C)$.
By a detailed case analysis we can show that, 
for each $p_{k}$ ($r \leq k \leq s$),
\begin{eqnarray*}
{\rm d}_{C}(p_{0}, p_{k}) + {\rm d}_{C}(p_{k}, p_{u}) \leq t 
& \Longleftrightarrow & |k| + |k - u| \leq t \\
& \Longleftrightarrow & \lceil (u - t) / 2 \rceil \leq k 
\leq \lfloor (u + t) / 2 \rfloor.
\end{eqnarray*}
Hence $M(p_{u}, t, C)$ is the part $p_{a} \ldots p_{b}$ of 
$p_{r} \ldots p_{s}$, where
\begin{equation}
a = \max \{r, \lceil (u - t) / 2 \rceil \}, 
b = \min \{s, \lfloor (u + t) / 2 \rfloor \}.
\label{equation:eq007}
\end{equation}
Note that $a \leq u \leq b$ and hence $p_{u}$ is in the 
part $p_{a} \ldots p_{b}$.

$C'$ is a consistent extension of $p_{a} \ldots p_{b}$ 
and hence $C'$ is of the form 
$w_{0} p_{a} \ldots p_{b} w_{1}$ with 
some (possibly empty) sequences of positions $w_{0}, w_{1}$, 
and ${\rm bc}_{C}(p_{a}) = {\rm bc}_{C'}(p_{a})$, 
${\rm bc}_{C}(p_{b}) = {\rm bc}_{C'}(p_{b})$.
$v'$ is either in $w_{0}$ or in $w_{1}$.
Suppose that $v'$ is in $w_{0}$.
Then $w_{0}$ is not empty, $p_{a-1}$ exists in 
$C = p_{r} \ldots p_{s}$, and $w_{0}$ ends with $p_{a-1}$.
From this we have
\begin{eqnarray*}
{\rm d}_{C}(p_{0}, p_{a-1}) + {\rm d}_{C}(p_{a-1}, p_{u}) 
& = & d_{C'}(p_{0}, p_{a-1}) + {\rm d}_{C'}(p_{a-1}, p_{u}) \\
& \leq & {\rm d}_{C'}(p_{0}, v') + {\rm d}_{C'}(v', p_{u}) \\
& \leq & t,
\end{eqnarray*}
contradicting $p_{a - 1} \not\in M(p_{u}, t, C)$.
Similarly we have a contradiction if $v'$ is in $w_{1}$.
\hfill $\Box$
%\end{pf}

\medskip

\subsection{An example of applications of 
Theorem \ref{theorem:th000}}
\label{subsection:example_application_consistency}

We show an example of applications of 
Theorem \ref{theorem:th000} in using 
the local map algorithm.

\medskip

\begin{ex}
\label{example:ex011}
\end{ex}

\vspace*{-\medskipamount}

In Example \ref{example:ex002}, for the configuration 
$C$ $(= C_{0})$ of g-2PATH 
shown in Fig. \ref{figure:fig011} 
we enumerated all $C'$ such that 
$C_{0} \equiv_{30} C'$.
Here we explain how we can enumerate these $C'$ using 
Theorem \ref{theorem:th000}.
Note that 
by Theorem \ref{theorem:th002} 
the condition (2) is not necessary in 
Theorem \ref{theorem:th000} and 
$C \equiv_{v,t}' C'$ if and only if 
$C'$ is a consistent extension of 
$M(v, t, C)$.

First we explain how to enumerate $C'$ such that 
$C_{0} \equiv_{30}' C'$.
For each $p_{u}$ in $C_{0} = p_{-11} \ldots p_{11}$ we have 
${\rm d}_{C_{0}}(p_{0}, p_{u}) \leq 30$.
Hence $C \equiv_{30}' C'$ if and only if 
$C_{0} \equiv_{30, p_{u}}' C'$ for some 
$p_{u}$, and hence if and only if 
$C'$ is a consistent extension of 
$M(p_{u}, 30, C_{0})$ for some $p_{u}$ 
($-11 \leq u \leq 11$).

\begin{figure}
\begin{center}
\includegraphics[scale=1.0]{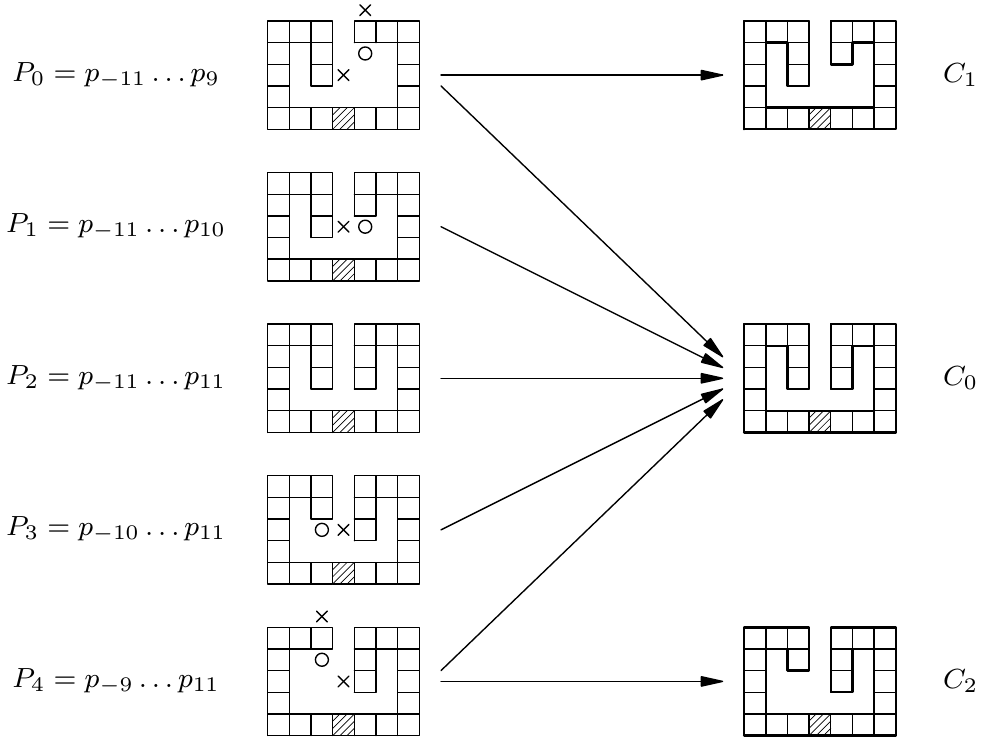}
\end{center}
\caption{Five paths $P_{0}, \ldots, P_{4}$ in $C_{0}$ and 
their three consistent extensions $C_{0}, C_{1}, C_{2}$.}
\label{figure:fig017}
\end{figure}

In the leftmost column in Table  \ref{table:tab000} 
we show values of $a, b$ defined by 
(\ref{equation:eq007}) 
for each $p_{u}$ in $C_{0}$ 
(the values such that $M(p_{u}, 30, C_{0}) = p_{a} \ldots p_{b}$).
\begin{table}[htbp]
\begin{center}
\includegraphics[scale=1.0]{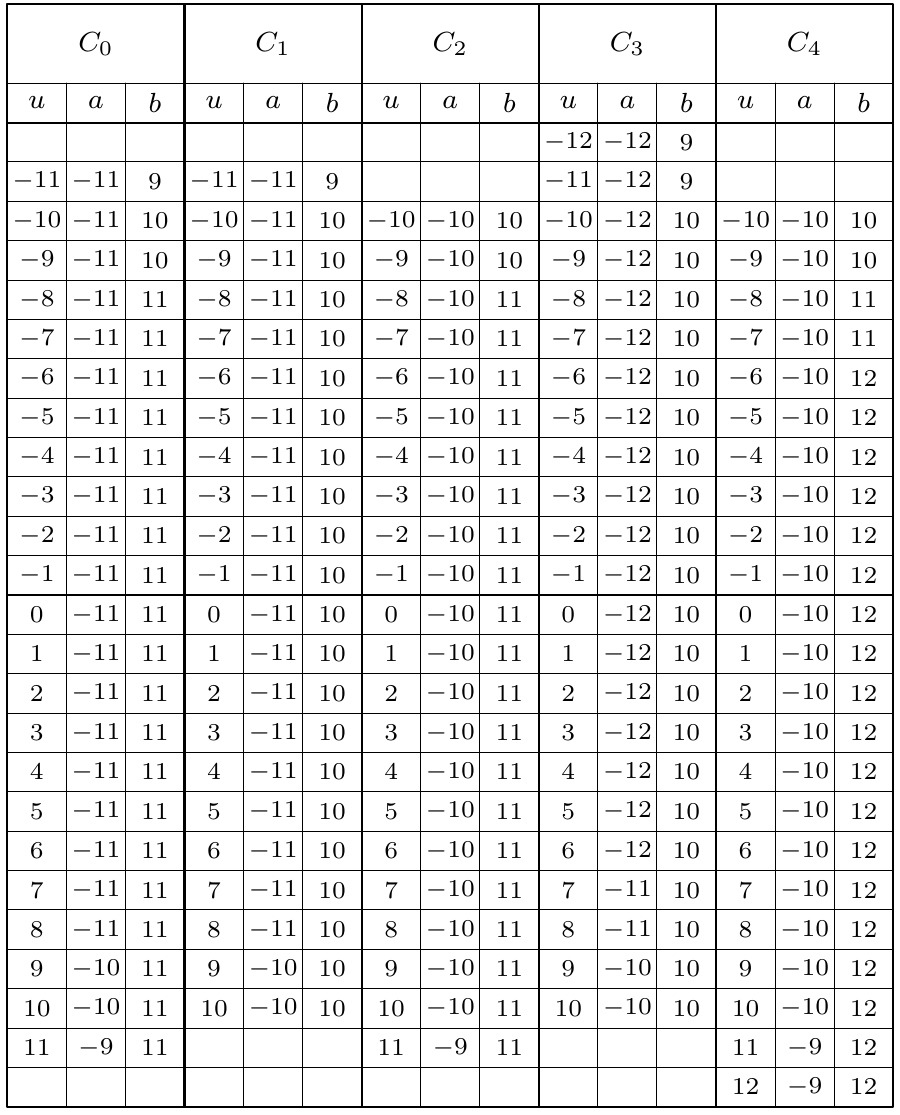}
\end{center}
\caption{The values of $a$, $b$ for $p_{u}$ in  $C_{k}$.}
\label{table:tab000}
\end{table}
This shows that $M(p_{u}, 30, C_{0})$ is one of the five paths 
$P_{0} = p_{-11} \ldots p_{9}$, 
$P_{1} = p_{-11} \ldots p_{10}$, 
$P_{2} = p_{-11} \ldots p_{11}$, 
$P_{3} = p_{-10} \ldots p_{11}$, 
$P_{4} = p_{-9} \ldots p_{11}$.
Hence $C_{0} \equiv_{30}' C'$ if and only if 
$C'$ is a consistent extension of one of these five 
paths.
In the left part of Fig. \ref{figure:fig017} we show these five paths 
and in the right part we list their consistent extensions.
$P_{0}$ has two consistent extensions $C_{1}$, $C_{0}$, 
each of $P_{1}, P_{2}, P_{3}$ has only one 
consistent extension $C_{0}$, and 
$P_{4}$ has two consistent extensions $C_{0}$, $C_{2}$.
Hence configurations $C'$ such that $C_{0} \equiv_{30}' C'$ 
are $C_{0}, C_{1}, C_{2}$.
Table \ref{table:tab000} also shows the values of 
$a, b$ for 
$C_{1}, C_{2}$ and so on, and using these values we can 
enumerate all $C'$ such that $C_{0} \equiv_{30} C'$, 
that is, $C_{0}, C_{1}, C_{2}, C_{3}, C_{4}$ shown in 
Fig. \ref{figure:fig012}.
In the columns for $C_{3}$, $C_{4}$, $p_{-12}$ and $p_{12}$ 
denote the positions $p_{-12} = p_{-11} + (1, 0)$, 
$p_{12} = p_{11} + (-1, 0)$ respectively that are not in $C$.
\hfill (End of Example \ref{example:ex011})

\section{The main result}
\label{section:main_result}

\subsection{Reflection partial solutions}
\label{subsection:reflection_partial_solutions}

Now we are ready to explain our main result.
For each configuration $C$ of g-2PATH we 
define a partial solution $A_{C}$ of g-2PATH.
Its domain contains $C$ and its firing time 
${\rm ft}(C, A_{C})$ for $C$ is represented by 
a simple formula (the formula (\ref{equation:eq008})).
Hence we have a simple formula that is an upper bound 
of ${\rm mft}_\text{g-2PATH}(C)$.

This upper bound is larger than 
${\rm mft}_\text{g-2PATH}(C)$ for some $C$ 
(Example \ref{example:ex003}).
However, for configurations $C$ that satisfy a condition 
which we call the {\it condition of noninterference of extensions}\/ 
the upper bound is the exact value of 
${\rm mft}_\text{g-2PATH}(C)$ (Theorem \ref{theorem:th005}).
Hence ${\rm mft}_\text{g-2PATH}(C)$ has 
a simple characterization for configurations 
satisfying the condition.

Suppose that $C = p_{r} \ldots p_{s}$ is a configuration of 
g-2PATH and is fixed.
We define some sets and values that depend on $C$.
The variables $i$, $j$ denote values such that 
$r \leq i \leq 0 \leq j \leq s$. 
\begin{eqnarray*}
W(i, j) & = & 
\{ (x_{0}, x_{1}) ~|~ 
\text{$x_{0}, x_{1}$ are sequences of positions and 
$x_{0} p_{i} \ldots p_{j} x_{1}$ is } \\
& & \quad\quad\quad \text{a consistent extension of $p_{i} \ldots p_{j}$ 
(as a subset of } \\
& & \quad\quad\quad \text{$p_{r} \ldots p_{s}$)} \}, \\
U(i, j) & = & \{x_{0} ~|~ \exists x_{1} [ (x_{0}, x_{1}) \in W(i, j) ] \}, \\
V(i, j) & = & \{x_{1} ~|~ \exists x_{0} [ (x_{0}, x_{1}) \in W(i, j) ] \}, \\
f(i, j) & = & \max \{|x_{0}| ~|~ x_{0} \in U(i, j) \}, \\
g(i, j) & = & \max \{|x_{1}| ~|~ x_{1} \in V(i, j) \}.
\end{eqnarray*}
Here $|w|$ denotes the length of a sequence $w$ of positions 
(the number of positions in $w$).
All of $W(i, j)$, $U(i, j)$, $V(i, j)$ are nonempty.

Let $\tilde{T}$ be defined by 
\begin{equation}
\tilde{T} = 
\min_{i, j} \max \{ - 2i + j + g(i, j), 2j - i + f(i, j) \}.
\label{equation:eq008}
\end{equation}
The values $f(i, j)$, $g(i, j)$ may be $\infty$ for 
some $i, j$.
However it is well-defined because 
$W(r, s) = \{(\epsilon, \epsilon)\}$, 
$U(r, s) = V(r, s) = \{\epsilon\}$ 
and hence $f(r, s) = g(r, s) = 0$
($\epsilon$ denotes the empty sequence).
We show that $\tilde{T}$ is an upper bound of 
${\rm mft}_\text{g-2PATH}(C)$.
Let $h(i, j)$ be the function
\[
h(i, j) = \max \{ -2i + j + g(i, j), 2j - i + f(i, j) \}
\]
so that $\tilde{T} = \min_{i,j} h(i, j)$.

\begin{thm}
\label{theorem:th003}
For any configuration $C$ of ${\rm g}\text{-}2{\rm PATH}$, 
\[
{\rm mft}_{{\rm g}\text{-}{\rm 2PATH}}(C) \leq \tilde{T}
= \min_{i, j} \max \{ - 2i + j + g(i, j), 2j - i + f(i, j) \}.
\]
\end{thm}

%\begin{pf}
\noindent
{\it Proof}. 
Let $i_{0}$, $j_{0}$ be values of $i, j$ that 
minimize $h(i, j)$.
(There may be more than one such pair $i_{0}, j_{0}$.)
We construct a partial solution $A_{C}$ of g-2PATH.
The domain of $A_{C}$ includes $C$ and $A_{C}$ fires 
$C$ at $\tilde{T}$.
This shows ${\rm mft}_\text{g-2PATH}(C) \leq \tilde{T}$.
Suppose that $C'$ is an arbitrary configuration of g-2PATH and 
copies of $A_{C}$ are placed at nodes of $C'$.

Let $y$, $y_{0}$, $y_{1}$ be the subsequences of $C$ defined by 
$y = p_{i_{0}} \ldots p_{j_{0}}$, 
$y_{0} = p_{i_{0}} \ldots p_{0}$, 
$y_{1} = p_{0} \ldots p_{j_{0}}$.
Note that $C'$ is a consistent extension of $y$ if and only if 
it is consistent extensions of both of $y_{0}$, $y_{1}$.
Intuitively a node of $C'$ fires at a time 
if and only if 
it knows that $C'$ is a consistent extension of $y$ 
and the current time is $\tilde{T}$.
To realize this, $A_{C}$ uses four signals 
${\rm R}$, ${\rm S}$, ${\rm R}'$, ${\rm S}'$.

The signal ${\rm R}$ starts at the general $p_{0}$ at 
time $0$ and proceeds to $p_{-1}$, $p_{-2}$, $\ldots$, $p_{i_{0}}$
to check whether $C'$ is a consistent extension of $y_{0}$.
If not, the signal ${\rm R}$ knows this at some point and vanishes.
If $C'$ is a consistent extension of $y_{0}$, the signal ${\rm R}$ knows 
this when it arrives at $p_{i_{0}}$ at time $-i_{0}$.
Then the signal ${\rm S}$ is generated at $p_{i_{0}}$ 
at time $- i_{0}$ and it is propagated to all nodes of $C'$.
Similarly, the signal ${\rm R}'$ checks whether 
$C'$ is a consistent extension of $y_{1}$.
If $C'$ is a consistent extension of $y_{1}$, 
the signal ${\rm S}'$ is generated at $p_{j_{0}}$ at 
time $j_{0}$ and is propagated to all nodes of $C'$.
A node fires at a time if and only if it has 
received both of ${\rm S}$, ${\rm S}'$ before or at 
the time and the current time is $\tilde{T}$.

If $C'$ is not a consistent extension of $y$, 
at least one of ${\rm S}$, ${\rm S}'$ is not 
generated and hence any node does not fire.
Suppose that $C'$ is a consistent extension of $y$ of the form 
$z_{0} y z_{1}$ with sequences $z_{0}$, $z_{1}$.
Then we have 
$|z_{0}| \leq f(i_{0}, j_{0})$, 
$|z_{1}| \leq g(i_{0}, j_{0})$.
${\rm S}$ is generated at $p_{i_{0}}$ at time $-i_{0}$.
A node in $z_{0}$ receives it before or at 
$-i_{0} + f(i_{0}, j_{0}) \leq 2j_{0} - i_{0} + f(i_{0}, j_{0}) \leq 
\tilde{T}$.
A node in $y z_{1}$ receives it before or at 
$-2i_{0} + j_{0} + g(i_{0}, j_{0}) \leq \tilde{T}$.
Similarly ${\rm S}'$ is generated at $p_{j_{0}}$ at 
time $j_{0}$ and any node in $z_{0} y z_{1}$ 
receives ${\rm S}'$ before or at $\tilde{T}$.
Therefore, any node of $C'$ 
receives both of ${\rm S}$, ${\rm S}'$ 
before or at $\tilde{T}$ and hence fires at time $\tilde{T}$.

$A_{C}$ is a partial solution of g-2PATH such 
that its domain is the set of all consistent extensions of 
$y$ (including $C$ itself) and it fires any configuration in its 
domain at $\tilde{T}$.
\hfill $\Box$
%\end{pf}

\medskip

We call the partial solution $A_{C}$ the {\it reflection partial 
solution}\/ for $C$ and denote it by $A_{{\rm ref}, C}$.
We use the term ``reflection'' because 
each of the two nodes $p_{i_{0}}$, $p_{j_{0}}$ plays 
the role of a half mirror that both reflects (the signals 
${\rm S}$, ${\rm S}'$ 
proceed back to the general $p_{0}$) 
and passes (${\rm S}$, ${\rm S}'$ continue to proceed to 
the terminal positions $p_{r}$, $p_{s}$) 
the wave coming from the general $p_{0}$ (the signals 
${\rm R}$, ${\rm R}'$).

Let $N_{{\rm ref}, C}$ denote the number of states of $A_{{\rm ref}, C}$.
We estimate the value.

$A_{{\rm ref}, C}$ simulates two finite automata.
One automaton simulates the signals ${\rm R}$, ${\rm S}$.
The generation and the move of the signal ${\rm R}$ are simulated 
with $-i_{0}$ states 
${\rm R}_{0}$, \ldots, ${\rm R}_{-i_{0}-1}$ 
and the generation and the propagation of the signal ${\rm S}$ 
are simulated with $\tilde{T} + i_{0} + 1$ states 
${\rm S}_{-i_{0}}$, \ldots, ${\rm S}_{\tilde{T}}$.
Together with the quiescent state ${\rm Q}$, 
the automaton has $\tilde{T} + 2$ states.
Another automaton simulates signals ${\rm R}'$, ${\rm S}'$ 
and has also $\tilde{T} + 2$ states 
${\rm R}'_{0}, \ldots, {\rm R}'_{j_{0}-1}, {\rm S}'_{j_{0}}, \ldots, 
{\rm S}'_{\tilde{T}}, {\rm Q}$.
Therefore, $A_{{\rm ref},C}$ has $(\tilde{T} + 2)^{2}$ states.

However we can reduce its size using the two facts: (1) the index 
$t$ of ${\rm R}_{t}$, ${\rm S}_{t}$ is the current time, 
(2) whether the state is ${\rm R}_{t}$ or ${\rm S}_{t}$ is 
determined by whether $t \leq -i_{0} - 1$ or $-i_{0} \leq t$, 
and similarly for ${\rm R}'_{t}$, ${\rm S}'_{t}$.
We can use 
$\{{\rm X}, {\rm Q}\} \times \{{\rm X}', {\rm Q}\} \times 
\{0, 1, \ldots, \tilde{T}, {\rm Q}\}$ as the set of states.
The first component represents whether the first automaton 
is in one of ${\rm R}_{0}, \ldots, {\rm S}_{\tilde{T}}$ or 
in ${\rm Q}$, and similarly for the second component.
The third component represents the current time 
($0, 1, \ldots, \tilde{T}$) or that the time is over 
$\tilde{T}$ (${\rm Q}$).
Hence we can reduce the size of $A_{{\rm ref}, C}$ 
from $(\tilde{T} + 2)^{2}$ to $2 \cdot 2 \cdot (\tilde{T} + 2)$ and 
\begin{equation}
N_{{\rm ref}, C} \leq 4 \tilde{T} + 8.
\label{equation:eq032}
\end{equation}

g-2PATH is a super-variation of 2PATH and 
$A_{{\rm ref}, C}$ can be used for 2PATH.
In this case the signals ${\rm R}$, ${\rm S}$ are not necessary 
(supposing that we represent configurations as $p_{0} \ldots p_{s}$), 
the set of states of 
$A_{{\rm ref},C}$ is $\{{\rm R}'_{0}, \ldots, {\rm R}'_{j_{0} - 1}$, 
${\rm S}'_{j_{0}}, \ldots, {\rm S}'_{\tilde{T}}$, ${\rm Q}\}$, 
and 
\begin{equation}
N_{{\rm ref},C} \leq \tilde{T} + 2.
\label{equation:eq031}
\end{equation}

Below we show two examples of configurations of g-2PATH.
For the first example (Example \ref{example:ex012}), 
$\tilde{T} = {\rm mft}_\text{g-2PATH}(C)$ and 
for the second example (Example \ref{example:ex003}), 
${\rm mft}_\text{g-2PATH}(C) < \tilde{T}$.

\medskip

\begin{figure}[htbp]
\begin{center}
\includegraphics[scale=1.0]{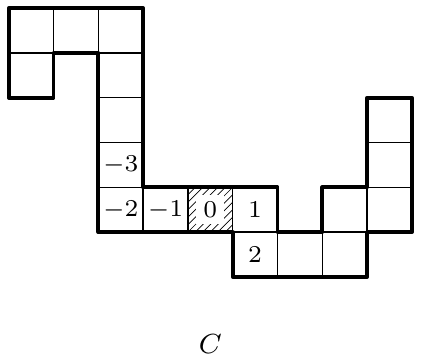}
\end{center}
\caption{An example of representations of paths.}
\label{figure:fig026}
\end{figure}

We introduce one method to represent paths in examples.
Let $C$ be the path shown in Fig. \ref{figure:fig026}.
We represent this path by 
${\rm SW}^{2}{\rm N}^{3}p_{-3}p_{-2}p_{-1}p_{0}p_{1}p_{2}
{\rm E}^{2}{\rm NEN}^{2}$.
The part ${\rm SW}^{2}{\rm N}^{3} = {\rm SWWNNN}$ 
represents the sequence of directions to go from 
$p_{-3}$ to the left terminal written in reverse order.
Similarly ${\rm E}^{2}{\rm NEN}^{2} = {\rm EENENN}$
is that for going from $p_{2}$ to the right terminal.
The letters ${\rm E}$, ${\rm N}$, ${\rm W}$, ${\rm S}$ 
represent the directions the east, the north, the west, 
the south, or the vectors $(1, 0)$, $(0, 1)$, $(-1, 0)$, $(0, -1)$, 
respectively.

\medskip

\begin{ex}
\label{example:ex012}
\end{ex}

\vspace*{-\medskipamount}

For the configuration $C$ of g-2PATH 
shown in Fig. \ref{figure:fig011} 
we showed ${\rm mft}_\text{g-2PATH}(C)$ $= 30$ 
(Example \ref{example:ex002}).
We construct $A_{{\rm ref}, C}$ and show that 
its firing time for $C$ is ${\rm mft}_\text{g-$2$PATH}(C)$ 
and hence ${\rm mft}_\text{g-$2$PATH}(C) = \tilde{T}$.

First we determine the values $f(i, j)$, $g(i, j)$.
For any $m$ ($\geq 0$) there is a path of the form 
${\rm N}^{m}{\rm E} p_{-8} \ldots p_{11}$.
Hence, for each $-8 \leq i$ we have $f(i, j) = \infty$.
Similarly, for each $j \leq 8$ we have $g(i, j) = \infty$.
Therefore $h(i, j)$ is finite only for 
$i \leq -9$, $9 \leq j$.

We have 
\begin{tabbing}
\quad\quad \= $f(-9, 9) = 3$, \quad \= $f(-9, 10) = 2$, 
           \quad \= $f(-9, 11) = 2$, \\
\> $f(-10, 9) = 2$, \> $f(-10, 10) = 1$, \> $f(-10, 11) = 1$, \\
\> $f(-11, 9) = 0$, \> $f(-11, 10) = 0$, \> $f(-11, 11) = 0$.
\end{tabbing}
For example, the consistent extension 
${\rm ESS} p_{-9} \ldots p_{9} {\rm S}$ 
of $p_{-9} \ldots p_{9}$ gives the value 
$f(-9, 9) = |{\rm ESS}| = 3$.
Values of $g(i, j)$ are determined by 
$g(i, j) = f(-j, -i)$ because $C$ is symmetric.
From these results, we know that 
the pair $i_{0} = -9$, $j_{0} = 9$ gives the minimum 
value $\tilde{T}$ of $h(i, j)$, and 
$\tilde{T} = h(-9, 9) = 
\max \{ 18 + 9 + 3, 18 + 9 + 3 \} = 30$.
Hence $A_{{\rm ref}, C}$ fires $C$ at $30$.

\begin{figure}[htbp]
\begin{center}
\includegraphics[scale=1.0]{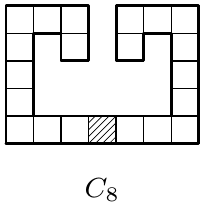}
\end{center}
\caption{The configuration $C_{8}$ in 
Example \ref{example:ex012}.}
\label{figure:fig018}
\end{figure}

The domain of $A_{{\rm ref}, C}$ is the set of all consistent extensions of 
the subsequence $p_{-9} \ldots p_{9}$ of $C$ and is 
the set $\{C_{0}, C_{1}, C_{2}, C_{3}, C_{4}, C_{8}\}$.
Here, the five configurations $C_{0}$, $C_{1}, \ldots, C_{4}$
are all of the configurations in the $\equiv_{30}$-equivalence 
class of $C$ ($= C_{0}$) shown in Fig. \ref{figure:fig012} and 
$C_{8}$ is the new configuration shown in Fig. \ref{figure:fig018}.
\hfill (End of Example \ref{example:ex012})

\begin{ex}
\label{example:ex003}
\end{ex}

\vspace*{-\medskipamount}

\begin{figure}[htbp]
\begin{center}
\includegraphics[scale=1.0]{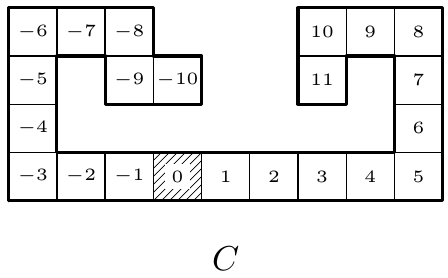}
\end{center}
\caption{The configuration $C$ of g-2PATH used in 
Example \ref{example:ex003}.}
\label{figure:fig019}
\end{figure}

Next we consider the configuration $C$ of g-2PATH shown in 
Fig. \ref{figure:fig019}.
For this $C$, first we show ${\rm mft}_\text{g-$2$PATH}(C) = 31$.
Then we construct $A_{{\rm ref}, C}$ and show that 
its firing time for $C$ is larger than ${\rm mft}_\text{g-$2$PATH}(C)$.

First we show ${\rm mft}_\text{g-2PATH}(C) = 31$. 
\begin{figure}[htbp]
\begin{center}
\includegraphics[scale=1.0]{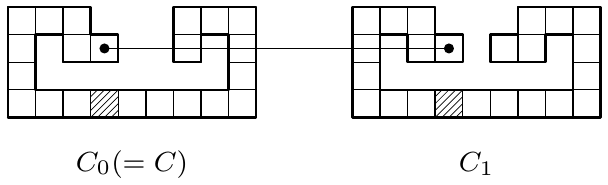}
\end{center}
\caption{All the configurations $C'$ such that $C \equiv_{31} C'$ 
in Example \ref{example:ex003}.}
\label{figure:fig020}
\end{figure}
Fig. \ref{figure:fig020} shows all the configurations $C'$ such that 
$C \equiv_{31} C'$, and ${\rm rad}(C') \leq 31$ for each of them.
Hence $31$ is unsafe for $C$ ($=C_{0}$).
\begin{figure}[htbp]
\begin{center}
\includegraphics[scale=1.0]{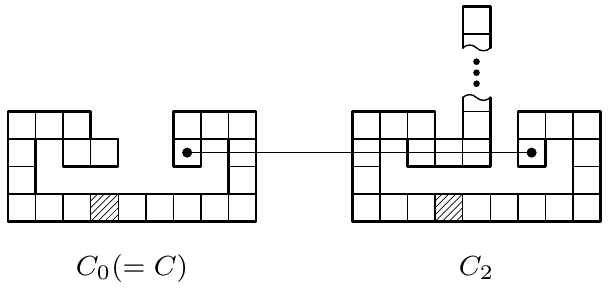}
\end{center}
\caption{Two configurations $C'$ such that $C \equiv_{30} C'$ in 
Example \ref{example:ex003}.}
\label{figure:fig021}
\end{figure}
Fig. \ref{figure:fig021} shows $C_{2}$ such that $C_{0} \equiv_{30} C_{2}$ and 
${\rm rad}(C_{2}) > 30$. (We select the extension of $C_{2}$ 
sufficiently long.)  Therefore $30$ is safe for $C$.  
Hence we have ${\rm mft}_\text{g-2PATH}(C) = 31$.

Next we construct $A_{{\rm ref}, C}$.
For any $m$ ($\geq 0$) there are paths of the forms 
${\rm N}^{m}{\rm E}^{2} p_{-9} \ldots p_{11}$ 
and 
$p_{-10} p_{-9} \ldots p_{9} {\rm W} {\rm N}^{m}$.
Hence $f(i, j) = \infty$ for any $-9 \leq i$ and $j$, 
and 
$g(i, j) = \infty$ for any $i$ and $j \leq 9$.
For $i = -10$ and $10 \leq j$ we have
$f(-10, 10) = 0$, $g(-10, 10) = 2$, $f(-10, 11) = 0$, $g(-10, 11) = 0$.
Both of the pairs $(i_{0}, j_{0}) = (-10, 10), (-10, 11)$ give the 
minimum value $\tilde{T}$ of $h(i, j)$:
\begin{eqnarray*}
\tilde{T} & = & h(-10, 10) = \max \{ 20 + 10 + 0, 20 + 10 + 2 \} \\
          & = & h(-10, 11) = \max \{ 22 + 10 + 0, 20 + 11 + 0 \} \\
          & = & 32.
\end{eqnarray*}
Therefore, each of the pairs $(-10, 10)$, $(-10, 11)$ 
defines $A_{{\rm ref},C}$ and its firing time is 
$\tilde{T} = 32$, larger than 
${\rm mft}_\text{g-$2$PATH}(C) = 31$.

The domain of $A_{{\rm ref}, C}$ determined by 
$(-10, 10)$ is $\{C_{0}, C_{1}\}$ and 
that determined by $(-10, 11)$ is $\{C_{0}\}$.
\hfill (End of Example \ref{example:ex003})

\subsection{The condition of noninterference of extensions}
\label{subsection:cni}

For a given fixed configuration $C = p_{r} \ldots p_{s}$ of 
g-2PATH 
we define three sets of pairs of indices:
\begin{eqnarray*}
I & = & \{ (i, j) ~|~ \text{$r \leq i - 1$, 
    $W(i, j)$ is an infinite set, 
    and} 
    \notag \\
  &   & \quad\quad\quad\quad\quad \text{$W(i - 1, j)$ is a finite set} \}, \\
J & = & \{ (i, j) ~|~ \text{$j + 1 \leq s$, 
    $W(i, j)$ is an infinite set, 
    and} \notag \\
  &   & \quad\quad\quad\quad\quad \text{$W(i, j + 1)$ is a finite set} \}, \\
K & = & \{ (i, j) ~|~ \text{$W(i, j)$ is a finite set} \}. 
\end{eqnarray*}
Note that although $K$ is always nonempty (because $W(r, s)$ is a 
finite set $\{(\epsilon, \epsilon)\}$ and $(r, s) \in K$), 
$I$, $J$ may be empty. 
$K$ and $I \cup J$ are disjoint but 
$I$ and $J$ may overlap.
By ${\rm NI}(i, j)$ we denote the statement 
``$W(i, j) = U(i, j) \times V(i, j)$'' (${\rm NI}$ is for 
{\it n}\/on-{\it i}\/nterference).

\begin{deff}
\label{definition:dfn000}
The {\it condition of noninterference 
of extensions} (abbreviated as CNI) for $C$ 
is the conjunction of the following three conditions.

\noindent
\quad The condition for $K$: for any $(i, j)$ in $K$, ${\rm NI}(i, j)$ 
is true.

\noindent
\quad The condition for $I$: for any $(i, j)$ in $I$, ${\rm NI}(i, j)$ 
is true.

\noindent
\quad The condition for $J$: for any $(i, j)$ in $J$, ${\rm NI}(i, j)$ 
is true.
\end{deff}

\noindent
The definition of the condition for $K$ given in 
Subsection \ref{subsection:main_result_and_implication} 
is slightly different from 
that given here but the two definitions are equivalent 
(because ${\rm NI}(i, j)$ is always true for $i = r$ or $j = s$).

\medskip

We use the following lemma in the proof of 
Theorem \ref{theorem:th005}.

\medskip

\begin{lem}
\label{lemma:lem004}
Suppose that $C = p_{r} \ldots p_{s}$ satisfies {\rm CNI}. 
Then {\rm (1)} if $(i, j) \in K$ and $i + 1 \leq 0$ then 
$V(i, j) = V(i+1, j)$, and 
{\rm (2)} if $(i, j) \in K$ and $0 \leq j-1$ then 
$U(i, j) = U(i, j-1)$.
\end{lem}

%\begin{pf}
\noindent
{\it Proof}. 
We prove only (1).
Suppose that $(i, j) \in K$ and $i + 1 \leq 0$.
${\rm NI}(i, j)$ is true.
Moreover, $(i + 1, j) \in I \cup K$ and hence 
${\rm NI}(i+1, j)$ is also true.
If $j = s$ then $V(i, j) = V(i+1, j)$ is true 
because both the sets are $\{ \epsilon \}$.
Hence we suppose that $j < s$.
We also suppose that $r \leq i-1$.
The proof for the case where $r = i$ is similar.
$V(i, j) \subseteq V(i+1, j)$ is evident.
To prove $V(i+1, j) \subseteq V(i, j)$, 
suppose that $p_{j+1} x$ is an element of $V(i+1, j)$.
$p_{i-1} p_{i} p_{i+1} \ldots p_{j} p_{j+1}$ is a configuration.
Hence $p_{i-1} p_{i}$ is in $U(i+1, j)$.
By ${\rm NI}(i+1, j)$, $(p_{i-1} p_{i}, p_{j+1} x)$ is 
in $W(i + 1, j)$ and 
the sequence 
$p_{i-1} p_{i} p_{i+1} \ldots p_{j} p_{j+1} x$ is 
a configuration.
Hence $p_{j+1} x$ is an element of $V(i, j)$.
\hfill $\Box$
%\end{pf}

\medskip

\begin{ex}
\label{example:ex004}
\end{ex}

\vspace*{-\medskipamount}

Any configuration of 
g-2PATH of the form $p_{0} \ldots p_{s}$ 
(a configuration of 2PATH) satisfies CNI.
This is obvious because 
any element of $W(0, j)$ is of the form 
$(\epsilon$, $y)$ for any $j$.
\hfill (End of Example \ref{example:ex004})

\begin{ex}
\label{example:ex007}
\end{ex}

\vspace*{-\medskipamount}

The configuration $C$ of g-2PATH 
shown in Fig. \ref{figure:fig011} does not satisfy CNI.
We determine the three sets $I, J, K$.
For the case $-8 \leq i$, $U(i, j)$ is an infinite set.
For the case $j \leq 8$, $V(i, j)$ is an infinite set.
For the case $i \leq -9$ and $9 \leq j$, both of $U(i, j)$, 
$V(i, j)$ are finite sets.
Hence
\begin{eqnarray*}
I & = & \{ (-8, 9), (-8, 10), (-8, 11) \}, \\
J & = & \{ (-11, 8), (-10, 8), (-9, 8) \}, \\
K & = & \{ (i, j) ~|~ -11 \leq i \leq -9, 9 \leq j \leq 11 \}.
\end{eqnarray*}

\begin{figure}
\begin{center}
\includegraphics[scale=1.0]{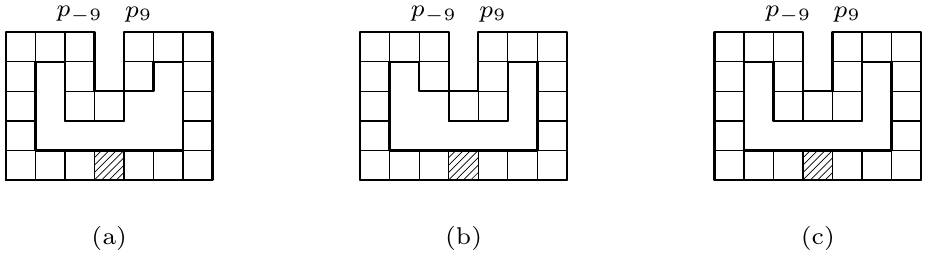}
\end{center}
\caption{Consistent extensions of sub-path $p_{-9} \ldots p_{9}$ 
of $C$ in Fig. \ref{figure:fig011} and 
their interference.}
\label{figure:fig023}
\end{figure}

For $(i, j) = (-9, 9) \in K$, 
there are two consistent extensions 
${\rm E} p_{-11} \ldots p_{9} p_{10}$, 
$p_{-10} \ldots p_{11} {\rm W}$
of 
$p_{-9} \ldots p_{9}$ 
(Fig. \ref{figure:fig023}(a), (b)).
Hence ${\rm E} p_{-11} p_{-10} \in U(-9, 9)$, 
$p_{10} p_{11} {\rm W} \in V(-9, 9)$.
However the sequence 
${\rm E} p_{-11} \ldots p_{11} {\rm W}$ 
is not a configuration because 
the position of ${\rm E}$ of ${\rm E} p_{-11}$ and 
the position of ${\rm W}$ of $p_{11} {\rm W}$ overlap 
(Fig. \ref{figure:fig023} (c)).
Hence 
$({\rm E} p_{-11} p_{-10}, p_{10} p_{11} {\rm W}) \not\in W(-9, 9)$ 
and $C$ does not satisfy the condition for $K$.
\hfill (End of Example \ref{example:ex007})

\medskip

\begin{ex}
\label{example:ex006}
\end{ex}

\vspace*{-\medskipamount}

The configuration $C$ shown in 
Fig. \ref{figure:fig019} also
does not satisfy CNI.
In this case, 
$I = \{(-9, 10), (-9, 11)\}$,
$J = \{(-10, 9)\}$,
$K = \{(-10, 10), (-10, 11)\}$.
$C$ satisfies the condition for $K$ because 
elements of $W(-10, 10)$, $W(-10, 11)$ are 
of the form $(\epsilon, y)$.
Both of $({\rm EE}, {\rm S})$, 
$({\rm E}, {\rm SW})$ are in $W(-9, 10)$ 
but $({\rm EE}, {\rm SW})$ is not in $W(-9, 10)$, 
and hence 
$W(-9, 10) \not= U(-9, 10) \times V(-9, 10)$ for 
$(-9, 10) \in I$.
Therefore $C$ does not satisfy the condition for $I$.
\hfill (End of Example \ref{example:ex006})

\subsection{A characterization of ${\rm mft}_\text{{\rm g-2PATH}}(C)$ for 
configurations satisfying CNI}
\label{subsection:characterization_for_cni_configurations}

\medskip

The following is the main result of this paper.

\begin{thm}
\label{theorem:th005}
If a configuration $C$ of 
${\rm g}\text{-}2{\rm PATH}$ satisfies 
the condition of noninterference of extensions then 
\[
{\rm mft}_{{\rm g}\text{-}{\rm 2PATH}}(C) = 
\min_{i, j} \max \{ - 2i + j + g(i, j), 2j - i + f(i, j) \}.
\]
\end{thm}

%\begin{pf}
\noindent
{\it Proof}. 
Let $\tilde{T}$ denote the right hand value of 
the equation in the theorem.
If $r = s = 0$ and hence 
$C$ is the configuration consisting only of the general $p_{0}$ 
then the theorem is true because 
${\rm mft}_\text{g-2PATH}(C) = 0$ 
(note that we use the boundary-sensitive model of FSSP 
and a solution can fire at time $0$) 
and $\tilde{T} = 0$.
Hence we assume that either $r < 0$ or $0 < s$.
Then we have $\tilde{T} > 0$ because 
for $i = j = 0$ either $f(i, j)$ or $g(i, j)$ 
is $\infty$ and hence 
the pair $(i, j)$ that minimizes $h(i, j)$ 
satisfies either $i < 0$ or $0 < j$ and 
$\tilde{T} = h(i, j) > 0$.
By Theorem \ref{theorem:th003} we know that 
${\rm mft}_\text{g-2PATH}(C) \leq \tilde{T}$.
Hence it is sufficient to prove 
$\tilde{T} \leq {\rm mft}_\text{g-2PATH}(C)$ and for this 
it is sufficient to show that 
$\tilde{T} - 1$ is safe for $C$ 
because ${\rm mft}_\text{g-2PATH}(C)$ is 
$\max\{ t ~|~ \text{$t$ is safe for $C$} \} + 1$.

Let $S(i, j)$ be the statement:
\begin{enumerate}
\item[$\bullet$] $(i, j) \in K$,
\item[$\bullet$] there is a consistent extension $C'$ of $p_{i} \ldots p_{j}$ 
(as a subset of $C$) 
of the form $C' = x_{0} p_{i} \ldots p_{j} x_{1}$ such that 
$|x_{0}| = f(i, j)$, $|x_{1}| = g(i, j)$, 
and $C \equiv_{\tilde{T} - 1} C'$.
\end{enumerate}
The statement $S(r, s)$ is true by the following reason.
We have $W(r, s) = \{(\epsilon, \epsilon)\}$ and hence 
$(r, s) \in K$.
Moreover $f(r, s) = g(r, s) = 0$ and hence the second condition 
of $S(r, s)$ is satisfied if we select $x_{0} = x_{1} = \epsilon$ 
and $C' = C$.
On the other hand, the statement $S(0, 0)$ is not true by the 
following reason.
We assume either $r < 0$ or $0 < s$.
If $r < 0$ then $U(0, 0)$ is an infinite set and 
if $0 < s$ then $V(0, 0)$ is an infinite set.
In either case $W(0, 0)$ is an infinite set and 
$(0, 0) \not\in K$.

In the remainder of this proof we show that 
if $S(i, j)$ is true then one of the following three cases 
is true.
\begin{enumerate}
\item[(1)] $i < 0$ and $S(i + 1, j)$ is true.
\item[(2)] $0 < j$ and $S(i, j - 1)$ is true.
\item[(3)] There is a configuration $C''$ such that 
$C \equiv_{\tilde{T} - 1} C''$ and 
${\rm rad}(C'') \geq \tilde{T}$.
\end{enumerate}
If we use this result repeatedly to find $(i, j)$ such 
that $S(i, j)$ is true and $j - i$ is smaller 
starting with $(i, j) = (r, s)$ then 
at some step the case (3) is true.
Then we have a configuration $C''$ such that 
$C \equiv_{\tilde{T} - 1} C''$ and ${\rm rad}(C'') 
\geq \tilde{T}$, and hence $\tilde{T} - 1$ is safe for $C$.

Suppose that $(i, j) \in K$, 
$C' = x_{0} p_{i} \ldots p_{j} x_{1}$ is a consistent extension 
of $p_{i} \ldots p_{j}$ such that 
$|x_{0}| = f(i, j)$, $|x_{1}| = g(i, j)$ and 
$C \equiv_{\tilde{T} - 1} C'$.
By the definition of $\tilde{T}$, 
$\tilde{T} \leq h(i, j) = \max \{ -2i + j + g(i, j), 2j - i + f(i, j) \}$, 
and hence either 
$\tilde{T} \leq -2i + j + g(i, j)$ or 
$\tilde{T} \leq 2j - i + f(i, j)$.
We consider the case $\tilde{T} \leq - 2i + j + g(i, j)$.
The proof for the other case is similar.

If $i = 0$ then we have 
${\rm rad}(C') \geq |p_{0} \ldots p_{j} x_{1}| - 1 
= - 2i + j + g(i, j) \geq \tilde{T}$ 
and we have the case (3) with $C'' = C'$.
Hence we assume that $i < 0$.
Let $v_{0}$ be the rightmost position in 
$p_{0} \ldots p_{j} x_{1}$.
If ${\rm d}_{C'}(p_{0}, v_{0}) \geq \tilde{T}$ then 
${\rm rad}(C') \geq \tilde{T}$ and we have the case (3) 
with $C'' = C'$.
Hence we assume that 
${\rm d}_{C'}(p_{0}, v_{0}) \leq \tilde{T} - 1$.

By $(i, j) \in K$, we know that $(i + 1, j) \in K \cup I$ 
and hence 
both of ${\rm NI}(i, j)$ and ${\rm NI}(i+1, j)$ are true.
By Lemma \ref{lemma:lem004} we have 
$V(i + 1, j) = V(i, j)$ and hence $|x_{1}| = g(i, j) = g(i+1, j)$.

Let $(x_{0}', x_{1}')$ be an arbitrary element of 
$W(i + 1, j)$.
Then $\tilde{C} = x_{0}' p_{i+1} \ldots p_{j} x_{1}'$ is 
a consistent extension of $p_{i+1} \ldots p_{j}$.
There is a sequence of positions $y$ such that 
$x_{0}' = y p_{i}$.
We have 
$(x_{0}', x_{1}) \in U(i + 1, j) \times V(i, j) 
= U(i + 1, j) \times V(i + 1, j) = 
W(i + 1, j)$.
Hence 
$C'' = x_{0}' p_{i+1} \ldots p_{j} x_{1}$ is a consistent 
extension of $p_{i+1} \ldots p_{j}$ and 
$C''$ is of the form 
$y p_{i} p_{i+1} \ldots p_{j} x_{1}$.

By our assumption ${\rm d}_{C'}(p_{0}, v_{0}) \leq \tilde{T}-1$, 
we have ${\rm d}_{C''}(p_{0}, v_{0}) \leq \tilde{T}-1$ and 
both of 
${\rm ai}(v_{0}, \tilde{T}-1, C')$, 
${\rm ai}(v_{0}, \tilde{T}-1, C'')$ 
are not ${\rm Q}$.
Moreover, 
${\rm d}_{C'}(p_{0}, p_{i}) + {\rm d}_{C'}(p_{i}, v_{0}) = 
{\rm d}_{C''}(p_{0}, p_{i}) + {\rm d}_{C''}(p_{i}, v_{0}) = 
-2i + j + g(i, j) \geq \tilde{T}$.
Hence the two sets 
$X' = \{v \in C' ~|~ d_{C'}(p_{0}, v) + 
d_{C'}(v, v_{0}) \leq \tilde{T}-1 \}$, 
$X'' = \{v \in C'' ~|~ d_{C''}(p_{0}, v) + 
d_{C''}(v, v_{0}) \leq \tilde{T}-1 \}$ 
are the same and both of them are in 
the part $p_{i+1} \ldots p_{j} x_{1}$ of 
$C'$, $C''$.
Moreover, $p_{i}$ is at the left of $p_{i+1}$ in both of $C'$, $C''$.
Therefore, ${\rm bc}_{C'}(v) = {\rm bc}_{C''}(v)$ for any node $v$ 
in the two sets $X', X''$.
Hence we have 
${\rm ai}(v_{0}, \tilde{T}-1, C') = 
{\rm ai}(v_{0}, \tilde{T}-1, C'') \not= {\rm Q}$ 
and 
$C' \equiv_{\tilde{T}-1}' C''$.

Finally we specify which element $(x_{0}', x_{1}')$ to select 
from $W(i + 1, j)$.
If $(i + 1, j) \in K$ then we can select $(x_{0}', x_{1}')$ 
so that $|x_{0}'| = f(i + 1, j)$.
Then we have the case (1).
If $(i + 1, j) \in I$, then $U(i + 1, j)$ is an infinite set 
because $V(i + 1, j)$ ($= V(i, j)$) is a finite set, 
and we can select $(x_{0}', x_{1}')$ so that 
$|x_{0}'| \geq \tilde{T} + 1$ and hence 
${\rm rad}(C'') \geq \tilde{T}$.
Then we have the case (3).
\begin{figure}
\begin{center}
\includegraphics[scale=1.0]{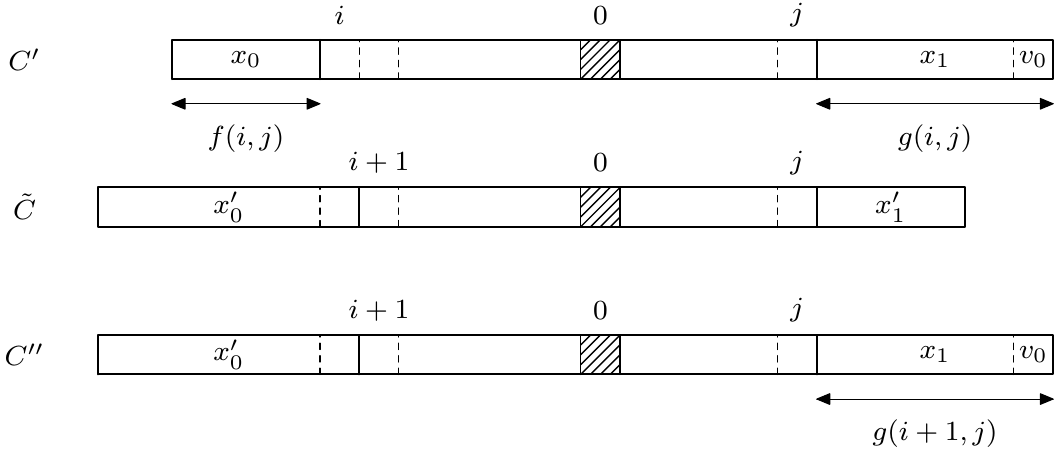}
\end{center}
\caption{Three configurations $C'$, $\tilde{C}$, $C''$ used 
in the proof of Theorem \ref{theorem:th005}.}
\label{figure:fig025}
\end{figure}
\hfill $\Box$
%\end{pf}

\medskip

For the configuration $C$ shown in Fig. \ref{figure:fig011} 
we showed that ${\rm mft}_\text{g-2PATH}(C) = \tilde{T}$ 
(Example \ref{example:ex012}) and 
$C$ does not satisfy CNI 
(Example \ref{example:ex007}).
Hence CNI is a sufficient condition for 
${\rm mft}_\text{g-2PATH}(C) = \tilde{T}$ 
but is not a necessary condition.

\section{Applications of the main result}
\label{section:application}

\subsection{A classification of configurations}
\label{subsection:types_of_configurations}

In this section we show examples of 
applications of our main result.

Recall that we call 
the parts $p_{r} \ldots p_{0}$ and $p_{0} \ldots p_{s}$ of 
a configuration $C = p_{r} \ldots p_{s}$ of g-2PATH  
the {\it left}\/ and 
the {\it right hands}\/ of $C$ respectively. 
When we show configurations of g-2PATH in figures 
we usually show them so that $0 < s$ and $p_{1} = (1, 0)$ 
(unless the configuration has only one node). 
Hence the right hand proceeds from the general 
to the east.
When we do not obey this rule we explicitly 
state the direction in the configuration.

We say that the left hand of $C$ is {\it free}\/ if 
either $r = 0$ or $r < 0$ and $f(r+1, s) = \infty$.
Otherwise (that is, $r < 0$ and $f(r+1, s)$ is finite) 
we say that the left hand of $C$ is 
{\it closed}\/.
Similarly we define the corresponding notions for right hands.

We classify configurations of $\text{g-2PATH}$ into 
three types:
\begin{description}
\item[Type I]: Configurations having two free hands.
\item[Type II]: Configurations having one free hand and one closed hand.
\item[Type III]: Configurations having two closed hands.
\end{description}

\begin{figure}[htbp]
\begin{center}
\includegraphics[scale=1.0]{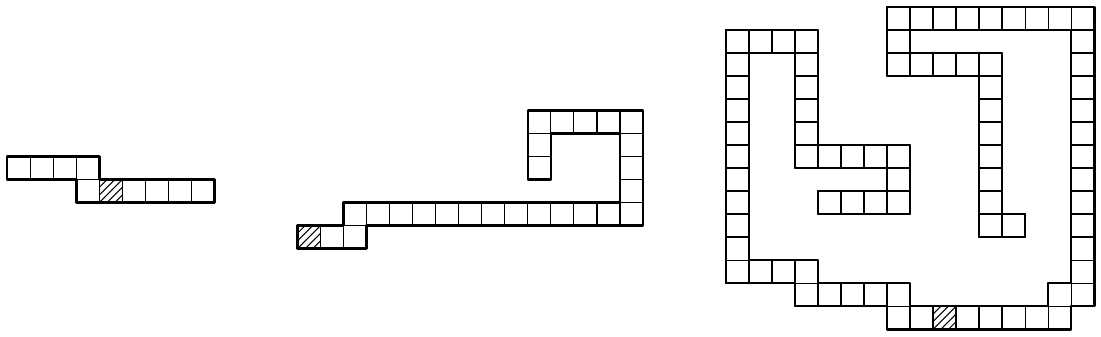}
\end{center}
\caption{Three configurations of Type I.} 
\label{figure:fig029}
\end{figure}
In Fig. \ref{figure:fig029} we show three examples of Type I
configurations.
\begin{figure}
\begin{center}
\includegraphics[scale=1.0]{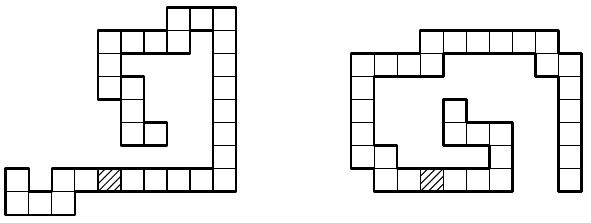}
\end{center}
\caption{Two configurations of Type II.}
\label{figure:fig057}
\end{figure}
In Fig. \ref{figure:fig057} we show two examples of 
Type II configurations.
Intuitively, a configuration of Type II constructs 
a bottle-shaped region surrounded by a part of the 
configuration.
One hand enters the bottle and cannot escape from it.
The other hand is out of the bottle and can be extended freely.
In Fig. \ref{figure:fig058} we show two examples of Type III
configurations.
\begin{figure}
\begin{center}
\includegraphics[scale=1.0]{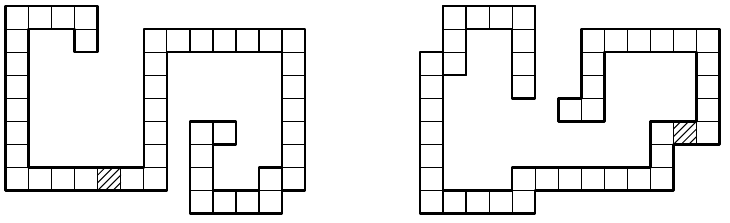}
\end{center}
\caption{Two configurations of Type III.}
\label{figure:fig058}
\end{figure}
Intuitively, there are two cases for a configuration 
of Type III.
In the first case, there are two bottles and two hands 
enter different bottles and cannot escape from them 
(the left configuration in Fig. \ref{figure:fig058}).
In the second case, there is one bottle and two hands 
enter the bottle and cannot escape from it 
(the right configuration in the figure).

\begin{lem}
\label{lemma:lem006}
If the left hand of a configuration $C = p_{r} \ldots p_{s}$ 
of $\text{\rm g-$2$PATH}$ is free, 
we have the following lower and upper bounds 
of ${\rm mft}_\text{\rm g-$2$PATH}(C)$:
\begin{equation}
-2r + s \leq {\rm mft}_\text{\rm g-$2$PATH}(C) \leq 
-r + s + \max \{ -r, s \}.
\label{equation:eq020}
\end{equation}
\end{lem}

%\begin{pf}
\noindent
{\it Proof}. 
We can construct a solution of g-2PATH 
that simulates a minimal-time solution of the generalized FSSP.
Its firing time (\ref{equation:eq001}) for $C$ is 
the upper bound in (\ref{equation:eq020})
($n = -r + s + 1$, $i = -r$).
Hence we have the upper bound. (This upper bound is 
true for any configuration of g-2PATH.)

For the case $r = 0$, the lower bound is obvious.
Suppose that $r < 0$.
Then we have $f(r+1, s) = \infty$ and 
there is a configuration $C'$ of the form 
$q_{m-1} \ldots q_{0} p_{r} p_{r+1} \ldots p_{s}$ 
such that ${\rm rad}(C') \geq -2r + s$.
For this $C'$ we have ${\rm ai}(p_{s}, -2r + s - 1, C) 
= {\rm ai}(p_{s}, -2r + s - 1, C') \not= {\rm Q}$ 
and hence $C \equiv_{-2r + s - 1, p_{s}}' C'$.
(Intuitively, $p_{s}$ cannot know the boundary condition of 
$p_{r}$ in time $-2r + s - 1$.)
This shows that the time $-2r + s - 1$ is safe for $C$ and 
we have the lower bound. 
\hfill $\Box$
%\end{pf}

\medskip

Using this lemma we can show two theorems.

\begin{thm}
\label{theorem:th007}
If a configuration $C = p_{r} \ldots p_{s}$ of 
${\rm g}\text{-}2{\rm PATH}$ 
is 
of Type I then its minimum firing time is represented as 
${\rm mft}_\text{{\rm g}\text{-}{\rm 2PATH}}(C) = -r + s + \max\{-r, s\}$.
\end{thm}

%\begin{pf}
\noindent
{\it Proof}. 
Both of the left and the right hands of $C$ are free.
Hence by Lemma \ref{lemma:lem006} 
both of $-2r + s$, $-r + 2s$ are lower bounds 
and hence 
$\max \{ -2r + s, -r + 2s \} = -r + s + \max \{-r, s\}$ 
is also a lower bound of ${\rm mft}_\text{g-$2$PATH}(C)$.
Hence we have the theorem.
\hfill $\Box$
%\end{pf}

\medskip

This theorem shows that the minimum firing time of a Type I configuration 
depends only on $r, s$.
Hence the minimum firing time of a configuration 
remains the same even if we modify its shape 
so long as it remains of Type I.
It also shows that g-2PATH is a conservative super-variation 
of the generalized FSSP.

\begin{thm}
\label{theorem:th008}
If $C = p_{r} \ldots p_{s}$ is a configuration of ${\rm g}\text{-}2{\rm PATH}$ 
of Type II having free left hand 
and $-r \geq s$, then its minimum firing time is 
represented as ${\rm mft}_{{\rm g}\text{-}2{\rm PATH}}(C) = -2r + s$.
\end{thm}

%\begin{pf}
\noindent
{\it Proof}. 
In this case, the upper bound 
and the lower bound in Lemma \ref{lemma:lem006} 
are the same value $-2r + s$.
\hfill $\Box$
%\end{pf}

\medskip

By these two theorems 
only configurations of Type II with $-r < s$ having 
free left hands and 
configurations of Type III are interesting.

\subsection{Configurations of Type II}
\label{subsection:typeii}

Let $C = p_{r} \ldots p_{s}$ be a configuration of 
$\text{g-2PATH}$ of Type II having free left hand.
If $r = 0$ then $C$ is a configuration of $\text{2PATH}$ 
and it satisfies CNI 
(Example \ref{example:ex004}) 
and we have the simple representation of 
${\rm mft}_\text{g-$2$PATH}(C)$ shown in 
Theorem \ref{theorem:th005}.
Hence we assume $r < 0$.  
In this case, $C$ always satisfies 
the conditions for $K$ and $J$ and 
the condition for $I$ is simplified.

\begin{thm}
\label{theorem:th009}
Suppose that $C = p_{r} \ldots p_{s}$ is 
a configuration of ${\rm g}\text{-}2{\rm PATH}$ of Type II 
having free left hand and $r < 0$.
Let $j_{0}$ be the smallest value of $j$ such that 
$0 \leq j$ and $W(r, j)$ is finite.
Then $C$ satisfies {\rm CNI} if and only if 
${\rm NI}(r+1, j_{0})$ is true.
\end{thm}

%\begin{pf}
\noindent
{\it Proof}. 
From the simple facts that 
$K = \{ (r, j) ~|~ j_{0} \leq j \leq s \}$, 
$I = \{ (r+1, j) ~|~ j_{0} \leq j \leq s\}$, 
$J = \{ (r, j_{0} - 1) \}$ if $0 < j_{0}$ and 
$J = \emptyset$ if $j_{0} = 0$, 
$C$ satisfies the conditions for $K$ and $J$ 
because elements of $W(r, j)$ are of the form 
$(\epsilon, y)$.
Hence CNI is equivalent to the condition for $I$.
However, we can easily show that 
the condition ${\rm NI}(r+1, j_{0})$ 
implies ${\rm NI}(r+1, j)$ for all 
$j_{0} \leq j \leq s$.
\hfill $\Box$
%\end{pf}

\medskip

\begin{ex}
\label{example:ex010}
\end{ex}

\vspace*{-\medskipamount}

\begin{figure}
\begin{center}
\includegraphics[scale=1.0]{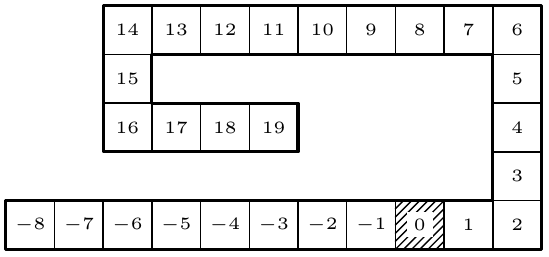}
\end{center}
\caption{The configuration of Type II used in 
Example \ref{example:ex010}.}
\label{figure:fig059}
\end{figure}

Consider the configuration $C = p_{-8} \ldots p_{19}$ 
of Type II shown in Fig. \ref{figure:fig059}.
The number $j_{0}$ defined 
in Theorem \ref{theorem:th009} 
is $16$.
We can show that ${\rm NI}(-7, 16)$ 
($= {\rm NI}(r+1, j_{0})$) is true as follows.
Any consistent extension of $p_{-7} \ldots p_{16}$ is of 
the form $x p_{-8} p_{-7} \ldots p_{16} p_{17} y$, 
$x p_{-8}$ is outside of the bottle surrounded by $p_{-6} \ldots p_{16}$, 
and $p_{17} y$ is in the bottle. 
Therefore $x p_{-8}$ and $p_{17} y$ can neither overlap nor touch.
Hence ${\rm NI}(-8, 17)$ is true and $C$ satisfies CNI.
The set $K$ is $\{ (-8, j) ~|~ 16 \leq j \leq 19 \}$.
$f(-8, j) = 0$ for any $(-8, j) \in K$ and 
$g(-8, j)$ is $-j + 22$ for $16 \leq j \leq 18$ and 
$0$ for $j = 19$.
By Theorem \ref{theorem:th005}, 
${\rm mft}_\text{g-$2$PATH}(C) = 
\min_{16 \leq j \leq 19} \max\{ 16 + j + g(-8, j), 2j + 8 \} = 40$.
The pair $(i, j)$ that minimizes the maximum is 
$(-8, 16)$.
\hfill (End of Example \ref{example:ex010})

\medskip

Finally we show two examples of large configurations of 
Type II.
The first $C_{0}$ satisfies CNI and the second $C_{1}$ does not.

\medskip

\begin{ex}
\label{example:ex005}
\end{ex}

\vspace*{-\medskipamount}

\begin{figure}[htbp]
\begin{center}
\includegraphics[scale=1.0]{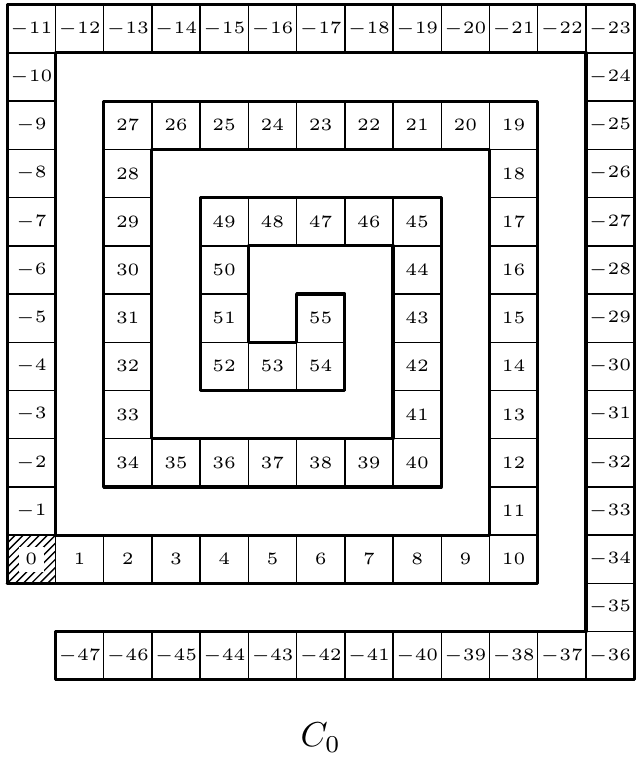}
\end{center}
\caption{The large configuration of Type II used in 
Example \ref{example:ex005}.  
This satisfies CNI.}
\label{figure:fig060}
\end{figure}

Consider the configuration $C_{0}$ shown in Fig. \ref{figure:fig060}.
This is of Type II and 
the value $j_{0}$ mentioned in Theorem 
\ref{theorem:th009} is $0$.
We can easily show that ${\rm NI}(-46, 0)$ is true 
and $C$ satisfies CNI by 
Theorem \ref{theorem:th009}.
Moreover we have 
$K = \{ (-47, j) ~|~ 0 \leq j \leq 55 \}$.

We determine the value ${\rm mft}_\text{g-2PATH}(C)$ 
by Theorem \ref{theorem:th005}.
The value $f(-47, j)$ is $0$ for any $j$.
We can determine the value $g(-47, j)$ for each $j$ by computer. 
\begin{figure}[htbp]
\begin{center}
\includegraphics[scale=1.0]{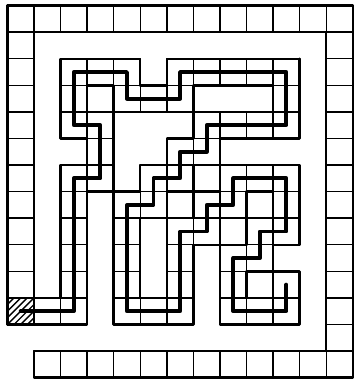}
\end{center}
\caption{A longest consistent right extension of 
$p_{-47} \ldots p_{0}$.}
\label{figure:fig062}
\end{figure}
Fig. \ref{figure:fig062} shows a longest consistent right 
extension of $p_{-47} \ldots p_{0}$ found by the exhaustive 
search by computer.
Its length is $59$ and this shows that $g(-47, 0) = 59$.
Similarly we can determine all the values of $g(-47, j)$.

\begin{table}[htbp]
\begin{center}
\includegraphics[scale=1.0]{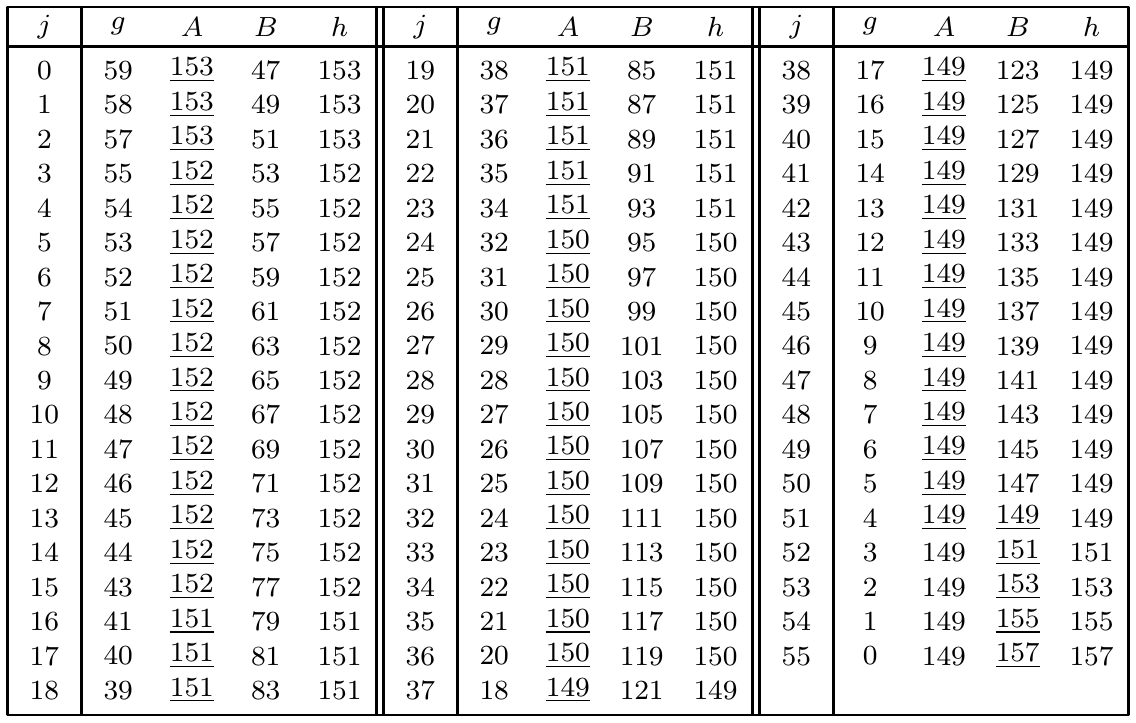}
\end{center}
\caption{Table of values of $g, A, B, h$ for 
$0 \leq j \leq 55$.}
\label{table:tab004}
\end{table}
In Table \ref{table:tab004} we show four values 
$g$, $A$, $B$, $h$ for each $j$.
Here, 
$g = g(-47, j)$, 
$A = -2 \cdot (-47) + j + g(-47, j) = j + 94 + g(-47, j)$, 
$B = 2j - (-47) + f(-47, j) = 2j + 47$, 
$h = h(-47, j) = \max \{ A, B \}$.
Of the two values $A$, $B$, we underline the larger one.
When $j$ increases by $1$, the value $g(-47, j)$ usually 
decreases by $1$ and the value $A$ remains the same.  
However, when the path selects a ``wrong'' direction (that is, 
a direction that is not into longest extensions), $g(-47, j)$ 
decreases by more than $1$ and $A$ decreases.
This happens four times, at 
$j = 2, 15, 23, 36$.
The value $B$ continues to increase by $2$.

By this table we have
\[
{\rm mft}_\text{g-2PATH}(C) 
= \min_{0 \leq j \leq 55} h(-47, j) = 149.
\]
The pairs $(-47, j)$ that minimize $h(-47, j)$ are 
$15$ pairs $(-47, 37)$, $(-47, 38)$, \ldots, $(-47, 51)$.

By Theorem \ref{theorem:th007}, the minimum firing time of 
a straight line with the same values of $r$, $s$ 
(that is, $r = -47$, $s = 55$) is 
$-r + s + \max \{-r, s\} = 47 + 55 + \max \{47, 55\} = 157$.
Therefore, the minimum firing time 
decreases by $8$ by bending 
a straight line to a coil of the form 
shown in Fig. \ref{figure:fig060}.
\hfill (End of Example \ref{example:ex005})

\begin{ex}
\label{example:ex013}
\end{ex}

\vspace*{-\medskipamount}

\begin{figure}[htbp]
\begin{center}
\includegraphics[scale=1.0]{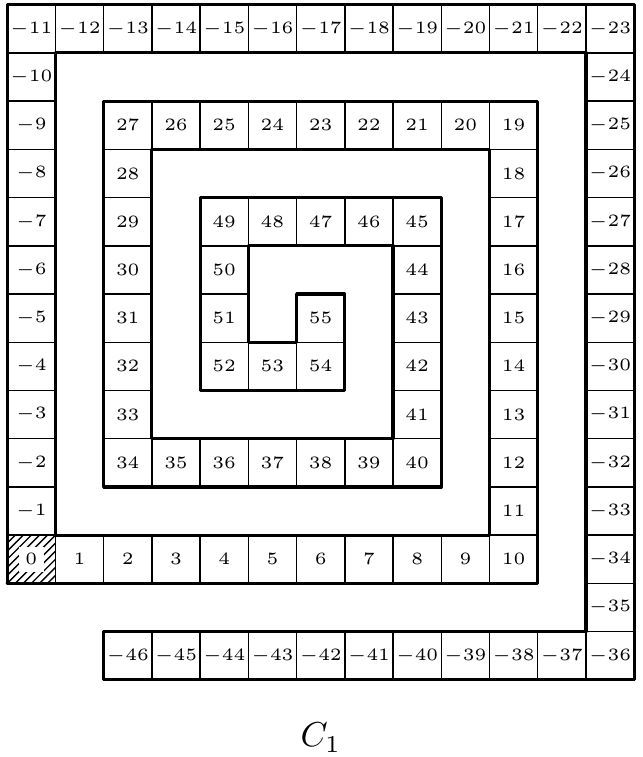}
\end{center}
\caption{The large configuration of Type II 
used in Example \ref{example:ex013}.
This does not satisfy CNI.}
\label{figure:fig061}
\end{figure}

Next, consider the configuration $C_{1}$ of Type II 
shown in Fig. \ref{figure:fig061}.
This configuration is obtained from $C_{0}$ by 
deleting only one node $p_{-47}$.
However, with this deletion, the configuration 
$C_{0}$ does not satisfy CNI as follows.

\begin{figure}[hbt]
\begin{center}
\includegraphics[scale=1.0]{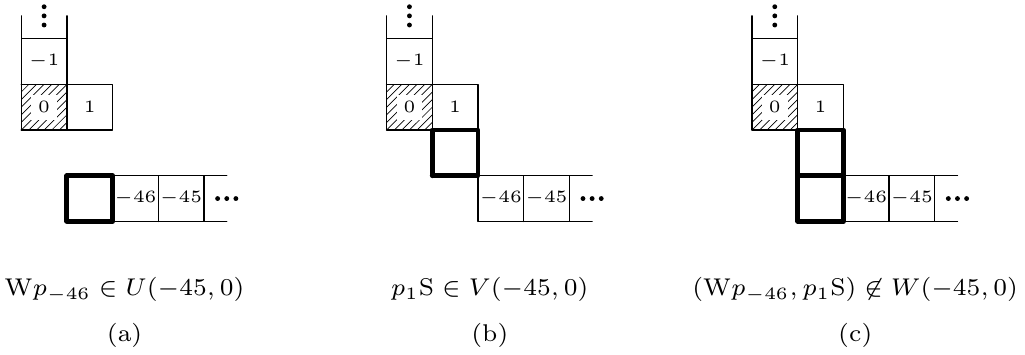}
\end{center}
\caption{Consistent extensions showing that 
${\rm NI}(-45, 0)$ is not true.}
\label{figure:fig063}
\end{figure}

The value $j_{0}$ mentioned in 
Theorem \ref{theorem:th009} is $0$.
Therefore, $C_{1}$ satisfies CNI if and only if 
${\rm NI}(-45, 0)$ is true.
Both of 
${\rm W} p_{-46} p_{-45} \ldots p_{0} p_{1}$, 
$p_{-46} p_{-45} \ldots p_{0} p_{1} {\rm S}$ 
are consistent extensions of 
$p_{-45} \ldots p_{0}$ 
(Fig. \ref{figure:fig063} (a), (b)).
However 
${\rm W} p_{-46} p_{-45} \ldots p_{0} p_{1} {\rm S}$ 
is not a configuration of g-2PATH (the figure (c)).
Hence, 
${\rm W} p_{-46} \in U(-45, 0)$, 
$p_{1} {\rm S} \in V(-45, 0)$, 
$({\rm W} p_{-46}, p_{1} {\rm S}) \not\in W(-45, 0)$, 
and ${\rm NI}(-45, 0)$ is not true. 
$C_{1}$ satisfies the condition for $K$ but 
does not satisfy that for $I$.

$C_{1}$ does not satisfy CNI and 
we cannot use our main result 
Theorem \ref{theorem:th005} to determine the value of 
${\rm mft}_\text{g-$2$PATH}(C_{1})$.
However we can prove 
${\rm mft}_\text{g-$2$PATH}(C_{1}) = 147$ as follows.

Let $C'$ be the configuration 
${\rm W \ldots W} p_{-46} p_{-45} \ldots p_{55}$.
The number of ${\rm W}$ is $101$ so that 
${\rm rad}(C') = 147$.
For this $C'$ we can show $C_{1} \equiv_{146, p_{55}}' C'$.
(Intuitively, $p_{55}$ cannot know the boundary condition of 
$p_{-46}$ in time $146$.)
Therefore $146$ is safe for $C_{1}$ and 
$147 \leq {\rm mft}_\text{g-$2$PATH}(C_{1})$.
Moreover Theorem \ref{theorem:th003} 
gives the upper bound 
${\rm mft}_\text{g-$2$PATH}(C_{1}) \leq \tilde{T} = 
\min_{0 \leq j \leq 55} \max \{92 + j + g(-46, j), 2j + 46 \} 
= 147$. 
The lower bound and the upper bound give the result 
${\rm mft}_\text{g-$2$PATH}(C_{1}) = 147$.
\hfill (End of Example \ref{example:ex013})

\subsection{Configurations of Type III}
\label{subsection:typeiii}

For configurations of Type III we show only some 
examples.

\medskip

\begin{ex}
\label{example:ex008}
\end{ex}

\vspace*{-\medskipamount}

\begin{figure}
\begin{center}
\includegraphics[scale=1.0]{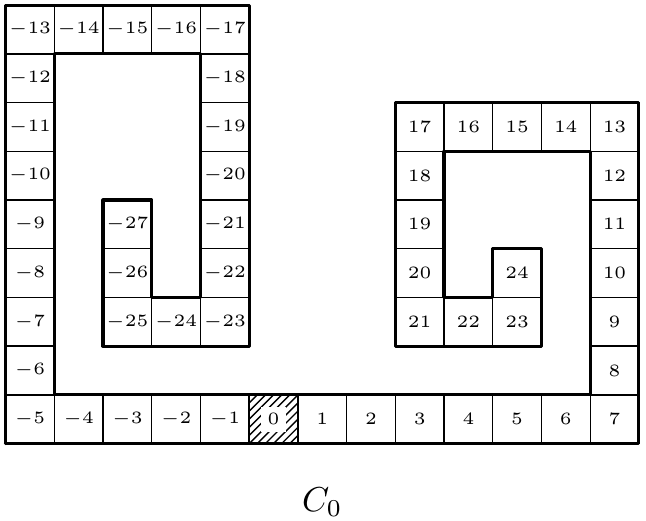}
\end{center}
\caption{A configuration of Type III used in 
Example \ref{example:ex008}.
This satisfies CNI.}
\label{figure:fig048}
\end{figure}

\begin{figure}
\begin{center}
\includegraphics[scale=1.0]{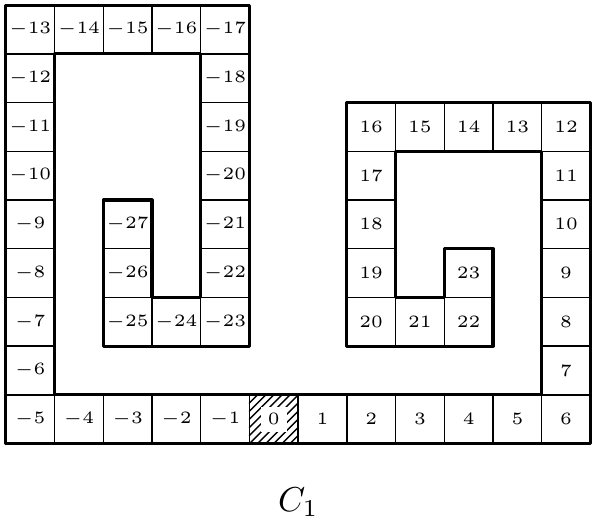}
\end{center}
\caption{Another configuration of Type III used in 
Example \ref{example:ex008}. 
This does not satisfy CNI.}
\label{figure:fig047}
\end{figure}

Consider the configurations $C_{0}$ and $C_{1}$ shown in 
Fig. \ref{figure:fig048} and Fig. \ref{figure:fig047}.
These configurations are of Type III.
We show that $C_{0}$ satisfies CNI and 
$C_{1}$ does not satisfy CNI.

First we analyze $C_{0}$.
If $-22 \leq i$ or $j \leq 20$ then $W(i, j)$ is infinite 
and $W(-23, 21)$ is finite.
Hence $I = \{(-22, j) ~|~ 21 \leq j \leq 24 \}$, 
$J = \{(i, 20) ~|~ -27 \leq i \leq -23 \}$, 
$K = \{ (i, j) ~|~ -27 \leq i \leq -23, 21 \leq j \leq 24 \}$.
We can easily show that $C_{0}$ satisfies CNI.

The values of $f(i, j)$, $g(i, j)$ for $(i, j) \in K$ are 
$f(i, j) = 29 + i$ for $-27 < i$ and 
$f(-27, j) = 0$, 
$g(i, j) = 25 - j$ for $j < 24$ and 
$g(i, 24) = 0$.
By Theorem \ref{theorem:th005}, 
\begin{eqnarray*}
{\rm mft}_\text{g-2PATH}(C_{0}) & = & \min_{(i, j) \in K} 
\max \{ -2i + j + g(i, j), 2j - i + f(i, j) \} \\
& = & \min_{-27 \leq i \leq -23, 21 \leq j \leq 24} 
    \max \{ -2i + j + g(i, j), 2j - i + f(i, j) \} \\	
& = & 71.
\end{eqnarray*}
The pair $(i, j) = (-23, 21)$ gives the minimum value $71$.
The minimum firing time decreases from $78$ 
($= 27 + 24 + \max \{27, 24\}$) 
to $71$ by 
bending a straight line as shown 
in Fig. \ref{figure:fig048}.

Next we analyze $C_{1}$.
If $-17 \leq i$ or $j \leq 15$ then $W(i, j)$ is infinite 
and $W(-18, 16)$ is finite.
Hence $K = \{ (i, j) ~|~ -27 \leq i \leq -18, 16 \leq j \leq 23 \}$.
It is easy to see that ${\rm NI}(-18, 16)$ is not true and 
$C_{1}$ does not satisfy the condition for $K$.
At present we are unable to determine the value 
${\rm mft}_\text{g-$2$PATH}(C_{1})$.

Intuitively, in $C_{0}$ there are two bottles, 
one surrounded by $p_{-23} \ldots p_{-1}$ and 
another surrounded by $p_{3} \ldots p_{21}$.
Consistent left and right extensions 
of $p_{-23} \ldots p_{21}$ enter different bottles 
and cannot interfere.
In $C_{1}$ there is one bottle surround by 
$p_{-18} \ldots p_{16}$.
Consistent left and right extensions 
of $p_{-18} \ldots p_{16}$ enter the unique bottle 
and can interfere freely.
\hfill (End of Example \ref{example:ex008})

\medskip

For configurations of Type III we have no result similar to 
Theorem \ref{theorem:th009}.
By studying many examples, we intuitively believe that, 
for each configuration $C$ of Type III, 
one of the following two is true:
\begin{enumerate}
\item[$\bullet$] The two hands ultimately enter disjoint 
two bottles and $C$ satisfies CNI (as $C_{0}$  in 
Fig. \ref{figure:fig048}).
\item[$\bullet$] The two hands ultimately enter one 
same bottle and $C$ does not satisfy the condition 
for $K$ (as $C_{1}$ in Fig. \ref{figure:fig047}).
\end{enumerate}
However, at present we do not know how to formalize 
intuitive notions such as ``bottles,'' ``a path enters a bottle,'' 
and so on, not to speak of proving the statement.

Moreover, our informal arguments using ``bottles'' 
do not apply to the three-dimensional analogue 
${\rm mft}_\text{g-3PATH}(C)$.
In the three-dimensional grid space we can construct 
very complicated structures (``gadgets'') with a path.
For example, in \cite{Goldstein_Kobayashi_SIAM_2005} we could 
prove that $\text{3PEP}$ (the three-dimensional analogue of 
$\text{2PEP}$) is NP-complete by simulating 
the Hamiltonian path problem with $\text{3PEP}$ 
and this simulation used bottles with three necks (holes) 
(see Figures 3.1 -- 3.7 in \cite{Goldstein_Kobayashi_SIAM_2005}).
This proof cannot be modified to the two-dimensional grid space 
and at present we cannot prove that $\text{2PEP}$ is NP-complete.

In mathematics there are many phenomena that are 
essentially different for the $2$-dimensional and the $3$-dimensional 
spaces, for example, recurrence vs transience of random walks 
in grid spaces, and planer graphs vs general graphs.
g-$2$PATH vs g-$3$PATH might be another example.

\medskip

\subsection{A simplified formula of ${\rm mft}_\text{g-2PATH}(C)$ 
for CNI-satisfying configurations of Type II}
\label{subsection:simplified_formula_for_cni_typeii}

In \cite{Kobayashi_TCS_2001} we showed 
the following representation of ${\rm mft}_\text{2PATH}(C)$.

\begin{thm}[\cite{Kobayashi_TCS_2001}]
\label{theorem:th010}
Suppose that $C = p_{0} \ldots p_{s}$ is a 
configuration of 
$2{\rm PATH}$ 
and let $j_{0}$ be the minimal value of $j$ 
such that 
$g(0, j) \leq j + 1$.
Then
\begin{align}
{\rm mft}_{\rm 2PATH}(C) & = 
\begin{cases}
2j_{0} + 1 & \text{if}\; g(0, j_{0}) = j_{0} + 1, \\
2j_{0}     & \text{if}\; g(0, j_{0}) < j_{0} + 1.
\end{cases}
\label{equation:eq010}
\end{align}
\end{thm}

\noindent
We have a similar result for g-2PATH.

\begin{thm}
\label{theorem:th011}
Suppose that $C = p_{r} \ldots p_{s}$ is a 
configuration of 
${\rm g}\text{-}2{\rm PATH}$ with a free left hand 
and $C$ satisfies {\rm CNI}.
When $-r \leq s$, let $j_{0}$ be the minimal value of $j$ 
such that 
$-r + g(r, j) \leq j + 1$.
Then
\begin{align}
\lefteqn{{\rm mft}_{{\rm g}\text{-}{\rm 2PATH}}(C)} \notag \\ 
& = 
\begin{cases}
-2r + s         & \text{if}\: -r > s, \\
-r + 2j_{0} + 1 & \text{if}\; -r \leq s \; \text{and} \; 
-r + g(r, j_{0}) = j_{0} + 1, \\
-r + 2j_{0}     & \text{if}\; -r \leq s \; \text{and} \; 
-r + g(r, j_{0}) < j_{0} + 1.
\end{cases}
\label{equation:eq011}
\end{align}
\end{thm}

%\begin{pf}
\noindent
{\it Proof}. 
By Theorem \ref{theorem:th005} and the fact that 
$f(i, j) = \infty$ for $r+1 \leq i \leq 0$ and 
$f(r, j) = 0$, 
we have the representation: 
\[
{\rm mft}_\text{g-2PATH}(C) 
= -r + \min_{j} \max \{ -r + j + g(r, j), 2j\}.
\]

Let $A(j), B(j)$ be defined by 
$A(j) = -r + j + g(r, j)$, $B(j) = 2j$ so that
${\rm mft}_\text{g-2PATH}(C) = -r + 
\min_{j} \max \{ A(j), B(j) \}$.
We have
$A(0) \geq A(1) \geq \ldots \geq A(s) = -r + s$, 
$0 = B(0) < B(1) < \ldots < B(s) = 2s$.
If $-r > s$ then $A(s) > B(s)$, 
${\rm mft}_\text{g-2PATH}(C) = -r + A(s) = -2r + s$, and 
the statement of the theorem is true.
Therefore we assume that $-r \leq s$.
Then $j_{0}$ is well defined, 
$A(j_{0}) \leq B(j_{0}) + 1$, 
and $A(j_{0} - 1) \geq B(j_{0} - 1) + 2$ if 
$0 < j_{0}$.
We consider two cases.

\medskip

\noindent
(Case 1) $A(j_{0}) = B(j_{0}) + 1$ 
(or equivalently, $-r + g(r, j_{0}) = j_{0} + 1$).
In this case $A(j_{0}) > B(j_{0})$ and 
$A(j_{0} + 1) \leq A(j_{0}) = B(j_{0}) + 1 = B(j_{0} + 1) - 1 
< B(j_{0} + 1)$ if $j_{0} < s$.
Hence $\max \{A(j_{0}), B(j_{0}) \} = A(j_{0})$, and 
$\max \{A(j_{0} + 1), B(j_{0} + 1) \} = B(j_{0} + 1) > A(j_{0})$ 
if $j_{0} < s$, and 
${\rm mft}_\text{g-2PATH}(C) = -r + A(j_{0}) = 
-r + B(j_{0}) + 1 = -r + 2j_{0} + 1$.

\medskip

\noindent
(Case 2) $A(j_{0}) < B(j_{0}) + 1$ 
(or equivalently, $-r + g(r, j_{0}) < j_{0} + 1$).
In this case we have $A(j_{0}) \leq B(j_{0})$ 
and $\max \{A(j_{0}), B(j_{0})\} = B(j_{0})$.
Hence, if $j_{0} = 0$ then 
${\rm mft}_\text{g-2PATH}(C) = -r + B(j_{0}) = -r + 2j_{0}$.
Suppose that $0 < j_{0}$.
Then $A(j_{0} - 1) \geq B(j_{0} - 1) + 2 = B(j_{0}) > B(j_{0} - 1)$ 
and $\max \{ A(j_{0} - 1), B(j_{0} - 1) \} = A(j_{0} - 1)$.
Consequently, ${\rm mft}_\text{g-2PATH}(C) = -r + 
\min \{A(j_{0} - 1), B(j_{0}) \} = -r + B(j_{0}) 
= -r + 2j_{0}$.
\hfill $\Box$
%\end{pf}

\medskip

In Example \ref{example:ex004} we showed that 
any configuration of g-2PATH of the form 
$p_{0} \ldots p_{s}$ (a configuration of 2PATH) 
satisfies CNI.
In this case (that is, $r = 0$), the formula 
(\ref{equation:eq011}) in Theorem \ref{theorem:th011} 
is reduced to the formula (\ref{equation:eq010}) 
in Theorem \ref{theorem:th010}.
This shows again that g-2PATH is a conservative 
supervariation of 2PATH.

\begin{ex}
\label{example:ex009}
\end{ex}

\vspace*{-\medskipamount}

We can determine the value ${\rm mft}_\text{g-2PATH}(C_{0})$ 
for the configuration $C_{0}$ in 
Example \ref{example:ex005} 
using Theorem \ref{theorem:th011}.
In this case, $r = -47$, $s = 55$, 
$j_{0} = 51$, $g(r, j_{0}) = 4$, and 
$-r + g(r, j_{0}) < j_{0} + 1$ (see Table 
\ref{table:tab003}).
Hence ${\rm mft}_\text{g-2PATH}(C) = -r + 2j_{0} 
= 47 + 2 \cdot 51 = 149$.

\begin{table}
\begin{center}
\begin{tabular}{|c||c|c|c|} \hline
$j$ & $g(r, j)$ & $-r + g(r, j)$ & $j + 1$ \\ \hline 
$\cdots$ & $\cdots$ & $\cdots$ & $\cdots$ \\
$49$ & $6$ & $53$ & $50$ \\
$50$ & $5$ & $52$ & $51$ \\
$51$ & $4$ & $51$ & $52$ \\
$52$ & $3$ & $50$ & $53$ \\
$53$ & $2$ & $49$ & $54$ \\ 
$\cdots$ & $\cdots$ & $\cdots$ & $\cdots$ \\ \hline
\end{tabular}
\caption{Values of $g(r, j)$, $-r + g(r, j)$, $j+1$ 
for Example \ref{example:ex009}.}
\label{table:tab003}
\end{center}
\end{table}
\hfill (End of Example \ref{example:ex009})

\section{Time and space analysis of the local map 
algorithm}
\label{section:time_space_analysis}

In this section we analyze the time and the space complexity 
of the local map algorithm 
to compute ${\rm mft}_\text{g-2PATH}(C)$.

We assume that we have fixed some standard way to represent 
a configuration $C$ by a word in some fixed alphabet.
We denote this representation by ${\rm des}(C)$.
We assume that $C = p_{r} \ldots p_{s}$ is a fixed configuration 
of g-2PATH for which we want to determine the value 
${\rm mft}_\text{g-2PATH}(C)$.

First we consider the time to decide CNI.

\begin{thm}
\label{theorem:th012}
For configurations $C$ of 
${\rm g}\text{-}2{\rm PATH}$ we can determine 
whether $C$ satisfies {\rm CNI} or not in 
polynomial time.
\end{thm}

\begin{figure}[htb]
\begin{center}
\includegraphics[scale=1.0]{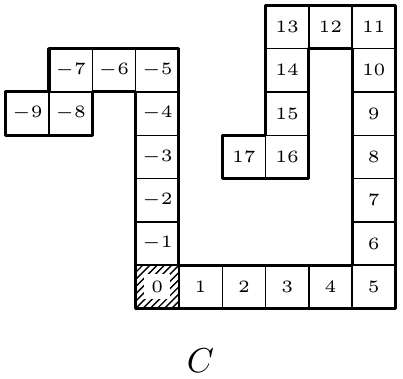}
\end{center}
\caption{A configuration $C$ of g-2PATH.}
\label{figure:fig055}
\end{figure}

\begin{figure}[htb]
\begin{center}
\includegraphics[scale=1.0]{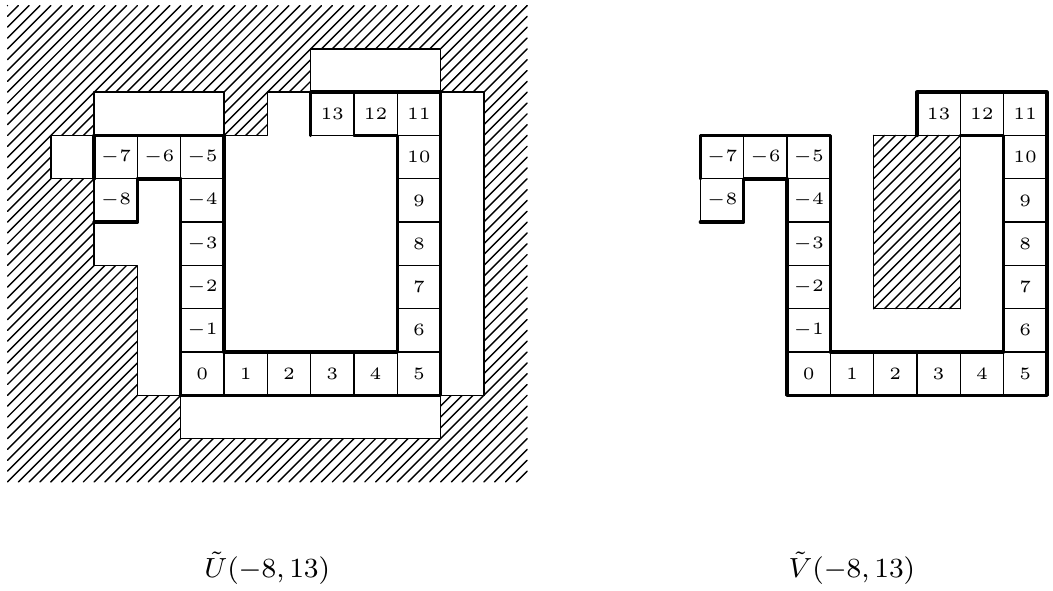}
\end{center}
\caption{The sets $\tilde{U}(-8,13)$ and $\tilde{V}(-8,13)$ 
for the configuration $C$ shown in 
Fig. \ref{figure:fig055}.}
\label{figure:fig056}
\end{figure}

%\begin{pf}
\noindent
{\it Proof}. 
We explain only the idea of the proof.
We define two sets of positions
$X = \{ (x, y) ~|~ |x|, |y| \leq \max \{-r, s\} + 3\}$, 
$X' = \{ (x, y) \in X ~|$ at least one of $|x|$, $|y|$ 
is $\max \{-r, s\} + 3\}$. 
$X$ is a square with its center at $(0, 0)$ and 
$X'$ is its boundary.
For $i, j$, we define 
$\tilde{U}(i, j)$ to be the set of positions that 
appear in some sequences of positions in $U(i, j)$, 
and define $\tilde{V}(i, j)$ similarly from 
$V(i, j)$.
In Fig. \ref{figure:fig056} we show 
$\tilde{U}(-8, 13)$, $\tilde{V}(-8, 13)$ with shadow for 
the configuration $C = p_{-9} \ldots p_{17}$ shown in 
Fig. \ref{figure:fig055}.
$\tilde{U}(i, j)$ and $\tilde{V}(i, j)$ are similar 
and we explain only on $\tilde{U}(i, j)$ when 
the explanation is also true for $\tilde{V}(i, j)$.

It is possible that $\tilde{U}(i, j)$ is an infinite set.
However we can easily show that 
$\tilde{U}(i, j)$ includes all positions in $\overline{X}$ 
if $\tilde{U}(i, j) \cap X'$ is nonempty and 
$\tilde{U}(i, j)$ includes no positions in $\overline{X}$ 
otherwise ($\overline{X} = \mathbf{Z}^{2} - X$).
Therefore the finite set $\tilde{U}(i, j) \cap X$ completely 
determines the possibly infinite set $\tilde{U}(i, j)$.
It is obvious that we can determine the set 
$\tilde{U}(i, j) \cap X$ in polynomial time.

We can decide whether $C$ satisfies CNI or not in 
polynomial time as follows.
A set $W(i, j)$ is finite if and only if 
both of $\tilde{U}(i, j)$, $\tilde{V}(i, j)$ are 
finite.
Hence we can decide it in polynomial time.
Therefore, we can determine the three sets $I$, $J$, $K$ 
in polynomial time.
Moreover, we can decide whether ${\rm NI}(i, j)$ is 
true or not in polynomial time 
because ${\rm NI}(i, j)$ is true if and only if 
the two sets $\tilde{U}(i, j)$, $\tilde{V}(i, j)$ 
neither overlap nor are adjacent.
For example, Fig. \ref{figure:fig056} shows that 
$\tilde{U}(-8, 13)$ and $\tilde{V}(-8, 13)$ 
neither overlap nor are adjacent, 
and hence ${\rm NI}(-8, 13)$ is true 
for the configuration 
$C$ shown in Fig. \ref{figure:fig055}.
In this way we can decide whether $C$ satisfies CNI or not 
in polynomial time.
\hfill $\Box$
%\end{pf}

\medskip

Next we consider the time to compute 
${\rm mft}_\text{g-2PATH}(C)$ for CNI-satisfying 
configurations $C$.

\medskip

\begin{thm}
\label{theorem:th013}
For configurations $C$ of ${\rm g}\text{-}2{\rm PATH}$ 
that satisfy {\rm CNI}
we can compute ${\rm mft}_{{\rm g}\text{-}{\rm 2PATH}}(C)$ in polynomial time 
using an {\rm NP} set as an oracle.
\end{thm}

%\begin{pf}
\noindent
{\it Proof}. 
By Theorem \ref{theorem:th005} 
the formula 
\[
\tilde{T} = \min_{i,j} \max \{-2i + j + g(i, j), 
                              2j - i + f(i, j) \}
\]
gives the value ${\rm mft}_\text{g-2PATH}(C)$. 
We show that for each $(i, j)$ we can determine 
the value $f(i, j)$ in polynomial time 
using an ${\rm NP}$ set as an oracle.
The proof for $g(i, j)$ is similar.

$f(i, j)$ is finite if and only if 
$\tilde{U}(i, j)$ is finite. 
($\tilde{U}(i, j)$ is the set defined in 
the proof of Theorem \ref{theorem:th012}.)
Hence we can determine whether $f(i, j)$ is finite 
or not in polynomial time. 
Suppose that $f(i, j)$ is finite.

In this case any consistent left extension of 
$p_{i} \ldots p_{j}$ is in the square $X$ 
defined in the proof of Theorem 
\ref{theorem:th012} and hence 
we have $f(i, j) \leq (2m + 1)^{2}$ 
(the area of the square $X$) 
with $m = \max \{-r, s\} + 3$. 
Let $L$ be the set
\begin{align*}
L = \{ & {\rm des}(C) \# 0^{-i} \# 0^{j} \# 0^{k} ~|~ 
r \leq i \leq 0, 0 \leq j \leq s, \\
& \quad\quad\quad\quad 0 \leq k \leq (2m + 1)^{2}, k \leq f(i, j) \}.
\end{align*}
Then we can determine the value $f(i, j)$ 
by deciding whether ${\rm des}(C) \# 0^{-i} \# 0^{j} \# 0^{k}$ is in $L$ 
or not for all $k$ such that $0 \leq k \leq (2m + 1)^{2}$.
This is possible in polynomial time of $n$ ($n = |{\rm des}(C)|$) 
if we can use $L$ as an oracle.

For the case $0 < k$, $r < i$, $j < s$, 
$k \leq f(i, j)$ is true if and only if 
there is a configuration of the form 
$q_{k-1} \ldots q_{1} p_{i-1} p_{i} \ldots p_{j} p_{j+1}$.
For other cases too we have similar statements.
This shows that $L$ in in ${\rm NP}$.
\hfill $\Box$
%\end{pf}

\medskip

\begin{cor}
\label{corollary:cor018}
We can compute ${\rm mft}_{\rm 2PATH}(C)$ 
in polynomial time using an ${\rm NP}$ set as an oracle.
\end{cor}

%\begin{pf}
\noindent
{\it Proof}. 
g-2PATH is a conservative super-variation of 
2PATH (\cite{Kobayashi_TCS_2001}).
Hence, in order to compute 
${\rm mft}_{\text{\rm 2PATH}}(C)$ 
it is sufficient to compute ${\rm mft}_\text{g-2PATH}(C)$.
\hfill $\Box$
%\end{pf}

\medskip

Finally we consider space complexity.

\begin{thm}
\label{theorem:th014}
We can compute all of 
${\rm mft}_{2{\rm REG}}(C)$, 
${\rm mft}_{{\rm g}\text{-}{\rm 2PATH}}(C)$, 
${\rm mft}_{2{\rm PATH}}(C)$ 
in polynomial space.
\end{thm}

%\begin{pf}
\noindent
{\it Proof}. 
First we consider ${\rm mft}_\text{g-$2$PATH}(C)$.
We have an upper bound $-r + s + \max \{-r, s\}$ (the value 
(\ref{equation:eq001}) with $n = -r + s + 1$, $i = -r$) 
for ${\rm mft}_\text{g-$2$PATH}(C)$ 
because we can simulate a minimal-time solution of 
the generalized FSSP on configurations of g-2PATH.
Hence it is sufficient to prove that the set 
\begin{equation}
L_\text{g-2PATH} = \{ {\rm des}(C) \# 0^{t} ~|~ 
0 \leq t \leq -r + s + \max\{-r, s\}, 
\; t < {\rm mft}_\text{g-2PATH}(C) \}
\label{equation:eq012}
\end{equation}
is in PSPACE.

$t < {\rm mft}_\text{g-2PATH}(C)$ if and only if 
$t$ is safe for $C$ and this is true if and only if 
there exist $C_{0}, \ldots, C_{u-1}$ such that $1 \leq u$, 
$C_{0} = C$, 
${\rm rad}(C_{k}) \leq t$ for $0 \leq k \leq u-2$, 
$t < {\rm rad}(C_{u-1})$, 
and $C_{k} \equiv_{t}' C_{k+1}$ for $0 \leq k \leq u-2$.
We may assume ${\rm rad}(C_{u-1}) \leq t+1$.
Therefore, we can accept the set $L_\text{g-2PATH}$ 
by the following nondeterministic Turing machine $M$.
Let ${\rm des}(C) \# 0^{t}$ be the input to $M$.

We use $\tilde{C}$ as a variable that has a configuration 
as its value.
The initial value of $\tilde{C}$ is the input configuration $C$.
Then $M$ repeats the following step until a halting condition 
is satisfied.
If $t < {\rm rad}(\tilde{C})$ then $M$ halts 
accepting the input. 
Otherwise, $M$ nondeterministically selects a configuration 
$C'$ such that ${\rm rad}(C') \leq t + 1$ and checks whether 
$\tilde{C} \equiv_{t}' C'$ or not.
(This is done by checking whether $\tilde{C} \equiv_{t, v}' C'$ or not 
for each $v$ in $\tilde{C}$.)
If $\tilde{C} \equiv_{t}' C'$ then the newly selected $C'$ 
becomes the new value of $\tilde{C}$ and $M$ proceeds to 
the next step.
Otherwise $M$ halts without accepting the input.

It is obvious that $M$ halts accepting the input 
${\rm des}(C) \# 0^{t}$ if and only if 
$t$ is safe for $C$ and hence the input is in $L_\text{g-2PATH}$.
Moreover $M$ is polynomial space because the space 
used by $M$ is only the space to write $\tilde{C}$, $C'$ and 
to check $\tilde{C} \equiv_{t}' C'$.
Hence $L_\text{g-2PATH}$ is in ${\rm NPSPACE}$, the class of 
sets accepted by polynomial-space nondeterministic Turing machines.
However we know that ${\rm NPSPACE} = {\rm PSPACE}$ and hence 
$L_\text{g-2PATH}$ is in ${\rm PSPACE}$.

The result for ${\rm mft}_\text{$2$PATH}(C)$ follows because 
$2$PATH is a conservative subvariation of g-$2$PATH.
The proof for ${\rm mft}_\text{$2$REG}(C)$ is the same 
except the upper bound for ${\rm mft}_\text{$2$REG}(C)$.
Nishitani and Honda (\cite{Nishitani_Honda}) constructed 
a $3r_{G} + 1$ time solution of the FSSP for undirected graphs 
($r_{G}$ denotes the radius of an undirected graph $G$).
We can use this as a solution of $2$REG and 
hence we have an upper bound $3 \cdot {\rm rad}(C) + 1$ 
for ${\rm mft}_\text{$2$REG}(C)$.
\hfill $\Box$
%\end{pf}

\medskip

We summarize known results on the time and space complexity of 
problems ${\rm MFT}_{\Gamma}$ 
(the problem to compute ${\rm mft}_{\Gamma}(C)$).
In this summary we use $k$ to denote $2$ or $3$.

\medskip

\noindent
{\bf Lower bounds:}
\begin{enumerate}
\item[$\bullet$] 
${\rm 2PEP} \leq_{\rm T}^{\rm p} {\rm MFT}_{\Gamma}$ 
for $\Gamma =$ 2PATH, g-2PATH, 2REG (\cite{Kobayashi_TCS_2001}), 
\item[$\bullet$] 
${\rm HAMPATH} \leq_{\rm T}^{\rm p} {\rm MFT}_{\Gamma}$ 
for $\Gamma =$ 3PATH, g-3PATH, 3REG 
(\cite{Goldstein_Kobayashi_SIAM_2005}).
\end{enumerate}

\noindent
{\bf Upper bounds:}
\begin{enumerate}
\item[$\bullet$] 
${\rm MFT}_{k\text{PATH}} \in \Delta_{2, {\rm F}}^{\rm p}$ 
(Corollary \ref{corollary:cor018}, \cite{Kobayashi_TCS_2001}), 
\item[$\bullet$]
${\rm MFT}_{\text{g-}k\text{PATH,CNI}} \in \Delta_{2,{\rm F}}^{\rm p}$ 
(Theorem \ref{theorem:th013}), 
\item[$\bullet$]
${\rm MFT}_{\text{g-}k\text{PATH}} \in \text{PSPACE}_{\rm F}$ 
(Theorem \ref{theorem:th014}), 
\item[$\bullet$]
${\rm MFT}_{k\text{REG}} \in \text{PSPACE}_{\rm F}$ 
(Theorem \ref{theorem:th014}).
\end{enumerate}

\noindent
Here, ${\rm MFT}_{\text{g-}k\text{PATH,CNI}}$ denotes
the problem ${\rm MFT}_{\text{g-}k\text{PATH}}$ 
that is given only CNI-satisfying configurations.
HAMPATH is the Hamiltonian path problem known to be 
NP-complete.
Using 
$\Delta_{2, {\rm F}}^{\rm p} \subseteq {\rm PSPACE}_{\rm F}  
\subseteq {\rm EXP}_{\rm F}$, 
we can replace the time upper bound 
$\Delta_{2, {\rm F}}^{\rm p}$ with the space 
upper bound ${\rm PSPACE}_{\rm F}$ and 
the space upper bound 
${\rm PSPACE}_{\rm F}$ with the 
time upper bound ${\rm EXP}_{\rm F}$.

\section{Numbers of states of solutions that fire $C$ 
at time ${\rm mft}_{\Gamma}(C)$}
\label{section:minimum_state_numbers}

\subsection{The problem to design small solutions}
\label{subsection:small_solutions}

In this section, we consider the problem to design, 
for each given configuration $C$, 
a solution $A$ that fires $C$ at time 
${\rm mft}_\Gamma(C)$.
Moreover, we try to find $A$ that has 
a small number of states.
Although this problem is seemingly not related 
to the problem of existence or nonexistence of 
minimal-time solutions of $\Gamma$, 
there is a close relation between 
the two problems (Theorem \ref{theorem:th016}).
In this section, by the {\it size}\/ of a finite automaton 
we mean the number of the states of the automaton.

We know that if we have a partial solution $A$ of size $s$ 
that fires $C$ at ${\rm mft}_{\Gamma}(C)$ we can construct 
a similar solution $A'$ of size at most $c_{\Gamma} s$.
Here, $c_{\Gamma}$ is a constant that depends on the variation $\Gamma$ 
but not on $C$.
The idea is that we fix one solution $A''$ of $\Gamma$ 
and $A'$ simulates both of $A$, $A''$ and fires as soon as 
one of them fires.
The constant $c_{\Gamma}$ is the size of $A''$.

For all of the six variations $k$REG, g-$k$PATH, $k$PATH 
($k = 2, 3$), as $A''$ we can use 
the $296$ state solution of FSSP for undirected graphs 
constructed by Nishitani and Honda (\cite{Nishitani_Honda}).
Hence we can use $c_{\Gamma} = 296$.
For the four variations g-$k$PATH, $k$PATH, 
as $A''$ we can use the $6$ state solution of 
the generalized FSSP constructed by Umeo et al. 
(\cite{Umeo_2006}).
It is not a minimal-time solution.
However its state transition table is 
symmetrical with respect to the right and the left directions.
Therefore we need no information for representing 
the orientation 
(which of the two non-open input terminals is 
the left input) of nodes when we simulate it on 
paths in the grid space 
$\mathbf{Z}^{k}$.
Hence we can simulate it directly on paths and 
we can use $c_{\Gamma} = 6$ 
for the four variations g-$k$PATH, $k$PATH.

Let $\Gamma$ be a variation of FSSP that has a solution 
and $C$ be a configuration of $\Gamma$.
By the {\it minimum solution size} of $C$, 
denoted by ${\rm mss}_{\Gamma}(C)$, we mean 
the minimum value of the size of a solution $A$ of $\Gamma$ 
that fires $C$ at time ${\rm mft}_{\Gamma}(C)$, that is, 
\begin{eqnarray*}
{\rm mss}_{\Gamma}(C) & = & 
\min_{A} \{ \text{the number of the states of}\; A ~|~ \\
& & \quad\quad A \;\text{is a solution of $\Gamma$ and} \;\;
{\rm ft}(C, A) = {\rm mft}_{\Gamma}(C) \}.
\end{eqnarray*}

\begin{thm}
\label{theorem:th016}
A variation $\Gamma$ of {\rm FSSP} has a minimal-time solution 
if and only if the value ${\rm mss}_{\Gamma}(C)$ is 
bounded from above, that is, 
there is a constant $c$ such that 
${\rm mss}_{\Gamma}(C) \leq c$ for any configuration $C$ 
of $\Gamma$.
\end{thm}

%\begin{pf}
\noindent
{\it Proof}. 
The ``only if'' part is evident and 
we show the ``if'' part.
Suppose that there is a constant $c$ such that 
${\rm mss}_{\Gamma}(C) \leq c$ for any $C$.
There are only a finite number of solutions of $\Gamma$ 
with at most $c$ states.
Let $\tilde{A}$ be the finite automaton that simulates 
all of these solutions and fires as soon as 
at least one of them fires.
Then $\tilde{A}$ is a solution of $\Gamma$ that fires $C$ 
at ${\rm mft}_\text{g-$2$PATH}(C)$.
\hfill $\Box$
%\end{pf}

\medskip

The following two statements are equivalent 
and we believe that both of them are true 
for $\Gamma = \text{$k$REG, g-$k$PATH, $k$PATH}$:
\begin{enumerate}
\item[$\bullet$] The variation $\Gamma$ has no minimal-time 
solutions.
\item[$\bullet$] ${\rm mss}_{\Gamma}(C)$ is not bounded 
from above.
\end{enumerate}
Although we have at least one strategy to show 
circumstantial evidences for the former, 
at present we completely lack 
any ideas how to start to prove the latter.

\subsection{${\rm mss}_{\Gamma}(C)$ for $k$REG and $k$PATH}
\label{subsection:mss_for_kreg_and_kpath}

For $k$REG, the local map partial solution $A_{{\rm lm},T}$ 
is the only method we know to construct a solution that 
fires $C$ at ${\rm mft}_\text{$k$REG}(C)$.  
(As the value of $T$ we use ${\rm mft}_\text{g-$2$PATH}(C)$.)
Hence the upper bound
\begin{equation}
{\rm mss}_\text{$2$REG}(C) \leq 2^{64T^{2} + O(T)}
\label{equation:eq024}
\end{equation}
by (\ref{equation:eq005}) 
is the only upper bound we have for 
${\rm mss}_\text{$2$REG}(C)$.

For example, suppose that ${\rm mft}_\text{$2$REG}(C) = 20$ 
and we construct a solution that fires $C$ at time $20$ 
using the partial solution $A_{{\rm lm},20}$.
Then size of the solution is at least 
$296 \cdot 2^{(1/3) \cdot (20 - 3)^{2}} = 
2.9547 \ldots \cdot 10^{31}$ by (\ref{equation:eq021})
and at most 
$296 \cdot (1 + 21 \cdot 41^{2} \cdot 2^{16 \cdot 41^{2}}) = 
3.3253 \ldots \cdot 10^{8103}$ 
by (\ref{equation:eq005}).

For $k$PATH, we have the reflection partial solution 
$A_{{\rm ref}, C}$ that fires $C$ at ${\rm mft}_\text{$k$PATH}(C)$.
Its size is $T + 2$ by (\ref{equation:eq031}), where 
$T = {\rm mft}_\text{$k$PATH}(C)$.
Hence we have the small upper bound
\begin{equation}
{\rm mss}_\text{$k$PATH}(C) \leq 6 \cdot (T + 2)
\label{equation:eq025}
\end{equation}
for ${\rm mss}_\text{$k$PATH}(C)$.

Thus, our upper bound for ${\rm mss}_{\Gamma}(C)$ is 
tremendously large ((\ref{equation:eq024})) for $2$REG and 
very small ((\ref{equation:eq025})) for $2$PATH.
In the following subsection we consider $\text{g-$k$PATH}$.

\subsection{${\rm mss}_{\Gamma}(C)$ for g-$k$PATH and 
consistency checking partial solutions}
\label{subsection:mss_for_gkpath}

For g-$k$PATH we can use local map partial solutions 
$A_{{\rm lm},C}$ 
to construct solutions that fire $C$ 
at ${\rm mft}_\text{g-$k$PATH}(C)$.
The upper bound
\begin{equation}
{\rm mss}_\text{g-$2$PATH}(C) \leq 2^{4T + O(\log T)}
\label{equation:eq026}
\end{equation}
based on the upper bound (\ref{equation:eq018}) 
is the only upper bound we have for 
${\rm mss}_\text{g-$2$PATH}(C)$.

For example, consider the configuration $C = p_{-11} \ldots p_{11}$ of 
g-$2$PATH shown in Fig. \ref{figure:fig011}.
For this configuration $T = {\rm mft}_\text{g-$2$PATH}(C) = 30$ 
by Example \ref{example:ex002}.
Hence 
the size of the solution is at least 
$6 \cdot ({}_{15} C _{7})^{2} = 2.4845 \ldots \cdot 10^{8}$ 
by (\ref{equation:eq028}) and 
at most 
$6 \cdot (1 + 31 \cdot 61^{2} \cdot ((4^{31} - 1) / 3) \cdot 16^{2}) 
= 2.7236 \ldots \cdot 10^{26}$ 
by (\ref{equation:eq018}).

The second partial solutions available for g-$k$PATH 
are reflection partial solutions 
$A_{{\rm ref}, C}$.
They give the small upper bound 
\begin{equation}
{\rm mss}_\text{g-$k$PATH}(C) \leq 6 \cdot (4T + 8)
\label{equation:eq027}
\end{equation}
by (\ref{equation:eq032}).
However this upper bound is true only for configurations $C$ 
such that the value $\tilde{T}$ defined by (\ref{equation:eq008}) 
is the value ${\rm mft}_\text{g-$k$PATH}(C)$.
CNI-satisfying configurations are such configurations 
(Theorem \ref{theorem:th005}).
The configuration $C = p_{-11} \ldots p_{11}$ considered above 
does not satisfy CNI (Example \ref{example:ex007}) 
but satisfies $\tilde{T} = {\rm mft}_\text{g-$2$PATH}(C) = 30$ 
(Example \ref{example:ex012}).
Hence the above upper bound  $6 \cdot (4 \cdot 30 + 8) = 768$ 
is true for this CNI-nonsatisfying $C$.

In this section we show a third method to construct 
a partial solution that fires $C$ at ${\rm mft}_{\Gamma}(C)$.
We call it the {\it consistency checking partial solution}\/ 
of $C$ and denote it by $A_{{\rm cc}, C}$.
We can construct $A_{{\rm cc}, C}$ for the four variations 
g-$k$PATH, $k$PATH ($k = 2, 3$). 
The size of $A_{{\rm cc}, C}$ is large 
((\ref{equation:eq015})).
However, in many cases we can modify it 
and reduce the size considerably.
We will explain $A_{{\rm cc}, C}$ only for g-2PATH.

Let $C$ be a configuration of g-2PATH for which 
we construct $A_{{\rm cc}, C}$, 
$T$ be the time ${\rm mft}_\text{g-2PATH}(C)$, and 
$C_{0}$ ($=C$), $C_{1}$, \ldots, $C_{m-1}$ be an enumeration 
of all $C'$ such that $C \equiv_{T} C'$.
Note that ${\rm rad}(C_{k}) \leq T$ for any $k$ because 
$T$ is unsafe for $C$.

Let $C_{k} = p_{r} \ldots p_{s}$ be one of 
$C_{0}, \ldots, C_{m-1}$ and $p_{u}$ be a node in $C_{k}$.
(Usually we have used $p_{r} \ldots p_{s}$ to denote 
the target configuration $C$ for which we determine 
the value ${\rm mft}_{\Gamma}(C)$.
However, temporarily we use this notation for $C_{k}$.)
For this pair of $C_{k}$ and $p_{u}$ we define 
a finite automaton $A_{0}(u,C_{k})$.
Suppose that $C'$ is an arbitrary configuration of g-2PATH and 
that copies of $A_{0}(u, C_{k})$ are placed at all nodes in $C'$.
Then, intuitively, a node $w$ in $C'$ fires if and only if the following 
two conditions are satisfied: 
(1) $C'$ is a consistent extension of 
$M(p_{u}, T, C_{k})$, and 
(2) $w$ knows it before or at time $T$.
If $w$ fires, it fires at time $T$.
A formal definition of $A_{0}(u, C_{k})$ is as follows.

By the formula (\ref{equation:eq007}) we know that 
$M(p_{u}, T, C_{k})$ is the part $p_{a} \ldots p_{b}$ 
of $C_{k}$, where
\[
a = \max \{r, \lceil (u - T) / 2 \rceil \}, 
b = \min \{s, \lfloor (u + T) / 2 \rfloor \}.
\]
Note that $a \leq 0 \leq b$ and $a \leq u \leq b$ because 
${\rm rad}(C_{k}) \leq T$ and hence 
$-T \leq u \leq T$.
Let $x_{0}, x_{1}$ be the sequences 
$x_{0} = p_{a} \ldots p_{0}$, 
$x_{1} = p_{0} \ldots p_{b}$.
Then $C'$ is a consistent extension of $M(p_{u}, T, C_{k})$ 
if and only if it is consistent extensions of both of 
$x_{0}$, $x_{1}$.

We can check whether $C'$ is a consistent extension of $x_{0}$ or not 
with the idea used in constructing $A_{{\rm ref}, C}$.
A signal ${\rm R}$ starts at $p_{0}$ at time $0$ and 
proceeds to $p_{a}$ along the sequence $p_{a} \ldots p_{0}$ 
(from right to left).
If $C'$ is a consistent extension of $x_{0}$ the signal ${\rm R}$ 
knows it at $p_{a}$ at time $-a$.
Then the signal ${\rm S}$ is generated at $p_{a}$ at time $-a$ and 
propagates to all nodes until time $T$.
After time $T$ the signal ${\rm S}$ vanishes.
Similarly we can check whether $C'$ is a consistent extension of 
$x_{1}$ or not using two signals ${\rm R'}$, ${\rm S'}$.
A node $w$ in $C'$ fires if and only if it has received both of 
${\rm S}$, ${\rm S'}$ before or at time $T$.
Then, a node $w$ in $C'$ fires if and only if 
both of the following two conditions are satisfied: 
(1) $C'$ is a consistent extension of $M(p_{u}, T, C_{k})$, and 
(2) $-a + {\rm d}_{C'}(p_{a}, w) \leq T$, 
$b + {\rm d}_{C'}(p_{b}, w) \leq T$.

Using this $A_{0}(u, C_{k})$ we define $A_{{\rm cc}, C}$ as follows.
$A_{{\rm cc}, C}$ simulates $A_{0}(u, C_{k})$ for all $C_{k}$ 
and all $u$.
When copies of $A_{{\rm cc}, C}$ are placed on all nodes of 
a configuration $C'$, a node $w$ of $C'$ fires if and only if 
there is a pair of $C_{k}$ and $u$ such that 
$A_{0}(u, C_{k})$ fires at $w$.

\begin{thm}
\label{theorem:th017}
The finite automaton $A_{{\rm cc}, C}$ satisfies the following conditions:
\begin{enumerate}
\item[{\rm (1)}] $A_{{\rm cc}, C}$ is a partial solution of 
${\rm g}\text{-}2{\rm PATH}$.
\item[{\rm (2)}] The domain of $A_{{\rm cc}, C}$ is 
$\{C_{0}, \ldots, C_{m-1}\}$.
\item[{\rm (3)}] $A_{{\rm cc}, C}$ fires $C'$ at time $T$ 
for any configuration $C'$ in its domain.
\end{enumerate}
\end{thm}

%\begin{pf}
\noindent
{\it Proof}. 
First we show that any $p_{u}$ in any $C_{k}$ fires with $A_{{\rm cc}, C}$.
Suppose that copies of $A_{0}(u, C_{k})$ are placed on $C' = C_{k}$.
Then $C'$ is a consistent extension of $M(p_{u}, T, C_{k})$.
Moreover, for $w = p_{u}$, we have 
$-a + {\rm d}_{C'}(p_{a}, w) = -a + {\rm d}_{C_{k}}(p_{a}, p_{u}) 
= -a + |u - a| = -a + (u - a) = 
u - 2a \leq T$ by $\lceil (u - T)/2 \rceil \leq a$.
Similarly we have 
$b + {\rm d}_{C'}(p_{b}, w) \leq T$.
Hence $A_{0}(u, C_{k})$ fires $w$, 
and consequently $A_{{\rm cc}, C}$ fires $w$.

Next we show that if a node $w$ in a configuration $C'$ fires with 
$A_{{\rm cc}, C}$ then $C'$ is one of $C_{0}, \ldots, C_{m-1}$.
If a node $w$ fires with $A_{{\rm cc}, C}$ then it fires 
with some $A_{0}(u, C_{k})$.
This means that $C'$ is a consistent extension of $M(p_{u}, T, C_{k})$ 
and hence $C_{k} \equiv_{T, p_{u}}' C'$ by 
Theorem \ref{theorem:th002}.
Therefore, $C'$ is one of $C_{0}, \ldots, C_{m-1}$.

It is obvious that if $A_{{\rm cc}, C}$ fires it fires at time $T$.

From these three facts follow the three statements (1) -- (3) 
of the theorem.
\hfill $\Box$
%\end{pf}

\medskip

This theorem implies that $A_{{\rm cc}, C}$ is a partial solution 
that fires $C$ ($=C_{0}$) at time $T = {\rm mft}_\text{g-2PATH}(C)$.

We estimate the size $N_{{\rm cc},C}$ of $A_{{\rm cc}, C}$ for g-2PATH.
$A_{0}(u, C')$ is the same as 
$A_{{\rm ref}, C'}$ that uses the values 
$j_{0} = a$, $i_{0} = b$ 
(see the construction of $A_{{\rm ref}, C}$ in 
Subsection \ref{subsection:reflection_partial_solutions}).
$A_{{\rm cc}, C}$ simulates $A_{0}(u, C')$ for all $C'$ 
such that $C \equiv_{T} C'$ and 
all $p_{u}$ in $C'$. The number of $p_{u}$ in $C'$ is 
at most $2T+1$ because ${\rm rad}(C') \leq T$.
The number of such $C'$ is at most 
$(1 + 4 + 4^{2} + \ldots + 4^{T})^{2} = (4^{T+1} - 1)^{2}/9$.
The set of states of $A_{0}(u, C')$ is 
$\{{\rm X}, {\rm Q}\} \times \{{\rm X}', {\rm Q}\} \times 
\{0, 1, \ldots, T-1, T, {\rm Q}\}$.
The third component represents the current time.
Hence the third components of all of 
$A_{0}(u, C')$ may be one common set.
Therefore we have 
\begin{equation}
N_{{\rm cc},C} \leq 
4^{(4^{T+1} - 1)^{2} / 9} (T + 2) = 2^{2^{4T + O(\log T)}}
\label{equation:eq015}
\end{equation}
and 
\begin{equation}
{\rm mss}_\text{g-2PATH}(C) \leq 2^{2^{4T + O(\log T)}}
\label{equation:eq016}
\end{equation}
($T = {\rm mft}_\text{g-2PATH}(C)$).
This upper bound (\ref{equation:eq015}) is larger than the upper bound 
(\ref{equation:eq026}) with $A_{{\rm lm}, T}$.
However, as we show later, in many cases we can modify $A_{{\rm cc}, C}$ 
and reduce its size considerably.

\begin{ex}
\label{example:ex000}
\end{ex}

\vspace*{-\medskipamount}

We construct $A_{{\rm cc}, C}$ 
for the configuration $C = p_{-11} \ldots p_{11}$ 
of g-2PATH shown in Fig. \ref{figure:fig011}.
In this case $T = {\rm mft}_\text{g-2PATH}(C) = 30$ 
(Example \ref{example:ex002}) 
and 
$C_{0} = p_{-11} \ldots p_{11}$, 
$C_{1} = p_{-11} \ldots p_{10}$, 
$C_{2} = p_{-10} \ldots p_{11}$, 
$C_{3} = p_{-12} \ldots p_{10}$, 
$C_{4} = p_{-10} \ldots p_{12}$ 
(Example \ref{example:ex011} and 
Fig. \ref{figure:fig012}).
Here both of $p_{-12}$ and $p_{12}$ denote the 
position $(0, 2)$.

$A_{{\rm cc},C}$ simulates 
$23 + 22 + 22 + 23 + 23 = 113$ finite automata 
$A_{0}(u, C_{k})$.
Hence $A_{{\rm cc},C}$ has 
$4^{113} (30 + 2) = 3.4508 \ldots \cdot 10^{69}$ states.
We explain how to modify this $A_{{\rm cc}, C}$ and reduce its size.

A finite automaton $A_{0}(u, C_{k})$ is specified by $u$ and $C_{k}$.
However, it is determined completely by $a$, $b$ 
(defined by (\ref{equation:eq007})) and $C_{k}$.
Therefore, we will denote $A_{0}(u, C_{k})$ by 
$A_{1}(a, b, C_{k})$.
In Table \ref{table:tab000} we showed the values of 
$a$, $b$ 
for each combination of $C_{k}$ and $p_{u}$ in $C_{k}$.
This table shows that the $23$ finite automata 
$A_{0}(-11, C_{0})$, $A_{0}(-10, C_{0})$, \ldots, 
$A_{0}(10, C_{0})$, $A_{0}(11, C_{0})$ for $C_{0}$ 
contain only $5$ finite automata 
$A_{1}(-11, 9, C_{0})$, 
$A_{1}(-11, 10, C_{0})$, 
$A_{1}(-11, 11, C_{0})$, 
$A_{1}(-10, 11, C_{0})$, 
$A_{1}(-9, 11, C_{0})$ 
that are different as $A_{1}(a, b, C_{0})$.
The same is true also for $C_{1}, \ldots, C_{4}$.
In total there are $5 + 3 + 3 + 4 + 4 = 19$ finite automata 
that are different as $A_{1}(a, b, C_{k})$.
We show them in Table \ref{table:tab002}.

Let $p_{r} \ldots p_{s}$ be $C_{k}$ and 
suppose that copies of $A_{1}(a, b, C_{k})$ 
are placed on positions of $C_{k}$.
Then a node $p_{z}$ fires if and only if 
${\rm d}_{C_{k}}(p_{a}, p_{z}) \leq T + a$ and 
${\rm d}_{C_{k}}(p_{b}, p_{z}) \leq T - b$.
However, this condition is equivalent to 
$a' \leq z \leq b'$, where $a'$, $b'$ are defined by
\[
a' = \max\{r, -T + 2b\}, b' = \min \{s, T + 2a\}.
\]
In other words, the nodes $p_{z}$ with $z$ in the 
interval $[a', b']$ fire.

\begin{table}[htbp]
\begin{center}
\begin{tabular}{|c|c|c|}
\hline
$A_{1}(a, b, C_{k})$ & $[a', b']$ & \\ \hline \hline
$A_{1}( -11,  9, C_{0})$ & $[-11,  8]$ & $\ast$ \\
$A_{1}( -11, 10, C_{0})$ & $[-10,  8]$ &        \\
$A_{1}( -11, 11, C_{0})$ & $[ -8,  8]$ &        \\
$A_{1}( -10, 11, C_{0})$ & $[ -8, 10]$ &        \\
$A_{1}(  -9, 11, C_{0})$ & $[ -8, 11]$ & $\ast$ \\ \hline
$A_{1}( -11,  9, C_{1})$ & $[-11,  8]$ & $\ast$ \\
$A_{1}( -11, 10, C_{1})$ & $[-10,  8]$ &        \\
$A_{1}( -10, 10, C_{1})$ & $[-10, 10]$ & $\ast$ \\ \hline
$A_{1}( -10, 10, C_{2})$ & $[-10, 10]$ & $\ast$ \\
$A_{1}( -10, 11, C_{2})$ & $[ -8, 10]$ &        \\
$A_{1}(  -9, 11, C_{2})$ & $[ -8, 11]$ & $\ast$ \\ \hline
$A_{1}( -12,  9, C_{3})$ & $[-12,  6]$ & $\ast$ \\
$A_{1}( -12, 10, C_{3})$ & $[-10,  6]$ &        \\
$A_{1}( -11, 10, C_{3})$ & $[-10,  8]$ &        \\
$A_{1}( -10, 10, C_{3})$ & $[-10, 10]$ & $\ast$ \\ \hline
$A_{1}( -10, 10, C_{4})$ & $[-10, 10]$ & $\ast$ \\
$A_{1}( -10, 11, C_{4})$ & $[ -8, 10]$ &        \\
$A_{1}( -10, 12, C_{4})$ & $[ -6, 10]$ &        \\
$A_{1}(  -9, 12, C_{4})$ & $[ -6, 12]$ & $\ast$ \\ \hline
\end{tabular}
\end{center}
\caption{The $19$ automata $A_{1}(a, b, C_{k})$ that 
are different as $A_{1}(a, b, C_{k})$.}
\label{table:tab002}
\end{table}

In Table \ref{table:tab002}, we show this interval $[a', b']$ 
for each of the $19$ finite automata.
From this table, we know that, if we use $A_{1}(a, b, C_{0})$ 
to fire nodes in $C_{0}$, only 
$A_{1}(-11, 9, C_{0})$ and $A_{1}(-9, 11, C_{0})$ are sufficient 
because the former fires $p_{-11} \ldots p_{8}$ and the latter 
fires $p_{-8} \ldots p_{11}$.
The same is true for $C_{1}, \ldots, C_{4}$.
Therefore, of the $19$ automata, only $10$ automata with 
the mark `{\rm *}' in the table are sufficient to fire all nodes of 
$C_{0}, \ldots, C_{4}$.

Moreover, some automata in these $10$ automata are the same automata.
For example, $A_{1}(-11, 9, C_{0})$ and $A_{1}(-11, 9, C_{1})$ 
are the same automata because the part $p_{-11} \ldots p_{9}$ are 
in both of $C_{0}$, $C_{1}$ and its boundary conditions in 
$C_{0}$ and $C_{1}$ are the same.
$A_{{\rm cc}, C}$ needs to simulate only the following $6$ 
finite automata:
\begin{tabbing}
\quad\quad\quad\quad \= $M_{0}$ \quad \= $A_{1}(-11,  9, C_{0})$, \\
                     \> $M_{1}$       \> $A_{1}( -9, 11, C_{0})$, \\
                     \> $M_{2}$       \> $A_{1}(-10, 10, C_{1})$, \\
                     \> $M_{3}$       \> $A_{1}(-10, 10, C_{2})$, \\
                     \> $M_{4}$       \> $A_{1}(-12,  9, C_{3})$, \\
                     \> $M_{5}$       \> $A_{1}( -9, 12, C_{4})$.
\end{tabbing}
Therefore we could reduce the size of 
$A_{{\rm cc}, C}$ from 
$3.4508 \ldots \cdot 10^{69}$ to 
$4^{6} (30 + 2) = 131,072$.
The size of the solution obtained from 
this partial solution is $6 \cdot 131,072 = 786,432$.

\begin{figure}
\begin{center}
\includegraphics[scale=1.0]{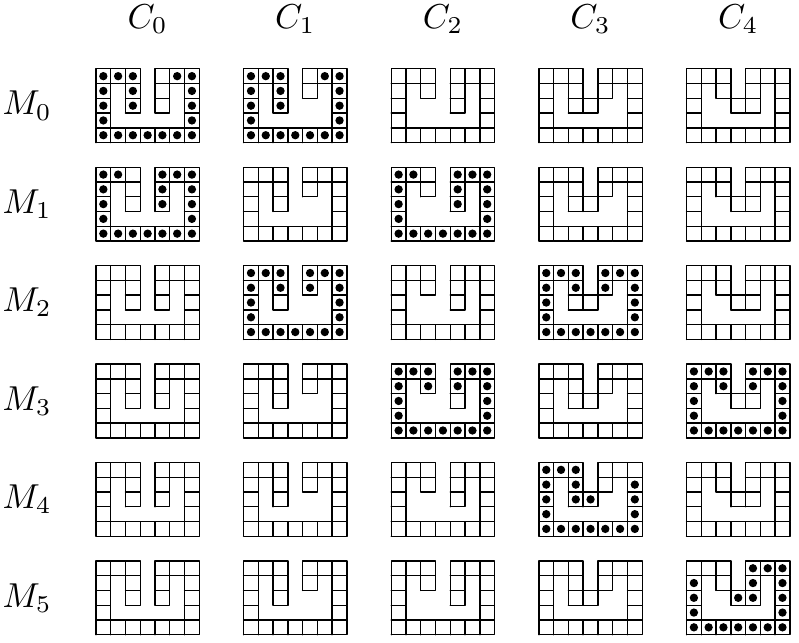}
\end{center}
\caption{Nodes that fire when $M_{i}$ is given to $C_{k}$ 
in Example \ref{example:ex002}.}
\label{figure:fig039}
\end{figure}

In Fig. \ref{figure:fig039}, for each $M_{i}$ and $C_{k}$ we show 
the nodes that fire when nodes of $C_{k}$ are copies of $M_{i}$ 
with dots.
Each node in each $C_{k}$ fires with at least one $M_{i}$.

Summarizing our results 
for the configuration $C = p_{-11} \ldots p_{11}$, 
$A_{{\rm lm}, T}$, $A_{{\rm ref},C}$, $A_{{\rm cc},C}$ 
give upper bounds $2.7236 \ldots \cdot 10^{26}$, 
$768$, and $786,432$ 
respectively for ${\rm mss}_\text{g-$k$PATH}(C)$.
Note that the domains of $A_{{\rm lm}, T}$ and 
$A_{{\rm cc}, C}$ are 
$\{C_{0}, \ldots, C_{4}\}$ (Fig. \ref{figure:fig012}) 
and that of $A_{{\rm ref}, C}$ is 
$\{C_{0}, \ldots, C_{4}, C_{8}\}$ (Fig. \ref{figure:fig018}).
Hence we may say that $A_{{\rm ref},C}$ is small 
because it does not try to exclude $C_{8}$ from its domain.
\hfill (End of Example \ref{example:ex000})

\medskip

Similarly, for the configuration $C = p_{-10} \ldots p_{11}$ 
shown in Fig. \ref{figure:fig019} (Examples \ref{example:ex003} 
and \ref{example:ex006}), 
$A_{{\rm lm},T}$ and $A_{{\rm cc},C}$ give upper bounds 
$7.3759 \ldots \cdot 10^{45}$ and 
$3,168$ respectively for ${\rm mss}_\text{g-$2$PATH}(C)$.
For this $C$ we have ${\rm mft}_\text{g-$2$PATH}(C) < \tilde{T}$ 
and $A_{{\rm ref},C}$ gives no upper bounds.

Finally we have a small comment on $k$REG and $A_{{\rm cc},C}$.
For $k$REG, $A_{{\rm lm}, T}$ is the only partial solution we know 
that fires $C$ at ${\rm mft}_\text{$k$REG}(C)$.
The idea we used to construct $A_{{\rm cc}, C}$ 
for a configuration $C$ of g-$k$PATH can be used in an essentially 
the same way to define a finite automaton $A_{{\rm cc}, C}$ 
for a configuration $C$ of $k$REG.
However, for this $A_{{\rm cc}, T}$ to be a partial solution, 
it is necessary for 
$T$, $C_{0}$, \ldots, $C_{m-1}$ to satisfy 
the condition:  for any $C_{k}$ and any $v$ in $C_{k}$, 
if $C'$ is a consistent extension of $M(v, T, C_{k})$ 
then $M(v, T, C') - M(v, T, C_{k}) = \emptyset$.
At present, this condition is true for all examples of $2$REG 
we have checked.
However, we are unable to prove that the condition is true 
for all configurations of $k$REG.

\section{Conclusions}
\label{section:conclusions}

We have the project to prove that 
the six variations $k$PATH, ${\rm g}\text{-}k{\rm PATH}$, $k$REG of FSSP 
($k = 2, 3$) have no minimal-time solutions 
under some complexity-theoretical assumptions 
using the following strategy.
For each $\Gamma$ of the variations, 
we select a complexity class ${\mathcal C}$ ($\supseteq {\rm NP}$) 
and a ${\mathcal C}$-complete set $\tilde{L}$ and 
try to achieve the following two goals:
\begin{description}
\item[Goal 1:] To show that $\tilde{L}$ is 
decidable in polynomial time using the oracle 
of ${\rm mft}_{\Gamma}(C)$.
\item[Goal 2:] To show that ${\rm mft}_{\Gamma}(C)$ 
is computable in polynomial time using a set in $\mathcal{C}$ 
as an oracle.
\end{description}
If we have achieved Goal 1, 
we have shown that if ${\rm P} \not= \mathcal{C}$ 
then $\Gamma$ has no minimal-time solutions.
Moreover, if we have achieved Goal 2 too, 
it means that this result cannot be improved further 
by our strategy.
We have several results along this line 
(\cite{Goldstein_Kobayashi_SIAM_2005,Kobayashi_TCS_2001}).
However we have achieved both of these two goals only for 
$3$PATH.

In this paper we focus on the two variations g-$k$PATH 
and as a necessary one step to achieve Goal 1 
we try to clarify what makes 
${\rm mft}_\text{g-$k$PATH}(C)$ more complex than 
${\rm mft}_\text{$k$PATH}(C)$ 
(if the former is really more complex than the latter).

We could show that the interference of left and right 
extensions of parts of configurations 
(a feature absent in $k$PATH and present in g-$k$PATH) 
is one of the factors that make 
${\rm mft}_\text{g-$k$PATH}(C)$ more complex than 
${\rm mft}_\text{$k$PATH}(C)$.
This implies that, if $\mathcal{C}$ is 
one of $\Sigma_{2}^{\rm p}$, $\Sigma_{3}^{\rm p}$, \ldots, 
${\rm PSPACE}$, we assume 
$\Delta_{2}^{\rm p} \not= \Sigma_{2}^{\rm p}$, 
and we try to prove nonexistence of minimal-time 
solutions of g-$k$PATH under the assumption 
${\rm P} \not= \mathcal{C}$ then we must utilize 
the interference of left and right extensions 
in the simulation in Goal 1.

The six variations considered in this paper are 
not widely studied.
As far as the author knows, 
papers 
\cite{%
Grasselli_1975, 
Kobayashi_Inf_Cntr_1977,
Kobayashi_TCS_1978,  
Nguyen_Hamacher_1974, 
Romani_1976%
}
before 1970s and 
\cite{Goldstein_Kobayashi_SIAM_2005, 
Kobayashi_TCS_2001, 
Kobayashi_Goldstein_UC_2005} 
after 2000s 
are all the papers on these problems 
(\cite{Roka_2000} considers different but similar variations).
However these variations deserve much more interests 
by researchers.

These variations are natural modifications of the two problems, 
(1) the original FSSP and (2) the generalized FSSP, for the 
two-dimensional and the three-dimensional 
grid spaces.
The problems to find minimal-time solutions for 
the two FSSP's (1), (2) were completely solved in 1960s 
by E. Goto (\cite{Goto_1962}), 
A. Waksman (\cite{Waksman_1966}), 
R. Balzer (\cite{Balzer_1967}), and 
F. R. Moore and G. G. Langdon (\cite{Moore_Langdon_1968}).
However, once thus modified for grid spaces, 
we have six problems that seem to be very difficult and deep.
The original two problems (1), (2) were 
purely automata-theoretic.
However, the modified six problems are ones where 
automata theory and complexity theory interact closely. 
They merit extensive study by researchers in both of 
these two areas.

The problems considered in this paper have one interesting 
feature.
In complexity theory we have many results on finite automata 
(see, for example, \cite{Garey_Johnson}). 
However, mostly these results are on decision problems 
concerning finite automata 
such as deciding equivalence of two finite automata.
On the contrary, in this paper we try to 
determine existence or nonexistence of 
some specific finite automata, 
a purely automata-theoretic problem, 
with the help of complexity theory.
As far as the author knows, our results on 
the six variations (\cite{Goldstein_Kobayashi_SIAM_2005, 
Kobayashi_TCS_2001}) 
are the first of this type.

In Section \ref{section:minimum_state_numbers} 
we introduced the problem to design a solution 
that fires $C$ at time ${\rm mft}_{\Gamma}(C)$ for each 
fixed given configuration $C$ 
and defined ${\rm mss}_{\Gamma}(C)$ to be the smallest 
size of such solutions.
To estimate this value ${\rm mss}_{\Gamma}(C)$ 
is also a challenging problem.

For all configurations of $k$PATH and some configurations 
of g-$k$PATH we know how to design small such solutions 
(the reflection partial solutions $A_{{\rm ref}, C}$).
Moreover, for many specific configurations we can design 
small solutions using ad hoc simplifications 
(for example, the consistency checking partial solutions 
$A_{{\rm cc}, C}$ for g-$k$PATH).
However, as a general method to design such solutions 
for all configurations of g-$k$PATH and $k$REG, 
the local map partial solution $A_{{\rm lm},T}$ is 
the only method we know and the size of solutions 
obtained by this method is at least $2^{(1/3)T^{2} - O(T)}$ for $k = 2$ 
(the value (\ref{equation:eq021}) multiplied by $296$)).
This value is $2.9547\ldots \cdot 10^{31}$ even for 
a modest size configuration $C$ of $2$REG such that 
${\rm mft}_\text{$2$REG}(C) = 20$.
We need general methods that give reasonably small values 
as upper bounds for ${\rm mss}_{\Gamma}(C)$.

To obtain nontrivial lower bounds for ${\rm mss}_{\Gamma}(C)$ 
will be very difficult because even a 
lower bound $\log^{*} n$ for ${\rm mss}_{\Gamma}(C)$ 
immediately implies nonexistence of minimal-time solutions of $\Gamma$ 
without any assumptions in complexity theory 
($n$ is the number of positions in $C$ and 
$\log^{*} n$ is the minimal number of $2$'s such that 
$2^{2^{.^{.^{.^{2}}}}} \geq n$).

\bibliographystyle{plain}
\bibliography{ms}

\medskip

\appendix

\noindent
{\Large\bf Appendix}

\section{Basic notions and notations in complexity theory}
\label{section:complexity}

In this appendix 
we summarize notions and notations 
in complexity theory 
(the theory of computational complexity) 
that are used in this paper.
For more details, see \cite{Balcazar, Hemaspaandra_Ogihara, Papadimitriou}, 
for example.

\medskip

\noindent
(I) Standard notions and notations

\medskip

${\rm P}$ is the class of sets (of words of some alphabet) 
that are decidable by 
polynomial-time deterministic Turing machines.
${\rm NP}$ is the class of sets that can be accepted by 
polynomial-time nondeterministic Turing machines.
${\rm PSPACE}$ and ${\rm NPSPACE}$ are similarly defined 
by replacing ``polynomial-time'' with ``poly\-nomial-space'' 
in the definitions of ${\rm P}$ and ${\rm NP}$.
We know that ${\rm PSPACE} = {\rm NPSPACE}$ and hence 
we use only the notation ${\rm PSPACE}$.
${\rm EXP}$ is the class of sets that are decidable 
by $2^{n^{c}}$-time deterministic Turing machines 
for some constant $c$ (\cite{Hemaspaandra_Ogihara}).
We know that ${\rm PSPACE} \subseteq {\rm EXP}$.

We define the classes of sets 
$\Sigma_{i}^{\rm p}$, 
$\Pi_{i}^{\rm p}$, $\Delta_{i}^{\rm p}$ ($0 \leq i$) 
(the {\it polynomial hierarchy}) 
by the induction on $i$.
First we define $\Sigma_{0}^{\rm p} = \Pi_{0}^{\rm p} = 
\Delta_{0}^{\rm p} = {\rm P}$.
For $i \geq 1$, a set $L$ is in the class $\Delta_{i}^{\rm p}$ 
if and only if there is a set $L'$ in $\Sigma_{i-1}^{\rm p}$ 
and a polynomial-time deterministic oracle Turing machine $M$ 
such that $M$ with the oracle $L'$ decides $L$.
$L$ is in $\Sigma_{i}^{\rm p}$ if and only if 
there is a set $L'$ in $\Sigma_{i-1}^{\rm p}$ and 
a polynomial-time nondeterministic oracle Turing machine $M$ 
such that $M$ with the oracle $L'$ accepts $L$.
$L$ is in $\Pi_{i}^{\rm p}$ if and only if its complement 
$\overline{L}$ is in $\Sigma_{i}^{\rm p}$.
Finally we define ${\rm PH} = \cup_{0 \leq i} \Sigma_{i}^{\rm p}$.
${\rm coNP}$ is the class of complements of sets in ${\rm NP}$.

We have $\Delta_{1}^{\rm p} = {\rm P}$, 
$\Sigma_{1}^{\rm p} = {\rm NP}$, 
$\Pi_{1}^{\rm p} = {\rm coNP}$, 
and 
\[
{\rm P} \subseteq {\rm NP} \subseteq \Delta_{2}^{\rm p} 
        \subseteq \Sigma_{2}^{\rm p} \subseteq \Delta_{3}^{\rm p} 
        \subseteq \Sigma_{3}^{\rm p} \subseteq \ldots 
        \subseteq {\rm PH} \subseteq {\rm PSPACE},
\]
\[
{\rm P} \subseteq {\rm coNP} \subseteq \Delta_{2}^{\rm p} 
        \subseteq \Pi_{2}^{\rm p} \subseteq \Delta_{3}^{\rm p} 
        \subseteq \Pi_{3}^{\rm p} \subseteq \ldots 
        \subseteq {\rm PH} \subseteq {\rm PSPACE}.
\]
We do not know whether ${\rm P} = {\rm PSPACE}$ or not.
Hence, for any two classes in these two sequences we do not 
know whether they are the same or not.
However we know that ${\rm P} \not= {\rm EXP}$.

The statement ${\rm P} \not= {\rm NP}$ is stronger than 
the statement ${\rm P} \not= \Delta_{2}^{\rm p}$ 
and hence a result of the form 
``${\rm P} \not= \Delta_{2}^{\rm p} \Longrightarrow \mathcal{A}$'' 
is better than (and is an improvement of) the result 
``${\rm P} \not= {\rm NP} \Longrightarrow \mathcal{A}$.''
We have similar comments on other statements 
${\rm P} \not= \Sigma_{2}^{\rm p}$,
${\rm P} \not= \Delta_{3}^{\rm p}$, \ldots, 
${\rm P} \not= {\rm PH}$, 
${\rm P} \not= {\rm PSPACE}$ too.
(Remember that, as we wrote in 
Subsection \ref{subsection:problems_and_motivations}, 
we say ``$\mathcal{A}$ is stronger than $\mathcal{B}$'' only to mean 
that $\mathcal{A} \Longrightarrow \mathcal{B}$ is true, 
not that both of $\mathcal{A} \Longrightarrow \mathcal{B}$ and 
$\mathcal{B} 
{\;\;\not\!\!\Longrightarrow} \mathcal{A}$ are true.)

For sets $L$, $L'$, we say that $L$ is polynomial-time many-one 
reducible to $L'$ and write $L \leq_{\rm m}^{\rm p} L'$ 
if there exists a function $f$ that is computable by 
a polynomial-time deterministic Turing machine such that 
$x \in L \Longleftrightarrow f(x) \in L'$ for any word $x$.
We say that $L$ is polynomial-time Turing reducible to $L'$ 
and write $L \leq_{\rm T}^{\rm p} L'$ 
if there exists a polynomial-time deterministic oracle 
Turing machine $M$ such that $M$ with the oracle $L'$ 
decides $L$.

For a class $\mathcal{C}$ and a set $\tilde{L}$, 
we say that $\tilde{L}$ is $\mathcal{C}$-{\it complete} if 
$\tilde{L} \in \mathcal{C}$ and 
$L \leq_{\rm m}^{\rm p} \tilde{L}$ for any $L \in \mathcal{C}$.
We say that $L$ is $\mathcal{C}$-{\it complete with 
polynomial-time Turing reducibility} 
if $\tilde{L} \in \mathcal{C}$ and 
$L \leq_{\rm T}^{\rm p} \tilde{L}$ for any $L \in \mathrm{C}$.
Many natural ${\rm NP}$-complete sets and ${\rm PSPACE}$-compete 
sets are known (\cite{Garey_Johnson}).
$\Sigma_{i}^{\rm p}$-complete sets are also known for small 
values of $i$ (\cite{Hemaspaandra_Ogihara}).
We know that if there exists a ${\rm PH}$-complete set 
then the polynomial hierarchy collapses 
(that is, ${\rm PH} = \Sigma_{i}^{\rm p}$ for some $i$).
Hence ${\rm PH}$ is unlikely to have complete sets.

\medskip

\noindent
(II) Notations that are used only in this paper

\medskip

We use some notations that are not standard.
For a class ${\mathcal C}$ that is defined by 
resource bounded deterministic Turing machines 
(possibly with oracles), 
${\mathcal C}_{\rm F}$ denotes the class of functions 
(taking words as their arguments and values) 
that are computable by Turing machines of the same type.
Classes of functions ${\rm P}_{\rm F}$, $\Delta_{i, {\rm F}}^{\rm p}$, 
${\rm PH}_{\rm F}$, ${\rm PSPACE}_{\rm F}$, ${\rm EXP}_{\rm F}$ 
are defined in this way.
For example, $\Delta_{2,{\rm F}}^{\rm p}$ is the class of 
functions computable by polynomial-time deterministic oracle Turing machines 
with an ${\rm NP}$ set oracle.

Let $F_{1}$, $F_{2}$ be computation problems 
(problems to compute values of functions 
such as ${\rm mft}_{\rm 2PATH}(C)$).
We write 
$F_{1} \leq_{\rm T}^{\rm p} F_{2}$ 
when the problem $F_{1}$ is solvable by 
a polynomial-time deterministic oracle Turing machine 
that uses the oracle for the problem $F_{2}$.
We write 
$F_{1} \equiv_{\rm T}^{\rm p} F_{2}$ when 
$F_{1} \leq_{\rm T}^{\rm p} F_{2}$ and 
$F_{2} \leq_{\rm T}^{\rm p} F_{1}$, 
and write 
$F_{1} <_{\rm T}^{\rm p} F_{2}$ when 
$F_{1} \leq_{\rm T}^{\rm p} F_{2}$ and 
$F_{2} \not\leq_{\rm T}^{\rm p} F_{1}$, 
We use these notations also for a pair of a set $L$ and 
a computation problem $F$, for example, 
$L \leq_{\rm T}^{\rm p} F$.

\section{The derivation of statements in 
Subsection \ref{subsection:main_result_and_implication}}
\label{section:facts_for_implication}

We show the derivation of some statements mentioned 
in the discussion on implications of our main result 
in Subsection \ref{subsection:main_result_and_implication}.

\medskip

\noindent
I. Proof that the assumption (1) implies 
the assumption (2) for the case $k = 3$.

Suppose that ${\rm MFT}_\text{g-$3$PATH} \leq_{\rm T}^{\rm p} 
L$ for some ${\rm NP}$ set $L$. 
Then ${\rm MFT}_\text{g-$3$PATH} \leq_{\rm T}^{\rm p} L 
\leq_{\rm T}^{\rm p} {\rm HAMPATH} \leq_{\rm T}^{\rm p} 
{\rm MFT}_\text{$3$PATH}$ 
(${\rm HAMPATH}$ is the Hamiltonian path problem 
(\cite{Goldstein_Kobayashi_SIAM_2005})) and 
this contradicts the assumption (1).

\medskip

\noindent
II. Derivation of the result (4) from the assumptions (1), (2).

It suffices to prove ${\rm MFT}_\text{g-$2$PATH} 
\not\leq_{\rm T}^{\rm p} {\rm MFT}_\text{g-$2$PATH,CNI}$. 
If ${\rm MFT}_\text{g-$2$PATH} \leq_{\rm T}^{\rm p} 
{\rm MFT}_\text{g-$2$PATH,CNI}$ 
then 
${\rm MFT}_\text{g-$2$PATH} \leq_{\rm T}^{\rm p} 
{\rm MFT}_\text{g-$2$PATH,CNI} \leq_{\rm T}^{\rm p} 
L$ for an ${\rm NP}$ set $L$ and 
this contradicts the assumption (2).

\medskip

\noindent
III. Derivation of the result (5) from the assumptions (1), (2).

It suffices to prove 
${\rm MFT}_\text{g-$3$PATH,CNI} 
\leq_{\rm T}^{\rm p} {\rm MFT}_\text{$3$PATH}$.
We have 
${\rm MFT}_\text{g-$3$PATH,CNI}$ $\leq_{\rm T}^{\rm p}$ 
$L \leq_{\rm T}^{\rm p} {\rm HAMPATH} 
\leq_{\rm T}^{\rm p} {\rm MFT}_\text{$3$PATH}$ 
for an ${\rm NP}$ set $L$ 
(\cite{Goldstein_Kobayashi_SIAM_2005}).

\section{Proof of Fact 7 in 
Section \ref{section:local_map_algorithm}}
\label{section:fact_7}

Suppose that ${\rm ai}(v, t, C)$ is safe.
If ${\rm ai}(v, t, C) = {\rm Q}$ then 
${\rm rad}(C) > t$ and hence $t$ is safe for $C$.
Suppose that 
${\rm ai}(v, t, C) \not= {\rm Q}$.
Then there exist $D_{0}, \ldots, D_{p-1}$, $v_{0}$, $w_{0}$, 
\ldots, $v_{p-1}$, $w_{p-1}$ mentioned in the definition of 
safeness of $\sigma = {\rm ai}(v, t, C)$.
Then we have $C \equiv_{t}' D_{0} \equiv_{t}' \ldots \equiv_{t}' 
D_{p-1}$ and ${\rm rad}(D_{p-1}) > t$.
Hence $t$ is safe for $C$.

Suppose that $t$ is safe for $C$.
Then there exist $C_{0}$, \ldots, $C_{m-1}$ such that 
$C = C_{0}$, $C_{0} \equiv_{t}' C_{1} \equiv_{t}' \ldots \equiv_{t}' 
C_{m-1}$, ${\rm rad}(C_{m-1}) > t$.
If $m = 1$ then ${\rm rad}(C) > t$ and there exists $v' \in C$ 
such that ${\rm ai}(v', t, C) = {\rm Q}$ and hence 
${\rm ai}(v', t, C)$ is safe.
Then ${\rm ai}(v , t, C)$ is also safe by Fact 6.
Suppose that $m \geq 2$. Then 
\begin{enumerate}
\item[$\bullet$] there exists $v_{1} \in C \cap C_{1}$ such 
that ${\rm ai}(v_{1}, t, C) = {\rm ai}(v_{1}, t, C_{1}) 
\not= {\rm Q}$, 
\item[$\bullet$] there exists $v_{2} \in C_{1} \cap C_{2}$ 
such that
${\rm ai}(v_{2}, t, C_{1}) = {\rm ai}(v_{2}, t, C_{2}) \not= {\rm Q}$, \\
\ldots
\item[$\bullet$] there exists $v_{m-1} \in C_{m-2} \cap C_{m-1}$ 
such that 
${\rm ai}(v_{m-1},$ $t,$ $C_{m-2})$ $=$ 
${\rm ai}(v_{m-1},$ $t,$ $C_{m-1})$ $\not=$ ${\rm Q}$,
\item[$\bullet$] there exists $w \in C_{m-1}$ such 
that ${\rm ai}(w, t, C_{m-1}) = {\rm Q}$ 
(and hence ${\rm rad}(C_{m-1})$ $>$ $t$).
\end{enumerate}
This means that ${\rm ai}(v_{1}, t, C)$ is safe.
(We use the list $C_{1}, \ldots, C_{m-1}$ as 
the list $D_{0}, \ldots, D_{p-1}$ in the definition of 
safeness (Fig. \ref{figure:fig006}).)
Then ${\rm ai}(v, t, C)$ is also safe by Fact 6.

\end{document}